%% file: Asgari_etal_2019.tex
\documentclass{aa}  

\usepackage{hyperref}
\usepackage{natbib}
\bibpunct{(}{)}{;}{a}{}{,}
\usepackage[T1]{fontenc}
\usepackage{ae,aecompl}
\usepackage{adjustbox}
\usepackage{multicol}        % Multi-column entries in tables
\usepackage{multirow}        % Multi-row entries in tables

\usepackage{epsfig,amstext,floatfig,alltt,fancyhdr,setspace}
\usepackage{graphicx}
\usepackage{lmodern} %correct
\usepackage{color}
\usepackage{ucs}
\usepackage{url}
\usepackage{mathrsfs}
\usepackage{amsmath} %correct
\usepackage{amssymb}
\usepackage{appendix}
\usepackage{varioref}
\usepackage{times}
\usepackage{enumerate}
\usepackage{magazin_abbrev}
\definecolor{purple}{RGB}{76, 0,153}

\def\ts{\thinspace}
\def \Eqt{Eq.\;}
\def \sect{Sect.\thinspace}
\def \fig{Fig.\thinspace}
\def \tab{Table\thinspace}
\def \App{Appendix\thinspace}

\def\D{{\cal D}}
\def\Om{{\Omega_{\rm m}}}
\def\xf{{x_{\rm f}}}
\def\yf{{y_{\rm f}}}
\def\xw{{x_{\rm b}}}
\def\yw{{y_{\rm b}}}
\def\Cov{\mathbb{C}}

\def\d{{\rm d}}

\def\MM{\tens{M}}

\hypersetup{draft} 
\begin{document} 

%%%%%%%%%%%%%%%%%%%%%%%%%%%%%%%%%%%%%%%%%%%%%%%%%%

%%%%%%%%%%%%%%%%%%% TITLE PAGE %%%%%%%%%%%%%%%%%%%

% Title of the paper, and the short title which is used in the headers.
% Keep the title short and informative.
\title{Consistent cosmic shear in the face of systematics:  a B-mode analysis of KiDS-450, DES-SV and CFHTLenS}

\author{Marika Asgari\inst{1}\fnmsep\thanks{E-mail: ma@roe.ac.uk}
		\and
		Catherine Heymans\inst{1}
		\and
		Hendrik Hildebrandt\inst{2}
	    \and
		Lance Miller\inst{3}
		\and
		Peter Schneider\inst{2}
		\and
		Alexandra Amon\inst{1,}\inst{4}
		\and
		Ami Choi\inst{5}
		\and
		Thomas Erben\inst{2}
		\and
		Christos Georgiou \inst{6}
		\and
		Joachim Harnois-Deraps\inst{1}
		\and
		Konrad Kuijken\inst{6}
          }
%\authorrunning{M. Asgari et al.}
   \institute{Scottish Universities Physics Alliance, Institute for Astronomy, University of Edinburgh, Royal Observatory,
Blackford Hill, Edinburgh, \; EH9 3HJ, U.K.
         \and
  Argelander-Institut für Astronomie, Auf dem Hügel 71, 53121 Bonn, Germany
         \and
Department of Physics, University of Oxford, Denys Wilkinson Building, Keble Road, Oxford OX1 3RH, UK
		\and
Kavli Institute for Particle Astrophysics and Cosmology and Department of Physics, Stanford University, Stanford, CA 94305, USA
	\and
Center for Cosmology and AstroParticle Physics, The Ohio State University, 191 West Woodruff Avenue, Columbus, OH 43210, USA
	\and
Leiden Observatory, Leiden University, Niels Bohrweg 2, 2333 CA Leiden, The Netherlands
             }

\date{Received XXX; accepted YYY}

\abstract{
We analyse three public cosmic shear surveys; the Kilo-Degree Survey (KiDS-450), the Dark Energy Survey (DES-SV) and the Canada France Hawaii Telescope Lensing Survey (CFHTLenS).   Adopting the `COSEBIs' statistic to cleanly and completely separate the lensing E-modes from the non-lensing B-modes, we detect B-modes in KiDS-450 and CFHTLenS at the level of $\sim 2.7\sigma$. For DES-SV we detect B-modes at the level of $2.8 \sigma$ in a non-tomographic analysis, increasing to a $5.5 \sigma$ B-mode detection in a tomographic analysis. In order to understand the origin of these detected B-modes we measure the B-mode signature of a range of different simulated systematics including PSF leakage, random but correlated PSF modelling errors, camera-based additive shear bias and photometric redshift selection bias.   We show that any correlation between photometric-noise and the relative orientation of the galaxy to the point-spread-function leads to an ellipticity selection bias in tomographic analyses. This work therefore introduces a new systematic for future lensing surveys to consider.   We find that the B-modes in DES-SV appear similar to a superposition of the B-mode signatures from all of the systematics simulated.  The KiDS-450 and CFHTLenS B-mode measurements show features that are consistent with a repeating additive shear bias. 
}


\titlerunning{Consistent cosmic shear in the face of systematics}
\authorrunning{M. Asgari et al.}
\maketitle
%%%%%%%%%%%%%%%%%%%%%%%%%%%%%%%%%%%%%%%%%%%%%%%%%

%%%%%%%%%%%%%%%%% BODY OF PAPER %%%%%%%%%%%%%%%%%

\section{Introduction}
\input{introduction.tex}
\section{Methods}
\input{Methods.tex}

\section{Data}
\input{Data.tex}
%%%
\section{Results: Survey E/B-modes}
\input{Results1.tex}
%%%%
\section{Modelling systematics}
\input{Systematics.tex}

%%%
\section{Results: Mock E/B-modes}
\input{Results2.tex}
%%%
\section{Discussion}
\input{Discussion.tex}

\section{Conclusions}
\input{Conclusions.tex}

\begin{acknowledgements}
We thank Joe Zuntz for his help with DES-SV data and {\sc cosmosis}. We also thank Benjamin Joachimi, Alex Hall, Tilman Troester, Michael Troxel and Angus Wright for useful discussions. CH, MA, AA and JHD acknowledge support from the European Research Council under grant number 647112.  This work was carried out in part at the Aspen Center for Physics, which is supported by the National Science Foundation grant PHY-1607611, where CH, HH and KK were also supported by a grant from the Simons Foundation. PS is supported by the Deutsche Forschungsgemeinschaft in the framework of the TR33 `The Dark Universe'. KK acknowledges support by the Alexander von Humboldt Foundation. HH is supported by Emmy Noether (Hi 1495/2-1) and Heisenberg grants (Hi 1495/5-1) of the Deutsche Forschungsgemeinschaft as well as an ERC Consolidator Grant (No. 770935). LM acknowledges support from STFC grant ST/N000919/1. AC acknowledges support from NASA grant 15-WFIRST15-0008.\\

Computations for the N-body simulations were performed in part on the Orcinus supercomputer at the WestGrid HPC consortium (www.westgrid.ca), in part on the GPC supercomputer at the SciNet HPC Consortium. SciNet is funded by: the Canada Foundation for Innovation under the auspices of Compute Canada; the Government of Ontario; Ontario Research Fund - Research Excellence; and the University of Toronto. \\
The KiDS-450 results in this paper are based on data products from observations made with ESO Telescopes at the La Silla Paranal Observatory under programme IDs 177.A-3016, 177.A-3017 and 177.A-3018, and on data products produced by Target/OmegaCEN, INAF-OACN, INAF-OAPD and the KiDS production team, on behalf of the KiDS consortium. \\
The CFHTLenS results in this paper are based on observations obtained with MegaPrime/MegaCam, a joint project of CFHT and CEA/IRFU, at the Canada-France-Hawaii Telescope (CFHT) which is operated by the National Research Council (NRC) of Canada, the Institut National des Sciences de l'Univers of the Centre National de la Recherche Scientifique (CNRS) of France, and the University of Hawaii. This research used the facilities of the Canadian Astronomy Data Centre operated by the National Research Council of Canada with the support of the Canadian Space Agency. CFHTLenS data processing was made possible thanks to significant computing support from the NSERC Research Tools and Instruments grant program.\\

This DES-SV results in this paper are based on public archival data from the Dark Energy Survey (DES). Funding for the DES Projects has been provided by the U.S. Department of Energy, the U.S. National Science Foundation, the Ministry of Science and Education of Spain, the Science and Technology Facilities Council of the United Kingdom, the Higher Education Funding Council for England, the National Center for Supercomputing Applications at the University of Illinois at Urbana-Champaign, the Kavli Institute of Cosmological Physics at the University of Chicago, the Center for Cosmology and Astro-Particle Physics at the Ohio State University, the Mitchell Institute for Fundamental Physics and Astronomy at Texas A\&M University, Financiadora de Estudos e Projetos, Funda{\c c}{\~a}o Carlos Chagas Filho de Amparo {\`a} Pesquisa do Estado do Rio de Janeiro, Conselho Nacional de Desenvolvimento Cient{\'i}fico e Tecnol{\'o}gico and the Minist{\'e}rio da Ci{\^e}ncia, Tecnologia e Inova{\c c}{\~a}o, the Deutsche Forschungsgemeinschaft, and the Collaborating Institutions in the Dark Energy Survey.
The Collaborating Institutions are Argonne National Laboratory, the University of California at Santa Cruz, the University of Cambridge, Centro de Investigaciones Energ{\'e}ticas, Medioambientales y Tecnol{\'o}gicas-Madrid, the University of Chicago, University College London, the DES-Brazil Consortium, the University of Edinburgh, the Eidgen{\"o}ssische Technische Hochschule (ETH) Z{\"u}rich,  Fermi National Accelerator Laboratory, the University of Illinois at Urbana-Champaign, the Institut de Ci{\`e}ncies de l'Espai (IEEC/CSIC), the Institut de F{\'i}sica d'Altes Energies, Lawrence Berkeley National Laboratory, the Ludwig-Maximilians Universit{\"a}t M{\"u}nchen and the associated Excellence Cluster Universe, the University of Michigan, the National Optical Astronomy Observatory, the University of Nottingham, The Ohio State University, the OzDES Membership Consortium, the University of Pennsylvania, the University of Portsmouth, SLAC National Accelerator Laboratory, Stanford University, the University of Sussex, and Texas A\&M University.
Based in part on observations at Cerro Tololo Inter-American Observatory, National Optical Astronomy Observatory, which is operated by the Association of Universities for Research in Astronomy (AURA) under a cooperative agreement with the National Science Foundation. \\

{\small {\it Author contributions:}  All authors contributed to the development and writing of this paper.  The authorship list is given in three groups:  the lead authors (MA, CH) followed by two alphabetical groups.  The first alphabetical group includes those who are key contributors to both the scientific analysis and the data products.  The second group covers those who have either made a significant contribution to the data products, or to the scientific analysis.}
\end{acknowledgements}


\bibliographystyle{aa}
\bibliography{COSEBIs} 
 
 %%%%%%%%%%%%%%%%%%%%%%%%%%%%%%%%%%%%%%%%%%%%%%%%%%

\appendix

\section{Binning bias with $\xi_{\pm}$}
\input{binning.tex}

\section{$\sigma_8 -\Omega_m$ degeneracy}
\input{Appendix_CCOSEBIs.tex}

\section{Optimising the COSEBIs B-mode null-test}
\input{ModelDistinction.tex}

\section{Supplementary data and figures}
\input{Appendix_Data.tex}

 %\onecolumn

% Don't change these lines
%\bsp	% typesetting comment
%\label{lastpage}

\end{document}

%% file: introduction.tex
\label{sec:Intro}

Weak gravitational lensing is recognised as a powerful probe of the large-scale structure of the Universe.   Its reach, however, will always be limited by the accuracy to which terrestrial and astrophysical contaminating signals can be controlled.   Known sources of astrophysical systematics include the intrinsic alignment of neighbouring galaxies {\citep[see][and references therein]{joachimi/etal:2015}} and the impact of baryon feedback when modelling the non-linear matter power spectrum \citep{semboloni/etal:2011} as well as the more subtle effect of the clustering of background `source' galaxies \citep{Schneider02}.  Known sources of terrestrial systematics arise from residual distortions resulting from uncertainty in the point-spread function (PSF) model \citep{hoekstra:2004}, biases in the adopted source redshift distributions \citep{Hildebrandt12}, object selection bias \citep{hirata/seljak:2003}, shear calibration bias \citep{heymans/etal:2006} and detector-level effects \citep{massey/etal:2014,Antilogus2014}.   As weak lensing surveys have grown in size, the list of known sources of error has also grown, with accompanying mitigation strategies \citep[see][]{mandelbaum:2017}.    This progress is impressive, but there will always be the possibility that hitherto unknown sources are contaminating the cosmic shear signals that we observe.   In this paper we therefore explore the sensitivity of the `COSEBIs' weak lensing statistic to blindly uncover a range of different contaminating signals. 

Complete Orthogonal Sets of E/B-Integrals, `COSEBIs', were defined by \citet{SEK10}.   They provide a complete set of filter functions which cleanly separate a measured cosmic shear signal into its curl-free (E-mode) and divergence-free (B-mode) distortion patterns over a finite angular range.   Weak lensing can only produce E-modes\footnote{Contributions beyond the first-order Born approximation \citep{Schneider98} and source clustering \citep{Schneider02} can produce insignificant levels of B-modes for the current generation of shear surveys.}, and as such any detected B-modes in the measured cosmic shear signal will have a non-lensing origin. The most popular statistic used in current cosmic shear analyses are the shear two-point correlation functions, $\xi_{\pm}$, (2PCFs) \citep{Jee2016,Joudaki_KiDS450, hildebrandt/etal:2017, troxel/etal:2017}.  As these direct measurements of the cosmic shear signal mix E and B modes, other methods are required in order to extract and identify any contaminating non-lensing signal through its B-mode distortion pattern.

A range of different statistics exist to filter E/B-modes in 2PCFs, for example, aperture mass statistics \citep{Schneider02dec}, $\xi_{\rm E/B}$ \citep{crittenden/etal:2002} and ring statistics \citep{schneider/kilbinger:2007}. The aperture mass statistics and $\xi_{\rm E/B}$ rely on knowing the 2PCFs at either very small angular separations (where the galaxy images blend) or very large angular scales, (beyond the surveyed area).   Using these statistics therefore results in biased estimates of E/B-modes. For the aperture mass statistic, the E/B-mode leakage is $\sim 10$ percent for a typical case \citep[][]{kilbinger/etal:2013}. Both ring and aperture mass statistics suffer from a loss of information due to their filtering method.

Alternatives to real-space estimators decompose the cosmic shear signal into its E and B-mode convergence power spectrum.  Quadratic estimators can be used  \citep{kohlinger/etal:2016}, but this method is sensitive to the modelling of the noise and is also challenging to use to estimate the power at large Fourier modes due to its computational speed \citep{kohlinger/etal:2017}.  Faster methods estimate `pseudo' power spectra where in an ideal case the E/B- power spectra can be easily separated.  Unfortunately the presence of masks mixes Fourier modes, and hence E/B-modes, making this method sensitive to the modelling of the mask \citep{Asgari16, Hikage18} .  Power spectra can also be estimated from 2PCFs, if the 2PCFs are known over all scales. In practice this is not feasible, hence the integrals over 2PCFs are truncated, which can produce biases in the estimates \citep{vanuitert/etal:2017}. Alternatives to band-power spectrum estimation from 2PCFs have also been suggested \citep{BeckerRozo14}, which attempt to minimize the information leakage from the out-of-range angular scales. {Another approach to power spectrum inference uses hierarchical Bayesian modelling \citep[]{alsing/etal:2016,alsing/etal:2017}. Although this method is not sensitive to masking effects, it is highly computationally expensive as it relays on estimating posterior density distributions for all combinations of power spectra, which in turn can produce inaccuracies in the analysis.}

In this paper we adopt the COSEBIs statistic as it is the only method that can cleanly, without loss of information, separate E and B-modes over a finite angular range from realistic lensing survey data. They are also efficient as a small number of COSEBIs modes ($\sim 5$ per tomographic redshift bin) can essentially capture the full cosmological information \citep{Asgari12}.   With data compression, using linear combinations of the tomographic COSEBIs modes that are most sensitive to the parameters to be estimated, the total number of data points can also be significantly decreased \citep{AS2014} . This compression then makes the method less sensitive to the accuracy to which the covariance matrix of the data can be estimated from numerical simulations. 

COSEBIs have been used to analyse the Canada-France-Hawaii Telescope Lensing Survey \citep[CFHTLenS,][]{kilbinger/etal:2013,asgari/etal:2017}, finding significant B-mode signals in the tomographic analysis that were not detected by a range of other systematic analyses \citep{Heymans12}.   The COSEBIs statistic is therefore more sensitive and stringent in detecting B-mode distortions.  It is not immediately apparent, however, how a COSEBIs B-mode detection can be used in order to uncover the origin of the observed non-lensing distortions.    In contrast, the $\xi_{\rm B}$ and aperture mass statistics are rather intuitive.  For example. a peak in the measured B-mode at the angular scale of the CCD chip can be readily associated with an issue on the chip-level.   It is also unclear how detected COSEBIs B-modes impact the cosmological parameters from the measured E-modes.   For example, is a significant high-order COSEBIs B-mode detection an issue, when all the cosmological information is contained in the first five COSEBIs E-modes?

By using a range of different simulated systematic errors and analysing three public weak lensing surveys, this paper explores how B-mode statistics can be used to diagnose data-related systematic errors as follows.   We describe the COSEBIs, $\xi_{\rm E/B}$ and compressed COSEBIs {(CCOSEBIs)} statistics as well as their covariance matrices in \sect\ref{sec:Method}.   In \sect\ref{sec:Data},  we introduce the three public weak lensing surveys that we analyse;  the Kilo-Degree Survey \citep[KiDS-450,][]{hildebrandt/etal:2017}, the science verification data from the Dark Energy Survey \citep[DES-SV,][]{DES2016} and CFHTLenS \citep[][]{Heymans13}, presenting a full B-mode analysis of these surveys in \sect\ref{sec:ResultsData}.   We then use mock weak lensing surveys to explore how the COSEBIs and $\xi_{\rm E/B}$ statistics respond to a range of different observationally motivated systematics, introduced in \sect\ref{sec:syssect}, with results presented in \sect\ref{sec:ResultsMock}.  We compare the results for the mocks and real data in \sect\ref{sec:Discussion} and conclude in \sect\ref{sec:Conclusions}. In \App\ref{app:binning} we discuss the biases that exist in published 2PCF analyses that arise from the angular binning of the 2PCFs.  We also show how these biases can be mitigated. \App\ref{app:CCOSEBIs} determines the $\sigma_8 - \Om$ degeneracy direction for a CCOSEBIs analysis of KiDS data. We discuss how to optimise B-mode null-tests using differing selections of the data vector in \App\ref{app:Model Distinction} and present supplementary material for the tomographic data analysis in \App\ref{sec:AppData}.

%% file: Methods.tex
\label{sec:Method}
The most familiar two-point statistics used in cosmic shear analysis are the shear two-point correlation functions, $\xi_\pm$, which
correlate $\gamma_\mathrm{t/{\times}}$, the tangential and cross components of shear, of two galaxies separated by an angle $\theta$ in the sky. They are defined as
\begin{align}
\xi_\pm(\theta) & =\langle\gamma_\mathrm{t} \gamma_\mathrm{t} \rangle (\theta)\pm 
\langle\gamma_\mathrm{\times} \gamma_\mathrm{\times} \rangle (\theta).
\end{align} 
In practice, galaxy ellipticities, $\epsilon$, are measured with differing accuracies, accounted for using weights, $w$. In this case, an unbiased estimator for $\xi_{\pm}$ is given by
\begin{equation}
\hat\xi_\pm(\theta)=\frac{\sum_{ab} w_a w_b
 \left[\epsilon_{\rm t}({\pmb x}_a)\epsilon_{\rm t}({\pmb x}_b)
 \pm\epsilon_{\times}({\pmb x}_a)\epsilon_{\times}({\pmb x}_b)\right]}
 {\sum_{ab} w_a w_b}\;,
\end{equation}
where the sum goes over all galaxy pairs in an angular bin labelled as $\theta$ (see \App\ref{app:binning} for binning choices). $w_a$ is the weight associated with the measured  ellipticity at ${\pmb x}_a$ and $\epsilon_{\rm t/\times}$ are the tangential and cross components of the measured ellipticity \citep{Schneider02dec}. Here the ellipticity is defined such that its expectation value is equal to the reduced shear, in absence of systematics \citep{Schramm95,Seitz97}.
If the ellipticity measurements require a multiplicative correction, $m$ \citep[see for example][]{Miller13}, then the correlation functions may be calibrated by dividing them with the following correction,
\begin{equation}
1+K(\theta)=\frac{\sum_{ab} w_a w_b(1+m_a)(1+m_b)}{\sum_{ab} w_a w_b} \, .
\end{equation}

Theoretically the 2PCFs can be calculated through their relation to the shear power spectrum, $P_\gamma$,
\begin{align}
 \label{eqn:xipm}
& \xi_+(\theta)=\int_0^\infty \frac{\mathrm{d}\ell\:\ell}{2\pi}\:
 \mathrm{J}_0(\ell\theta)\:P_\gamma(\ell)\;,\\ \nonumber
& \xi_-(\theta)=\int_0^\infty \frac{\mathrm{d}\ell\:\ell}{2\pi}\:
 \mathrm{J}_4(\ell\theta)\:P_\gamma(\ell)\;,
\end{align} 
where $\ell$ is the Fourier conjugate of $\theta$ and $\rm{J}_0$ and $\rm{J}_4$ are the ordinary Bessel functions of zeroth and fourth order {\citep[see][for example]{Kaiser92}}. The shear power spectrum
is in turn related to the three-dimensional matter power spectrum. This relation can be simplified assuming a flat-sky and Limber approximation \citep{Kaiser98}, although, these approximations start to fail for small $\ell$-modes {(corresponding to large scales)}. Various approximations and corrections are investigated in \cite{Kilbinger17}. 
Their "hybrid" case which is used in most of the recent cosmic shear data analysis \citep[see also][]{LoverdeAfshordi08}, can be written for redshift bins $i$ and $j$ as follows,
\begin{align}
 P_\gamma^{ij}(\ell)= \frac{9H_0^4\Omega_\mathrm{m}^2}{4c^4}
 \int_0^{\chi_\mathrm{h}}\mathrm{d}\chi\:\frac{g^i(\chi)g^j(\chi)}{a^2}\:P_\delta
 \left(\frac{\ell+1/2}{f_\mathrm{K}(\chi)},\chi\right)\;,
\end{align} 
where $H_0$ is the Hubble constant, $\Om$ is the matter density parameter, $c$ is the speed of light in vacuum, 
$a$ is the scale factor normalized to one at the present, $P_\delta$ is the 3D matter power spectrum and $\chi$ is the comoving radial coordinate. The geometric factor for redshift bin $i$, $g^i(\chi)$, is given by
\begin{equation}
g^i(\chi) = \int_{\chi}^{\chi_\mathrm{h}} \mathrm{d}\chi'\: p^i_{\chi}(\chi')\:
\frac{f_\mathrm{K}(\chi'-\chi)}{f_\mathrm{K}(\chi')}\;,
\label{eqn:gchi}
\end{equation} 
where $\chi_\mathrm{h}$ is the comoving horizon scale, $p^i_{\chi}(\chi)$ is the probability density of sources in comoving distance for redshift bin $i$ and $f_\mathrm{K}(\chi)$ is the comoving angular diameter distance, which is equal to $\chi$ for a Universe with flat spatial geometry.

{The correlation functions calculated using \Eqt\ref{eqn:xipm} need to be binned in $\theta$ before they can be compared to the measurement.}
%When we compare the theory and the estimated $\xi_\pm$, it is important to treat them in the same way. 
As we usually compress the data by binning $\xi_\pm$ into broad $\theta$-bins, we should apply the same binning to the theory, to take the functional form of the 2PCFs over the angular bin into account. {Additionally, the number of galaxy pairs is roughly proportional to their angular separation. Therefore in the binned data, 2PCFs values for larger $\theta$ contribute a larger weight to the mean signal in the bin.} In  \App\ref{app:binning} we calculate the biases introduced by binning 2PCFs data, showing that using a point estimate for the expected values of $\xi_\pm$ can produce biases of up to $\pm 10\%$ for the angular range and binning adopted in \citet{hildebrandt/etal:2017} and \citet{Troxel18}.

{In practice we need to modify the relation between the 2PCFs and shear power spectrum in \Eqt\ref{eqn:xipm}, to  accommodate any B-mode power spectra that may exist in the data,}
\begin{align}
\label{eq:xipmPower}
 \xi_+(\theta) &=\int_0^{\infty} \frac{\d \ell\, \ell}{2\pi} \rm{J}_0(\theta\ell) [P_{\rm E}(\ell)+P_{\rm B}(\ell)]\;, \\ \nonumber
 \xi_-(\theta) &=\int_0^{\infty} \frac{\d \ell\, \ell}{2\pi}  \rm{J}_4(\theta\ell) [P_{\rm E}(\ell)-P_{\rm B}(\ell)] \;,
\end{align}
where\footnote{neglecting small contributions from source clustering and higher order effects} $P_\gamma(\ell)=P_{\rm E}(\ell)$
and $P_{\rm B}(\ell)$ is the B-mode power spectrum {\citep[]{Schneider02}. In the following subsections we introduce three methods that separate E/B-modes in cosmic shear data: $\xi_{\rm E/B}$, COSEBIs and compressed COSEBIs. COSEBIs, being the most robust two-point statistic method, will be used as our primary E/B-mode separation method.}

\subsection{E/B-mode 2PCFs}
The correlation functions, $\xi_\pm$, can be separated into E/B-modes following \cite{crittenden/etal:2002} and \cite{Schneider02}, where 
\begin{equation}
\label{eq:xiEB}
\xi_{\rm E}=\frac{\xi_+(\theta)+\xi'(\theta)}{2}~~~~{\rm and}~~~~\xi_{\rm B}=\frac{\xi_+(\theta)-\xi'(\theta)}{2} \;,
\end{equation}
with
\begin{equation}
\label{eq:xiPrime}
\xi'(\theta)= \xi_-(\theta)+4\int_\theta^\infty \frac{\d\vartheta}{\vartheta}\xi_-(\vartheta)-12\theta^2\int_\theta^\infty\frac{\d\vartheta}{\vartheta^3}\xi_-(\vartheta)\;.
\end{equation}
The above definition makes $\xi_{\rm E/B}$ pure E/B-modes and hence we can write
\begin{align}
\label{eq:xiEBPower}
 \xi_{\rm E}(\theta) &=\int_0^{\infty} \frac{\d \ell\, \ell}{2\pi}  \rm{J}_0(\theta\ell) P_{\rm E}(\ell)\;, \\ \nonumber
 \xi_{\rm B}(\theta) &=\int_0^{\infty} \frac{\d \ell\, \ell}{2\pi}  \rm{J}_0(\theta\ell) P_{\rm B}(\ell)\;.
\end{align}
From equations \ref{eq:xipmPower}, \ref{eq:xiEB} and \ref{eq:xiEBPower} we can immediately 
see that for a B-mode free case $\xi_{\rm E}=\xi_+(\theta)$.

In \cite{Schneider02}, $\xi_{\rm E/B}(\theta)$ is denoted $\xi_{\rm E+/B+}(\theta)$ as they also provide an alternative definition for E/B two point correlation functions, $\xi_{\rm E-/B-}(\theta)$, in terms of integrals over $\xi_+(\vartheta)$. In that case the integrals are taken from $\vartheta=0$ up to $\vartheta=\theta$, instead. Although in both cases the integral is taken over a range of angular separations that are not observable, it is preferable to use equation \ref{eq:xiEB} since, at least for a 
B-mode free case, $\xi_-(\theta)/\theta$ is very small for large $\theta$ ($\xi_-(\theta)\propto \theta^{-3}$ at large scales).  In this case we can truncate the integrals in equation \ref{eq:xiPrime} without needing to extrapolate to infinitely large $\vartheta$. However, we may lose some B-mode information by this truncation, as there is no guarantee that the B-mode signal is negligible for large angular scales. One way to extend the integral to large angular scales that are not available in the data is to use the theoretical value of  $\xi_-(\theta)$ for these angular ranges. In this paper we use measurements over an angular range of $[0.5',300']$ and a theoretical $\xi_-(\theta)$ from $\theta=300'$ out to $\theta=1000'$. We find that {the inclusion of the theoretical extension of the integral} 
has less than $5\%$ effect on the largest angular bin (used in KiDS-450) centred at $~50'$ and drops to subpercent level for $\theta\lesssim 20'$.

\subsection{COSEBIs}
\label{sec:COSEBIs}
COSEBIs  (Complete Orthogonal Sets of E/B-Integrals) modes live neither in Fourier nor real space. The filter functions for COSEBIs form sets of basis functions which transform 2PCFs and shear power spectra to the COSEBIs modes. The two sets of COSEBIs basis functions are the Lin- and Log-COSEBIs filters, 
which are written in terms of polynomials in $\vartheta$ and $\ln(\vartheta)$ in real space, respectively.
In this analysis we use the Log-COSEBIs, as they require fewer modes compared to the Lin-COSEBIs 
to capture essentially all the cosmological information (see \citealt{SEK10} for a single redshift bin and \citealt{Asgari12} for the tomographic case). 

\begin{figure}
     \includegraphics[width=\hsize]{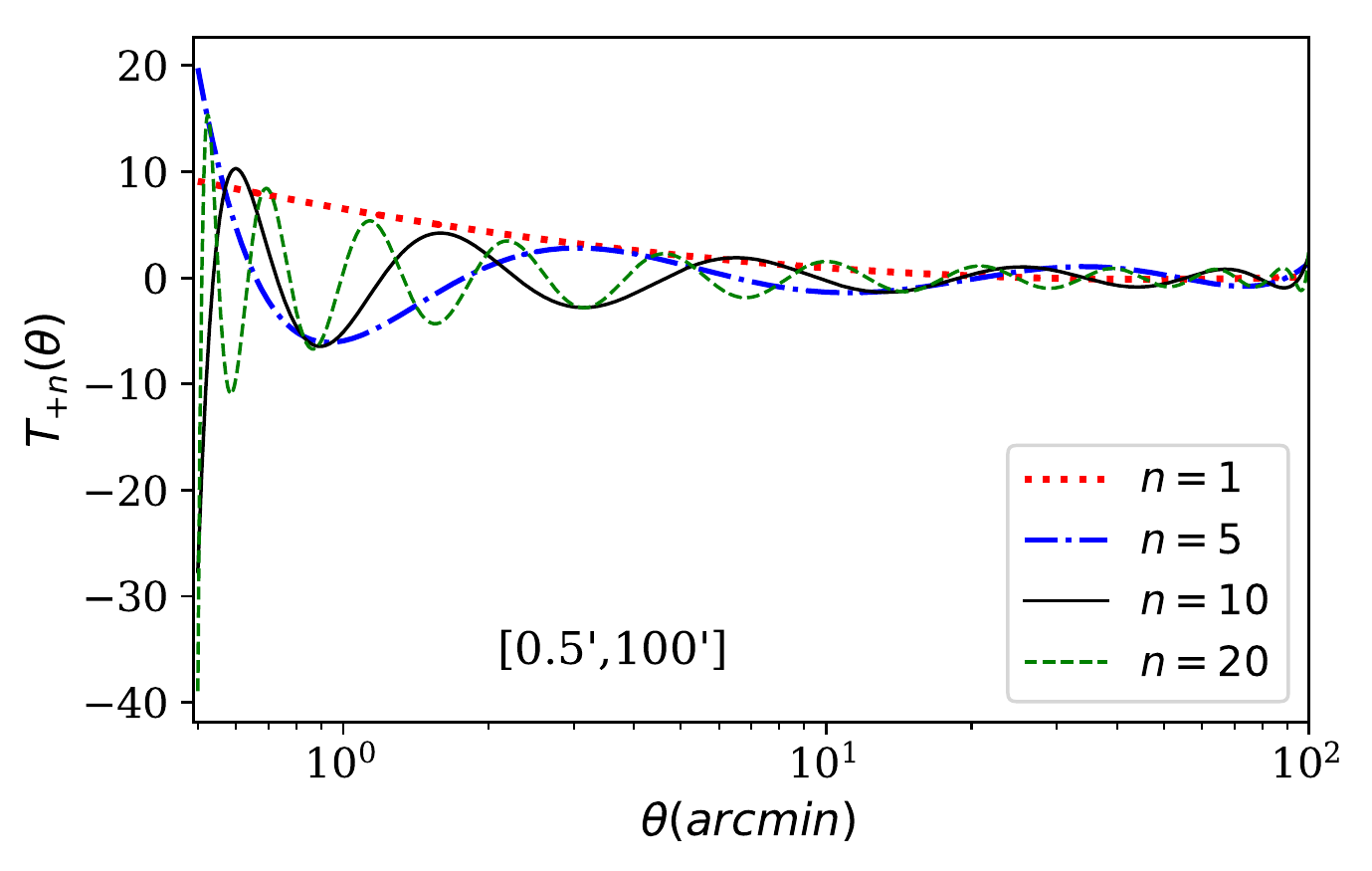} \\
     \includegraphics[width=\hsize]{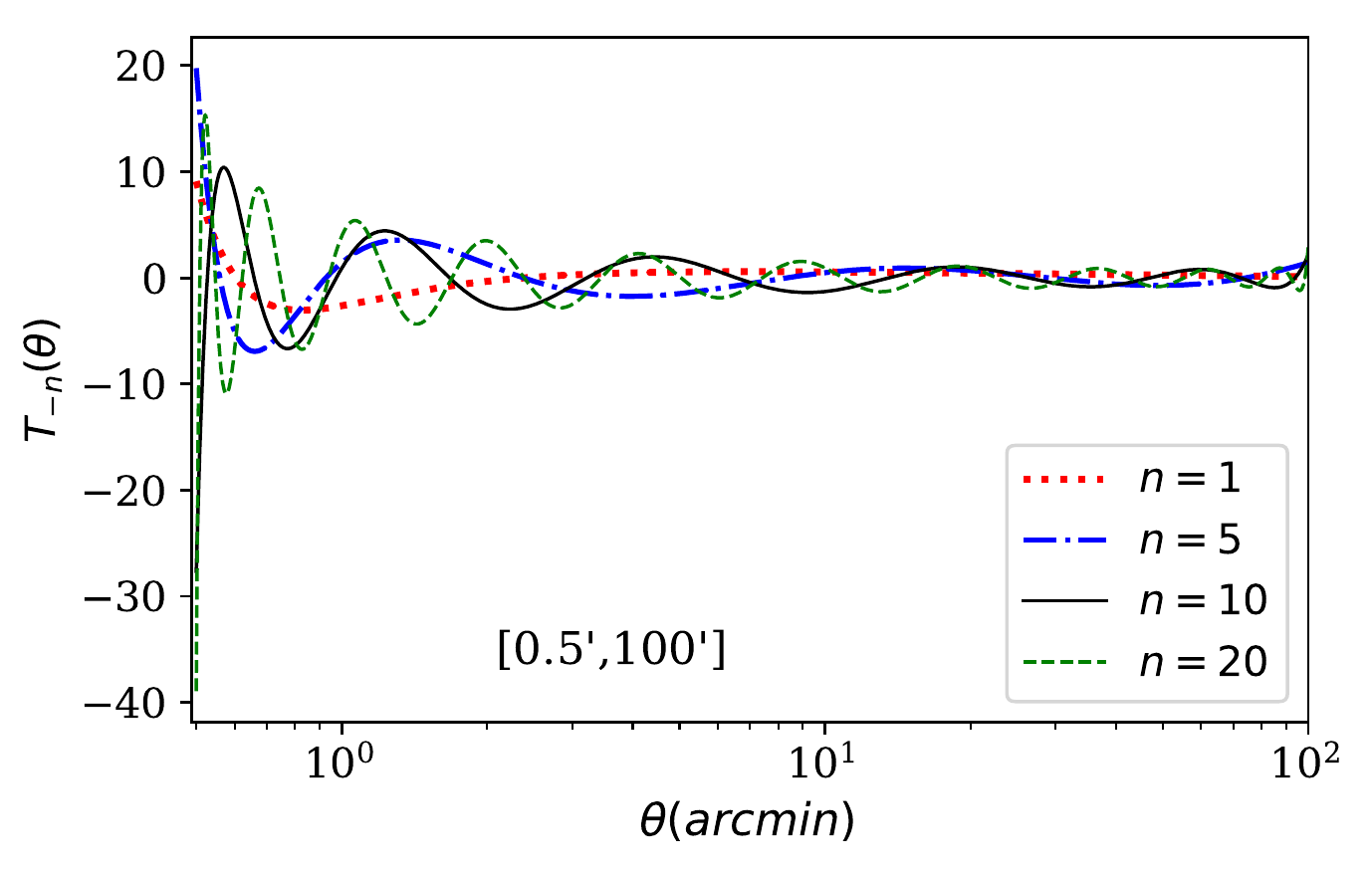}
     \caption{\small{Log-COSEBIs filter functions, $T_{\pm n}(\theta)$. These filter functions convert $\xi_\pm$ to COSEBIs E and B modes through equations~\ref{eq:EnReal} and~\ref{eq:BnReal}.  We show four example $n$-modes for each filter for the angular separation range of $[0.5', 100']$. By definition $T_{\pm n}(\theta)$ are equal to zero outside of the range of their support.}}
     \label{fig:Tpm}
 \end{figure}

The COSEBIs can be written in terms of the 2PCFs as
\begin{align}
\label{eq:EnReal}
 E_n^{(ij)} &= \frac{1}{2} \int_{\theta_{\rm min}}^{\theta_{\rm max}}
 \d\vartheta\,\vartheta\: 
 [T_{+n}(\vartheta)\,\xi^{(ij)}_+(\vartheta) +
 T_{-n}(\vartheta)\,\xi^{(ij)}_-(\vartheta)]\;, \\
 \label{eq:BnReal}
 B_n^{(ij)} &= \frac{1}{2} \int_{\theta_{\rm min}}^{\theta_{\rm
     max}}\d\vartheta\,\vartheta\: 
 [T_{+n}(\vartheta)\,\xi^{(ij)}_+(\vartheta) -
 T_{-n}(\vartheta)\,\xi^{(ij)}_-(\vartheta)]\;,
\end{align} 
where $E_n^{(ij)}$ and $B_n^{(ij)}$ are the E and B-mode COSEBIs for redshift bins $i$ and $j$, and $n$, a natural number, is the order of the COSEBIs modes.  $T_{\pm n}(\vartheta)$ are the COSEBIs filter functions, \citep[given in equations 28 to 37 in][]{SEK10}.   These are oscillatory functions with $n+1$ roots in their range of support, as shown in \fig\ref{fig:Tpm}.  
{Therefore, the COSEBIs modes with larger $n$ values have more oscillations in their range of support and can pick up features in the 2PCFs that appear as smaller-scale variations, compared to the modes with small $n$ and few oscillations. These small $n$-modes are more sensitive to the larger-scale variations in the input $\xi_\pm$ or the overall behaviour of these functions.}

\begin{figure}
     \includegraphics[width=\hsize]{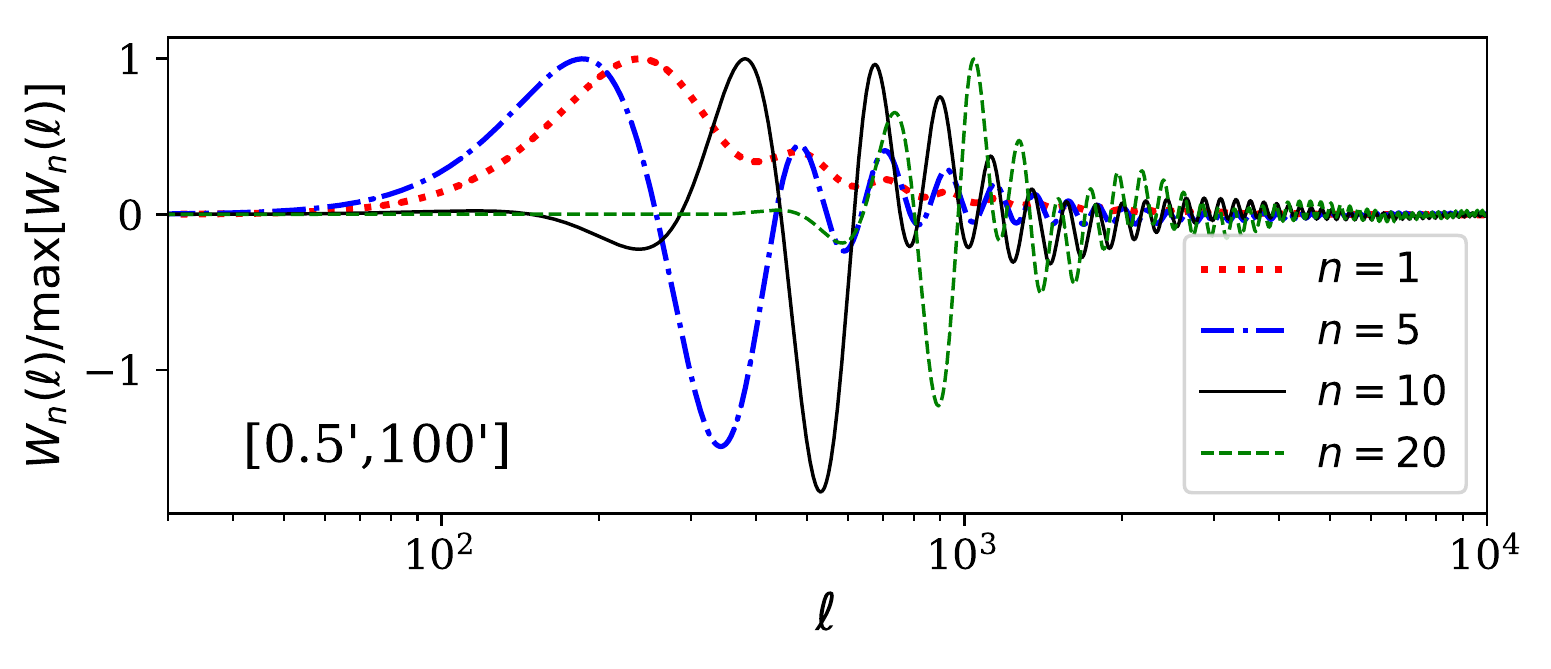}
     \caption{\small{Log-COSEBIs weight functions, $W_{n}(\ell)$, normalized to their maximum value. These weight functions convert E and B shear power spectra to COSEBIs modes through equation~\ref{eq:EnBnFourier}. Four example $n$-modes are shown for the angular range of $[0.5',100']$.  }}
     \label{fig:Wn}
 \end{figure}

The E/B-COSEBIs can also be expressed as a function of the convergence power spectra,
\begin{align}
\label{eq:EnBnFourier}
E_n^{(ij)} &= \int_0^{\infty}
\frac{\d\ell\,\ell}{2\pi}P^{(ij)}_{\mathrm{E}}(\ell)\,W_n(\ell)\;,\\ \nonumber
B_n^{(ij)} &= \int_0^{\infty}
\frac{\d\ell\,\ell}{2\pi}P^{(ij)}_{\mathrm{B}}(\ell)\,W_n(\ell)\;,
\end{align} 
where $P^{(ij)}_{\mathrm{E(B)}}$ are the E(B)-mode convergence power spectra and
the $W_n(\ell)$ are the Hankel transforms of $T_{\pm n}(\vartheta)$,
\begin{align}
\label{Wn}
W_n(\ell) & =  \int_{\vartheta_{\rm{min}}}^{\vartheta_{\rm{max}}}\d\vartheta\:
\vartheta\:T_{+n}(\vartheta) \rm{J}_0(\ell\vartheta)\;, \nonumber \\ 
& = \int_{\vartheta_{\rm{min}}}^{\vartheta_{\rm{max}}}\d\vartheta\:
\vartheta\:T_{-n} (\vartheta) \rm{J}_4(\ell\vartheta)\;.
\end{align} 
Figure\ts\ref{fig:Wn} shows the $W_n(\ell)$ functions corresponding to the $T_{\pm n}(\theta)$ filters shown in \fig\ref{fig:Tpm}. 
The first peak in $W_n(\ell)$ is set by the value of $\vartheta_{\rm max}$ and $n$. As can be seen, the higher order $W_n$ pick up more power from larger $\ell$.  We use \Eqt \ref{eq:EnBnFourier} to calculate the theoretical value of the E-mode COSEBIs as theories, in general, give their predictions in terms of the power spectrum.
However, in practice the shear 2PCFs are more straightforward to measure from data, hence
\Eqt \ref{eq:EnReal} and \Eqt \ref{eq:BnReal} are used to calculate the E/B-mode COSEBIs from 
data and simulations. To evaluate these integrals in the angular range of $[0.5',100']$ we use $4\times 10^5$ linear angular bins \citep[see][for a discussion on optimising the number of bins for this type of analysis]{asgari/etal:2017}.   
 
\subsection{CCOSEBIs}
\label{sec:CCOSEBIs}
We use the data compression method of \cite{AS2014} to explore the effect of systematics on cosmological parameter estimation, as this method is informed by the sensitivity of the data to the parameters. This method, which can be applied to any statistic,  reduces the number of data points, which is important to minimise errors when estimating covariance matrices from simulations \citep[see][for example]{Hartlap07}. 

To compress COSEBIs we need to have an estimate for their inverse covariance matrix (see \sect\ref{sec:Cov}), as well as their first and second-order derivatives with respect to the parameters to be measured. We then linearly combine the COSEBIs modes using the sensitivity of each mode to the given parameter(s) as their coefficient. For the first-order compressed E-COSEBIs we have,
\begin{equation}
E^{\rm c}_\mu=\sum_{n,m=1}^{n_{\rm max}} \frac{\partial E_m}{\partial \mu} (\Cov^{-1})_{mn} E_n\;,
\label{eqn:Ec}
\end{equation}
where $\mu$ is a cosmological parameter, $\Cov^{-1}$ is the inverse covariance matrix of $E_n$ and $n_{\rm max}$ is the number of COSEBIs modes considered in the compression. This first order compression is equivalent to a Karhunen-Loeve compression where the covariance matrix is known \citep[see][for example]{Tegmark97}, but using the first order compression alone can result in a loss of information when the covariance matrix estimate is inaccurate.   We therefore follow \cite{AS2014} by adding the following second-order compressed quantities to the data,
\begin{equation}
E^{\rm c}_{\mu\nu}=\sum_{n,m=1}^{n_{\rm max}} \frac{\partial^2 E_m}{\partial \mu\partial\nu} (\Cov^{-1})_{mn} E_n\;,
\end{equation}
where $\nu$ is a second cosmological parameter and second order derivatives of $E_n$ are taken. 
In short, we can write both first and second-order CCOSEBIs as the following matrix equation,
\begin{equation}
\label{eqn:EcBc}
\boldsymbol{E}^{\rm c}=\boldsymbol{\Gamma}\boldsymbol{E}~~~~~{\rm and} ~~~~~ \boldsymbol{B}^c=\boldsymbol{\Gamma}\boldsymbol{B}\;,
\end{equation}
where the elements of the compression matrix, $\boldsymbol{\Gamma}$, are formed from combinations of the derivatives of $E_n$ with respect to the parameters and their inverse covariance matrix.   

\subsection{Covariance matrix}
\label{sec:Cov}

To quantify the significance of the B-modes measured from the data or in the simulations, we need to know the covariance matrix of the data vector. Aside from currently negligible physical effects that can produce B-modes (discussed in \sect\ref{sec:Intro}) and in the absence of systematics, we expect any observed B-modes to be consistent with random noise arising from galaxy shape-noise.  Therefore, to calculate the B-mode covariance, we assume that they are only due to noise and find the covariance matrix for each of the above statistics.
 
Assuming a field of galaxies with ellipticities randomly picked from a Gaussian with zero mean and $\sigma^2_{\epsilon}$ variance, we can write the covariance of the 2PCF as
\begin{align}
\label{eq:CovGauss2PCFs}
\langle \xi_\pm^{ij,\rm noise}(\theta) \xi_\pm^{kl,\rm noise}(\vartheta) \rangle
=\frac{\sigma_{\epsilon}^4}{2 N^{ij}_{\rm pair}(\theta)}\delta_{\theta\vartheta}[\delta_{ik}\delta_{jl}+\delta_{il}\delta_{jk}]\;,
\end{align}
where $N^{ij}_{\rm pair}(\theta)$ is the number of galaxy pairs, in redshift bin pair $ij$, within an angular separation bin with the label $\theta$. The Kronecker symbols, $\delta_{ij}$ and $\delta_{\theta\vartheta}$, are equal to unity if their arguments are equal and are otherwise zero \citep[for example see Eq.\,34 in][]{JoachimiSE08}.

An approximation for $N_{\rm pair}(\theta)$ can be determined by calculating the number of pairs in an infinite field, scaled by the true finite field area, $A$, where
\begin{equation}
\label{eqn:Npairapprox}
 N_{\rm pair}^{ij}(\theta)^{\rm approx} = 2\pi A\, \theta\,\Delta\theta\, \bar{n}^i_{\rm gal}\bar{n}^j_{\rm gal} \, ,
\end{equation}
and $\bar{n}^i_{\rm gal}$ is the mean number density of galaxies in redshift bin $i$.   This approximation fails, however, as it does not account for intricate small-scale survey geometry, source clustering or any variable depth effects.  Furthermore, as we get closer to the field size, it does not account for the pairs of galaxies which are lost due to the discontinuities in the observed field \citep[see for example][]{JoachimiSE08}.
As the significance of any measured B-modes is determined entirely by the shot-noise, we therefore choose to use a direct measurement of $N_{\rm pair}(\theta)$ from the data.  We follow the method of \cite{Schneider02dec}, who determine the full covariance matrix for 2PCFs for a weighted ellipticity field, to find the shape-noise-only term of the covariance matrix, with the number of galaxy pairs given as
 \begin{equation}
  \label{eqn:Npair}
 N_{\rm pair}(\theta)= \frac{(\sum_{ab} w_a w_b)^2}{\sum_{ab} w_a^2 w_b^2}\;.
 \end{equation}
Here the sums are over galaxies in the given angular separation bin. Determining $N_{\rm pair}$ from \Eqt\eqref{eqn:Npair} instead of the approximation in  \Eqt\eqref{eqn:Npairapprox}, enlarges the covariance at large scales where there are fewer pairs of galaxies due to geometry effects.  On small scales where variable depth and source clustering become important, the covariance is decreased.  

Inserting \Eqt\eqref{eq:CovGauss2PCFs} into the following expression for the COSEBIs covariance \citep[][]{SEK10}
\begin{align}
C^{ij,kl}_{mn}&=\frac{1}{4}\int_{\theta_{\rm min}}^{\theta_{\rm max}} \d\theta\,\theta\,\int_{\theta_{\rm min}}^{\theta_{\rm max}} \d\theta'\,\theta'\\ \nonumber
&\times \sum_{\mu\nu={+,-}} T_{\mu m}(\theta)T_{\nu n}(\theta') C^{ij,kl}_{\mu\nu}(\theta,\theta')\;,
\end{align}
where $C_{\pm\pm}(\theta,\theta')$ is the covariance of $\xi_\pm$, we find the B-mode covariance for COSEBIs, 
\begin{align}
\label{eq:COSEBIsCov_dirac}
C^{ij,kl}_{mn}&= \frac{\sigma^4_{\epsilon}}{8}\int_{\theta_{\rm min}}^{\theta_{\rm max}} \d\theta\,\theta\,\int_{\theta_{\rm min}}^{\theta_{\rm max}} \frac{\d\theta'\,\theta'}{n^{ij}_{\rm pair}(\theta')} [\delta_{ik}\delta_{jl}+\delta_{il}\delta_{jk}]\\ \nonumber 
&\times\delta_{\rm D}(\theta-\theta') [T_{+m}(\theta)T_{+n}(\theta')+T_{-m}(\theta)T_{-n}(\theta')]\;,
\end{align}
where $n^{ij}_{\rm pair}(\theta)\, \d \theta=N^{ij}_{\rm pair}(\theta) $, $\delta_{\rm D}$ is the Dirac delta function and we have used $\delta_{\theta\theta'}=\delta_{\rm D}(\theta-\theta')\,\Delta\theta$ to remove the Kronecker symbol. Taking the inner integral in \Eqt\eqref{eq:COSEBIsCov_dirac} results in,
\begin{align}
\label{eq:COSEBIsCov}
C^{ij,kl}_{mn}&= \frac{\sigma^4_{\epsilon}}{8} [\delta_{ik}\delta_{jl}+\delta_{il}\delta_{jk}]\int_{\theta_{\rm min}}^{\theta_{\rm max}} \frac{\d\theta\,\theta^2}{n^{ij}_{\rm pair}(\theta)}\\ \nonumber 
&\times [T_{+m}(\theta)T_{+n}(\theta)+T_{-m}(\theta)T_{-n}(\theta)]\;.
\end{align}
We calculate the COSEBIs B-mode covariance using trapezoidal integration with fine $\theta$-bins and verified that these equations accurately predict the noise-only covariance, {by analysing 1000 shape-noise-only mock simulations. We find that for a $100$ deg$^2$ field the noise term for the COSEBIs covariance is underestimated by $~30\%$ if we use $N^{\rm approx}_{\rm pair}$ from \Eqt\eqref{eqn:Npairapprox}, while using \Eqt\eqref{eqn:Npair} recovers the measured covariance from the simulation.} 

%for a discrete $\theta$,
%%
%\begin{align}
%\label{eq:COSEBIsCov}
%C^{ij,kl}_{mn}&= \frac{(\Delta\theta)^2\sigma^4_{\epsilon}}{8}[\delta_{ik}\delta_{jl}+\delta_{il}\delta_{jk}]\\ \nonumber
%&\times\sum_{\theta} \frac{\theta^2}{N^{ij}_{\rm pair}(\theta)}
%[T_{+m}(\theta)T_{+n}(\theta)+T_{-m}(\theta)T_{-n}(\theta)]\;.
%\end{align}
%
The corresponding covariance for CCOSEBIs is simply equal to the COSEBIs covariance sandwiched between two compression matrices,
\begin{equation}
\label{eq:CCOSEBIsCov}
 \Cov^{\rm c}=\boldsymbol{\Gamma} \Cov \boldsymbol{\Gamma}^{\rm t}\;,
\end{equation}
where $^{\rm t}$ denotes a transposed matrix\footnote{The transpose is applied to the right hand $\Gamma$, since $\Gamma$ is a matrix with $p$, the number of cosmological parameters, rows and $n_{\rm max}$ columns.}. 

The covariance matrix of $\xi_{B}$ can also be calculated from \Eqt\eqref{eq:CovGauss2PCFs},
\begin{equation}
\label{eq:xiBCov}
 \langle \xi_B^{ij}(\theta) \xi_B^{kl}(\vartheta) \rangle
=\frac{\sigma_{\epsilon}^4}{4 N^{ij}_{\rm pair}(\theta)}\delta_{\theta\vartheta}[\delta_{ik}\delta_{jl}+\delta_{il}\delta_{jk}]\;.
\end{equation}
Note that the only difference between Eqs.\,\eqref{eq:xiBCov} and \eqref{eq:CovGauss2PCFs} is a factor of 2, which arises from the fact that $\xi_\pm$ depends on both E/B-modes and their associated noise, while $\xi_{\rm E/B}$ only depends on a single component, as can be seen in Eqs.\,\eqref{eq:xiEBPower} and \eqref{eq:xipmPower}. As a result, $\xi_{\rm E/B}$ is only sensitive to the noise components that resemble E/B-modes. The power spectrum of the noise can be equally divided into an E-mode and a B-mode component, and as such the noise covariance for $\xi_{\rm E/B}$ is half the amplitude of the corresponding covariance for 2PCFs.

In addition to B-modes, we show E-mode measurements for the data with error bars calculated assuming Gaussian covariances. We choose not to include the non-Gaussian and super sample terms in the error calculation which primarily affect the off-diagonal terms of the covariance matrix\footnote{\cite{Semboloni2007} find the transition between the Gaussian and non-Gaussian terms occurs at $\theta\sim 20'$.  At this scale the cosmic variance and mixed term roughly double the size of the error bars. At $\theta\sim 1'$ the non-Gaussian term is an order of magnitude larger than the Gaussian cosmic variance term, but as the noise term is dominant here the effect of the non-Gaussian term on the error bars is only $\sim 10\%$. }. As we do not analyse the E-modes in a quantitative way in this study, and use the E-mode covariances solely for plotting purposes, our chosen Gaussian treatment of the covariance is sufficient. We can write the Gaussian covariance for the E-modes in terms of three contributors,
\begin{equation}
\Cov= {\rm Cosmic\; variance}+ {\rm Mixed}+{\rm Noise},\;
\end{equation}
where the Mixed term depends on both cosmology and noise. The Noise term here is estimated in the same manner as the B-modes covariance, (Eqs.\,\ref{eq:COSEBIsCov} to \ref{eq:xiBCov}), taking all the survey effects into account. For the other two contributions, however, we assume a simple survey geometry and follow Eqs.\,(53) and (54) in \cite{JoachimiSE08}  for the covariance of power spectra and correlation functions\footnote{These two terms are the same for $\xi_+$ and $\xi_{\rm E}$.}, respectively. The Gaussian mixed and cosmic variance terms for COSEBIs covariance are given in equation 11 in  \cite{Asgari12} for the tomographic case.

%% file: Data.tex
\label{sec:Data}
We use three sets of cosmic shear catalogues that are in the public domain, KiDS-450, DES-SV and CFHTLenS. Our focus in this paper is the analysis of their B-mode signal, but we also compare the corresponding measured E-mode signals to theoretical predictions, based on the published best fitting cosmological parameters from each survey, as given in \tab\ref{tab:CosmoParam}.   This allows for the level of B-modes to be assessed, relative to the E-modes, but we leave a full E-mode cosmological parameter analysis to a future paper. 

\begin{table}
\centering
\caption{\small{The published best-fitting cosmological parameters for the surveys \citep[KiDS-450, CFHTLenS and DES-SV:][]{hildebrandt/etal:2017, Heymans13, DES15_CP}, and the simulation \citep[SLICS,][]{Harnois-Deraps/etal:2018}, that we use in this paper. $\sigma_8$ is the standard deviation of perturbations in a sphere of radius $8 h^{-1} {\rm Mpc}$ today.  $n_{\rm s}$ is the spectral index of the primordial power spectrum. $\Om$ and $\Omega_{\rm b}$ are the matter and baryon density parameters, respectively and $h$ is the dimensionless Hubble parameter. The underlying cosmology for all cases is a flat $\Lambda$CDM model with Gaussian initial perturbations.  The final column shows $A_{\rm IA}$, the amplitude of the intrinsic galaxy alignment model. DES-SV best-fit parameters are provided by {\cite{PC_Joe}}.}}
\label{tab:CosmoParam}
\resizebox{\columnwidth}{!}{
\begin{tabular}{ c   c  c  c  c  c c }
%   \cline{2-7}
         &     $\sigma_8$       & $\Om$        & $n_{\rm s}$         & $h$               & $\Omega_\mathrm{b}$  & $A_{\rm IA}$\\
  \hline
  KiDS-450 & 0.849        &  0.2478                               & 1.09                          & 0.747            & 0.0400    & 1.1 \\
  \hline
  CFHTLenS & 0.794        & 0.255                                 & 0.967                        & 0.717            & 0.0437    & $-1.18$\\
  \hline
  DES-SV & 0.745            & 0.378                                 & 0.96                          & 0.405            & 0.0440   &   2.07 \\
  \hline
    SLICS &  0.826            &  0.2905                              &  0.969                       & 0.6898          & 0.0473   & 0\\
   \hline
\end{tabular}
}
\end{table}

The theoretical predictions are calculated using {\sc cosmosis} \citep{cosmosis}\footnote{{\sc cosmosis}: bitbucket.org/joezuntz/cosmosis} with linear matter power spectra calculated with {\sc camb} \citep{camb2000,camb12}\footnote{{\sc camb}: http://camb.info}. \cite{Takahashi12} is used to model the nonlinear evolution of the matter power spectrum.  A Limber approximation is employed to estimate the lensing power spectrum as described in \sect\ref{sec:Method}.  For the intrinsic alignment of galaxies we adopt the non-linear model from \cite{Bridle07}\footnote{\texttt{bk\char`_corrected} in {\sc cosmosis}}, which is equivalent to the models used in the analysis of all three surveys.  The 2PCFs are measured from the data and the simulations using {\sc athena}\footnote{{\sc athena}: www.cosmostat.org/software/athena} \citep[]{KilbingerAthena14}.

\subsection*{CFHTLenS}
\cite{Heymans12} present the Canada-France Hawaii Telescope Lensing Survey (CFHTLenS), a completed survey with 154 square degrees of observed data in 5 photometric bands.  The public data products that we analyse here are processed by {\sc theli} \citep{Erben13}, with galaxy ellipticities measured using \emph{lens}fit \citep{Miller13} and photometric redshifts determined using the Bayesian photometric redshift code {\sc bpz} \citep{BPZ2000, Hildebrandt12}.    

The 2PCFs cosmic shear analysis for CFHTLenS is presented in \cite{kilbinger/etal:2013} and \cite{Heymans13}.  As summarised in \citet{Kilbinger17}, however, several improvements have been recognised since these publications, in particular with respect to the calibration of the photometric redshifts \citep[see for example][]{Choi15, joudaki/etal:2017} and the shear measurements \citep[see][]{Kuijken15,fenech-conti/etal:2017}.  The resulting uncertainty in these calibrations will impact the E-mode cosmological parameter constraints from this survey.  As our focus is on a B-mode analysis however, which is independent of these calibration corrections,  we choose to use the redshift distributions and calibration corrections adopted by \cite{Heymans13} for this study.

We follow \cite{Heymans13} by dividing the data into six photometric redshift bins: $ z_1\in(0.2,0.39]$, $z_2\in(0.39,0.58]$, $z_3\in(0.58,0.72]$, $z_4\in(0.72,0.86]$, $z_5\in(0.86,1.02]$ and $z_6\in(1.02,1.3]$, also including a single bin case that uses the full range of $z\in(0.2,1.3]$.  In \cite{asgari/etal:2017}, we analysed CFHTLenS using COSEBIs to find a significant level of B-modes.  We extend this analysis to explore higher modes in COSEBIs, in addition to $\xi_{\rm E/B}$, and we use an exact noise covariance (\Eqt\ref{eqn:Npair}) in contrast to our earlier work which used \Eqt\eqref{eqn:Npairapprox}.

\subsection*{KiDS-450}
The Kilo-Degree Survey (KiDS) will collect 1350 square degrees and in combination with VIKING (VISTA Kilo-degree Infrared Galaxy survey) will present data in nine photometric bands (see \citealt{Kuijken15} and \citealt{dejong/etal:2017}).   We analyse the data products released for the first 450 square degrees (KiDS-450), that has been processed by {\sc theli} \citep{Erben13} and Astro-WISE \citep{begeman/etal:2013}.   Galaxy ellipticities are measured with \emph{lens}fit \citep{Miller13} and calibrated using the image simulations described in \cite{fenech-conti/etal:2017}. The 4-band photometric redshifts are calibrated using external overlapping spectroscopic surveys \citep{hildebrandt/etal:2017} and galaxies are binned into tomographic bins using {\sc bpz}.

The KiDS-450 2PCFs cosmic shear analysis is shown in \cite{hildebrandt/etal:2017}  and \cite{ Joudaki_KiDS450, Joudaki_KiDS_2dFLenS}, with complementary cosmic shear power spectrum analyses calculated using quadratic estimators in \citet{kohlinger/etal:2017}, and integrals over 2PCFs in \citet{vanuitert/etal:2017}. All these analyses reported significant but low-level traces of B-modes in the data.  

As in the KiDS-450 cosmic shear analyses we divide the data into four photometric redshift bins: $z_1\in(0.1,0.3]$, $z_2\in(0.3,0.5]$, $z_3\in(0.5,0.7]$ and $z_4\in(0.7,0.9]$, including a single bin case that uses the full range of $z\in(0.1,0.9]$.

\subsection*{DES-SV}
 
\cite{DES2005} introduce the Dark Energy Survey (DES) project which will produce 5000 square degrees of gravitational lensing data in five bands.   The science verification data also known as DES-SV \footnote{DES-SV: http://des.ncsa.illinois.edu/releases/sva1} is the public dataset that we analyse here. The galaxy ellipticities in DES-SV are measured using {\sc ngmix} \citep{Jarvis2016} and photometric redshifts are determined using a machine learning-based pipeline, {\sc skynet} \citep{Bonnett/etal:2016}.

\citet{Becker15} present the primary cosmic shear analysis of the DES-SV data using 2PCFs along with cosmic shear power spectrum measurements \citep[also see][for the analysis of the first 1300 square degrees of DES data]{troxel/etal:2017}.   Fourier space B-mode measurements detected no significant B-modes on scales $\ell < 2500$. 

We divide the data into three photometric redshift bins following \cite{Becker15}: $z_1\in(0.3,0.55)$, $z_2\in(0.55,0.83)$ and $z_3\in(0.83,1.3)$ and also consider a single bin case that uses the full range of  $z\in(0.3,1.3)$.  In order to compare our measured E-mode signal to the published best-fitting cosmological parameters, listed in \tab\ref{tab:CosmoParam}, we also take into consideration the best-fitting DES-SV shear calibration and photometric redshift biases in our predictions, which \citet{DES15_CP} include as nuisance parameters in their fit.  For our single bin analysis of DES-SV data we adopt zero bias for the photometric redshift and the same value as the first tomographic bin for the shear calibration bias, which is similar to the average of the biases measured for the three bins  (see \tab\ref{tab:calibDES}  in \App\ref{sec:AppData}).

%% file: Results1.tex
\begin{figure*}
   \begin{center}
     \begin{tabular}{c}
     \resizebox{180mm}{!}{\includegraphics{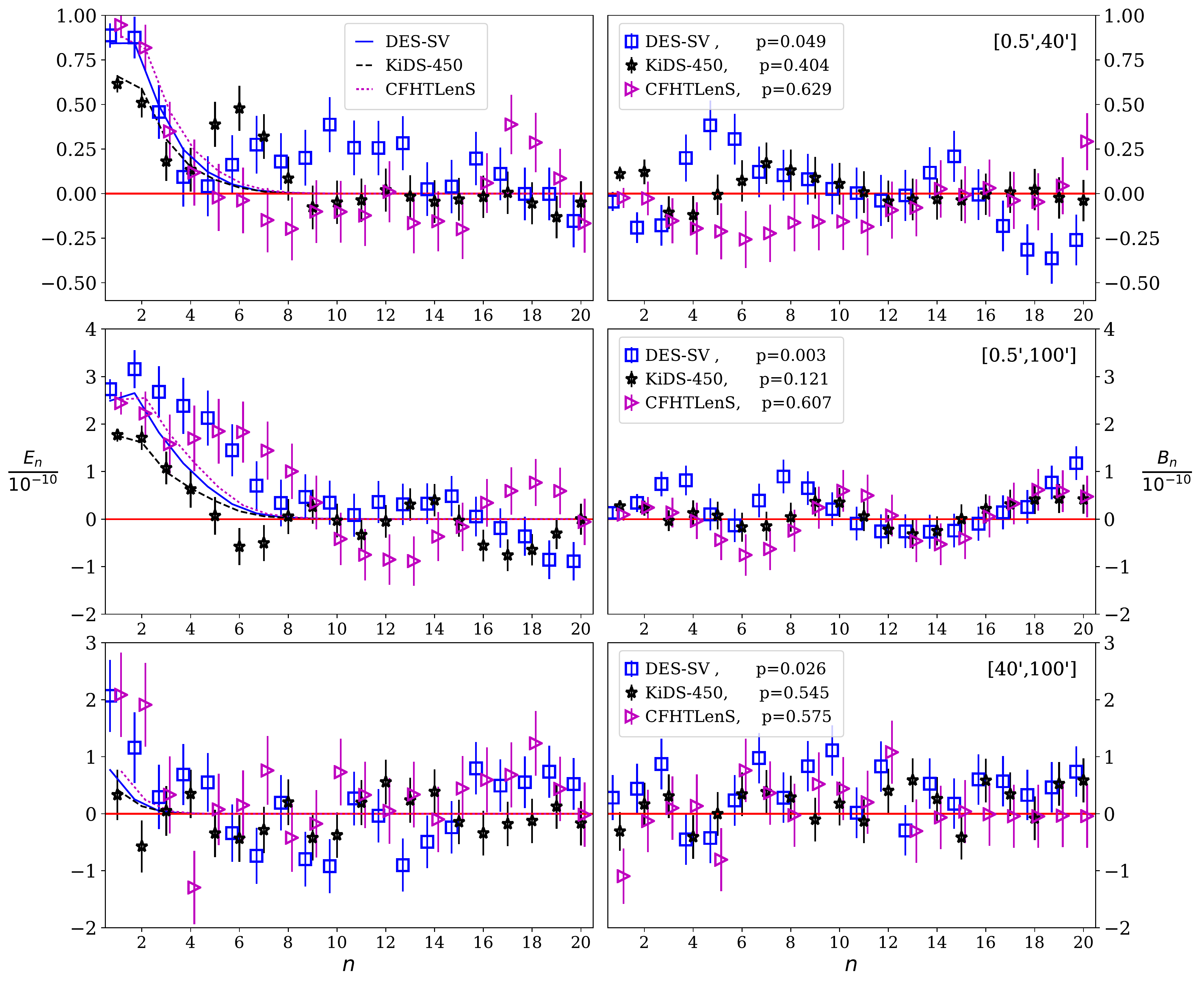}}
     \end{tabular}
   \end{center}
     \caption{\small{COSEBIs E-modes (left) and B-modes (right) for a single broad redshift bin. Results for DES-SV are shown with blue squares, KiDS-450 with black stars and CFHTLenS with magenta triangles. The angular ranges are shown for each row in the upper right corner. In addition, the significance of the B-modes is shown as $p$-values for each survey and angular range. E-mode predictions are calculated using the best fitting cosmological parameter values given in \tab\ref{tab:CosmoParam} for DES-SV (solid), KiDS-450 (dashed) and CFHTLenS (dotted). Note that COSEBIs modes are discrete and the theory values are connected to each other only as a visual aid. A zero-line is also shown for reference.}}
\label{fig:COSEBIsdata1bin}
 \end{figure*}

\label{sec:ResultsData}
 In this section we present the measured COSEBIs {({\Eqt\ref{eq:EnReal} and \Eqt\ref{eq:BnReal})}, CCOSEBIs (\Eqt\ref{eqn:EcBc}) and $\xi_{E/B}$ (\Eqt\ref{eq:xiEB})} for KiDS-450, DES-SV and CFHTLenS.  In \fig\ref{fig:COSEBIsdata1bin} we show the COSEBIs measurement for a single redshift bin encompassing the full range of redshifts adopted by each survey.   For the COSEBIs statistics we need to choose an angular range and throughout this paper we show results for three sets of angular ranges: the full angular range: $[0.5',100']$, large scales: $[40',100']$  and small scales: $[0.5',40']$.   These were chosen to span both the survey-adopted $\xi_+(\theta)$ angular ranges: KiDS-450 ($0.5' < \theta_+ < 72'$), DES-SV ($ 2' \lesssim \theta_+ \lesssim 60'$), CFHTLenS ($1.5' < \theta_+ < 35'$), whilst also probing some of the larger angular scales used in the corresponding $\xi_-(\theta)$ analysis: KiDS-450 ($8.6' < \theta_- < 300'$), DES-SV ($ 24.5' \lesssim \theta_- \lesssim 245.5'$) and CFHTLenS ($1.5' < \theta_- < 35'$). The large scale cut for $\xi_+(\theta)$ is generally employed to avoid biasing the results, when a constant additive bias term (c-term) is present in the shear catalogues. The same large scale angular-cut is not applied to $\xi_-(\theta)$, since this statistic is not sensitive to a constant c-term. COSEBIs share this insensitivity with  $\xi_-(\theta)$ and hence any measured COSEBIs B-modes that use scales beyond the maximum $\theta_+$ range are not a result of a constant c-term.

Each row in \fig\ref{fig:COSEBIsdata1bin} corresponds to one angular range, as denoted in the right panels, with E-modes on the left and B-modes on the right.   The different symbols show the results for DES-SV (squares), KiDS-450 (stars) and CFHTLenS (triangles).   Overlaid are the theoretical predictions, given the published best-fitting survey cosmological parameters from \tab\ref{tab:CosmoParam}.  We show these E-mode predictions as curves for ease of comparison, even though the COSEBIs modes are discrete.   As the COSEBIs modes are correlated to their neighbouring modes \citep[see][for plots of the covariance matrices]{Asgari12, asgari/etal:2017},  we caution that the goodness-of-fit to the model should not be deduced by simply looking at the graphs, a practice commonly known as `$\chi$-by-eye'.  Any goodness-of-fit exercise must take into account the significant correlations between the points. 

{Throughout this paper we truncate the COSEBIs measurements at $n=20$. In principle the COSEBIs can be estimated for an infinite number of modes. However as can be seen in the left hand panels of  \fig\ref{fig:COSEBIsdata1bin} the E-mode predictions are equivalent to zero for $n\gtrsim 7$. Therefore, we do not expect to gain any cosmological information from these modes. On the other hand the signal from systematic effects does not necessarily follow the same behaviour. We expect a significant signal at larger $n$-modes for certain systematics (see \sect\ref{sec:ResultsMock}). As a result we choose $n=20$ as our maximum $n$-mode, which encompasses modes that are important for both cosmological and systematic analyses. Future analyses may however want to extend their diagnostic B-mode analysis to even higher $n$-modes, depending on the signature of the systematic that they are searching for.}

Focusing first on the E-mode measurements (left panels of \fig\ref{fig:COSEBIsdata1bin}) we expect to measure signal in the lower $n$-modes and none for the modes $n \gtrsim 8$, as seen in the theoretical predictions. This arises from the fact that both the 2PCFs and shear power spectrum are relatively smooth functions with a few features that are captured, almost entirely, by the first few COSEBIs modes. Any significant detection of high-order COSEBIs modes indicates high-frequency variations in the 2PCFs, which are unexpected in a $\Lambda$CDM cosmology and therefore indicative of systematics. We remind the reader that our E-mode errors, which include both sampling variance assuming a Gaussian shear field and shot noise, will be slightly underestimated as we have not included the sub-dominant super sample and non-Gaussian contributions to the sampling variance terms (see \sect\ref{sec:Cov}).

Turning to the B-mode measurements (right panels of \fig\ref{fig:COSEBIsdata1bin}), we determine the significance of the measured B-modes using `$p$-values', for each dataset and angular range, listed in \tab\ref{tab:pvalue}. The $p$-value is equal to the probability of randomly producing a B-mode that is equally or more significant than the measured B-mode signal, given the model that B-modes are equal to zero and their distribution is Gaussian (see \App\ref{app:Model Distinction} for the mathematical definition of $p$-value).  This model is appropriate for B-modes generated from random noise. {The degrees-of-freedom here is equal to the number of COSEBIs modes (20 modes in the single redshift bin case), as the model has no free parameters to be fitted.}
 The $p$-values take into account the correlations between the COSEBIs modes. Our error analysis for the B-modes is accurate, taking into account the weighted number of galaxy pairs in each dataset. We consider the B-modes to be significant when the measured $p$-values are $p<0.01$ (highlighted in bold), corresponding to greater than $2.3 \sigma$ detection of B-modes.  We find that the B-modes of KiDS-450 and CFHTLenS are consistent with zero, finding $p > 0.1$ in all cases.  DES-SV, however, shows significant $2.8 \sigma$ B-modes with $p = 0.0026$,  when the full angular range is considered. {We discuss the complexity of linking B-mode features with E-mode features in \sect\ref{sec:ResultsMock}.}

In \tab\ref{tab:pvalue} we also list the significance of the measured COSEBIs B-modes for a tomographic analysis of the three angular ranges, using the survey-defined photometric redshift bins (see \sect\ref{sec:Data}).  The COSEBIs tomographic measurements for each survey, adopting the full angular range, are shown in \App\ref{sec:AppData}.  For all angular ranges, we find no significant COSEBIs B-modes for KiDS-450.  In contrast, for DES-SV data we find a $4.0\sigma$ detection of B-modes for the large-scale angular range that includes angular scales used in the DES-SV cosmic shear analysis.   For the full angular range, including small-scale information that was excluded from the DES-SV cosmic shear analysis, the significance of the detection increases to $5.5\sigma$.   For CFHTLenS we find a significant B-mode detection for small scales, but not at large scales.  This result is in contrast to \cite{asgari/etal:2017} who found significant CFHTLenS B-modes for large, but not small scales.  We do however recover this result if we limit our $p$-value analysis to the first 7 COSEBIs modes adopted by \cite{asgari/etal:2017}. This demonstrates that the $p$-values are sensitive to the choice of modes considered in the analysis, motivating the study of how different systematics impact different COSEBIs {E/B-modes} in \sect\ref{sec:ResultsMock}.

\begin{table}
\centering
\caption{\small{The probability of zero B-mode contamination for each survey, given the measured COSEBIs B-modes.   Results are tabulated for three different angular ranges, including the tomographic and broad single redshift bin analysis.   All $p$-values that are smaller than $0.01$ are shown in bold, corresponding to a greater than $2.3 \sigma$ B-mode detection. }}
\label{tab:pvalue}
\resizebox{\columnwidth}{!}{
\begin{tabular}{  c  c  c  c   }
%   \cline{2-7}
                  & $[0.5',40']$        &  $[0.5',100']$         &  $[40',100']$           \\
  \hline
      DES-SV, \;\;\; Single bin&  $0.049$       & $\boldsymbol{2.6\times 10^{-3}}$             & $0.026$\\
      DES-SV,       Tomography &  $\boldsymbol{9.9\times 10^{-7}}$       &   $\boldsymbol{1.5\times 10^{-8}}$           & $\boldsymbol{3.8\times 10^{-5}}$\\
  \hline
  	 KiDS-450,\;\;\; Single bin & $0.40$      & $0.12$             & $0.55$\\
  	 KiDS-450,       Tomography & $0.94$      & $0.61$             & $0.77$\\
  	 \hline
  	 CFHTLenS,\;\;\; Single bin & $0.63$      & $0.61$            & $0.58$\\
  	 CFHTLenS,       Tomography & $\boldsymbol{2.5\times 10^{-3}}$      & $0.047$            & $0.037$\\
\end{tabular}
}
\end{table} 

\begin{table}
\centering
\caption{\small{Same as \tab\ref{tab:pvalue} but for CCOSEBIs. }}
\label{tab:pvalueCCOSEBIs}
\resizebox{\columnwidth}{!}{
\begin{tabular}{  c  c  c  c   }
%   \cline{2-7}
                  & $[0.5',40']$        &  $[0.5',100']$         &  $[40',100']$           \\
  \hline
      DES-SV, \;\;\; Single bin &  $\boldsymbol{3.3\times 10^{-3}}$       & $\boldsymbol{1.1\times 10^{-3}}$             & $0.17$\\
      DES-SV,       Tomography &  $0.029$       &   $0.014$           & $\boldsymbol{2.6\times 10^{-3}}$\\
  \hline
  	 KiDS-450,\;\;\; Single bin & $\boldsymbol{4.8\times 10^{-3}}$      & $\boldsymbol{3.0\times 10^{-3}}$             & $0.56$\\
  	 KiDS-450,       Tomography & $0.013$      & $\boldsymbol{3.3\times 10^{-3}}$             & $0.51$\\
  	 \hline
  	 CFHTLenS,\;\;\; Single bin & $0.62$      & $0.55$            & $0.068$\\
  	 CFHTLenS,       Tomography & $0.70$      & $0.90$            & $0.026$\\
\end{tabular}
}
\end{table}

\begin{figure*}
   \begin{center}
     \begin{tabular}{c}
     \resizebox{70mm}{!}{\includegraphics{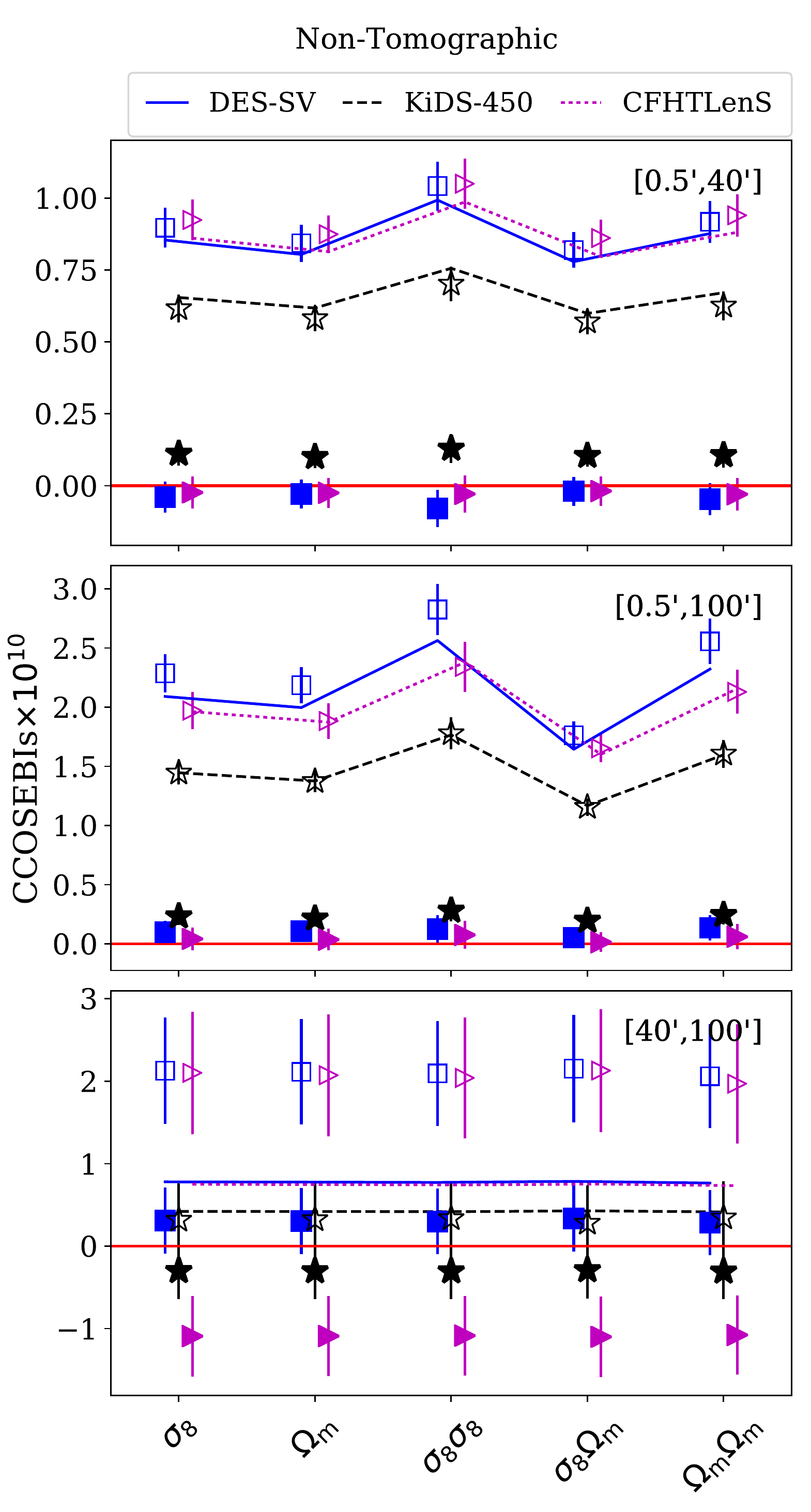}}
     \resizebox{72mm}{!}{\includegraphics{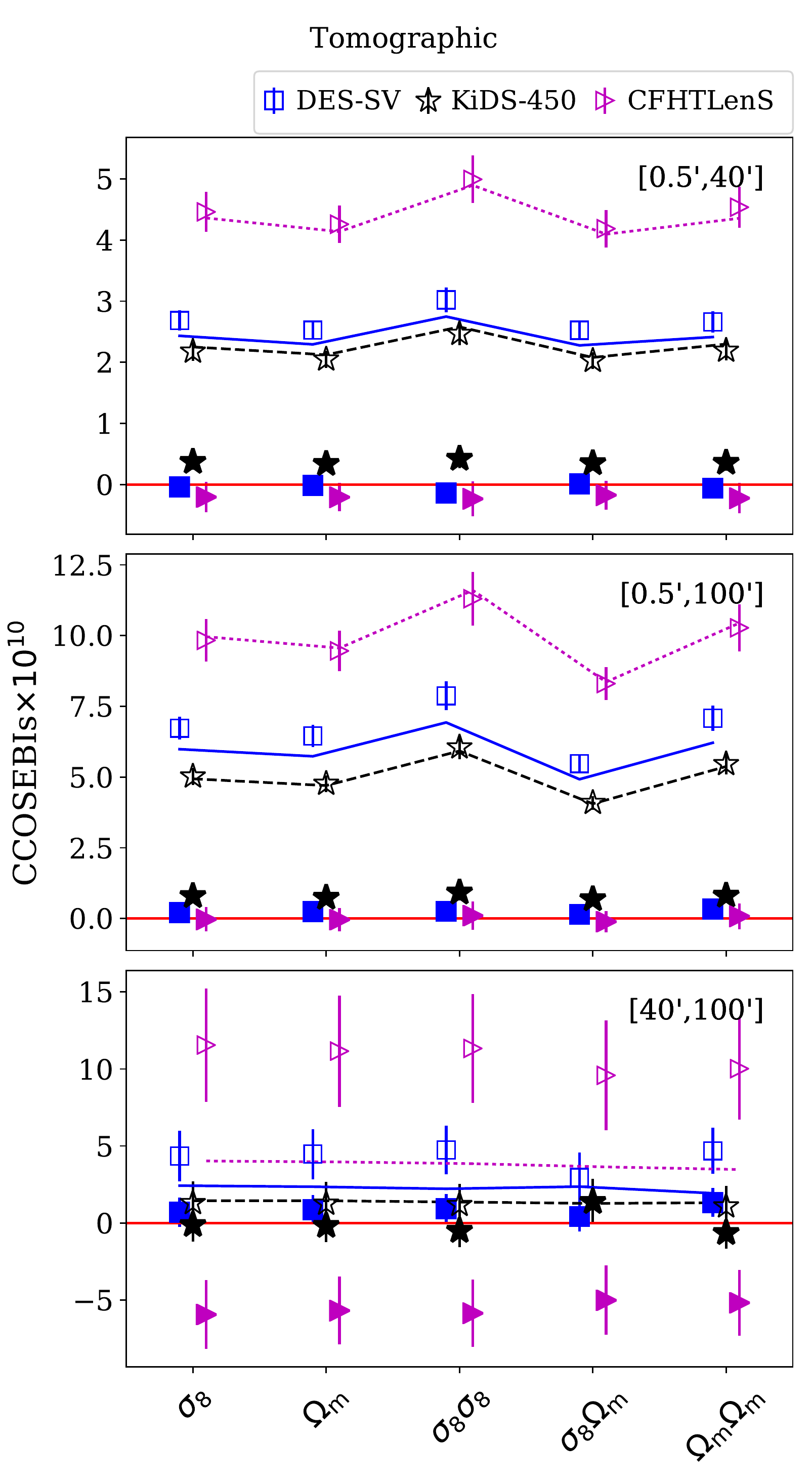}}
     \end{tabular}
   \end{center}
     \caption{\small{CCOSEBIs E and B-modes for non-tomographic (left) and tomographic (right) analyses. The E-modes are shown as empty symbols, with the B-modes shown as filled symbols, for DES-SV (blue squares), KiDS-450 (black stars) and CFHTLenS (magenta triangles). The analysis is conducted over three different angular ranges, denoted in the upper right corner of each panel.   The CCOSEBIs mode is indicated on the horizontal-axis.  E-mode predictions are calculated using the best fitting cosmological parameter values given in \tab\ref{tab:CosmoParam} for DES-SV (solid), KiDS-450 (dashed) and CFHTLenS (dotted). A zero-line is also shown for reference.}}
     \label{fig:CCOSEBIsdata}
 \end{figure*}

In \fig\ref{fig:CCOSEBIsdata} we show the measured compressed COSEBIs, where the COSEBIs modes are combined to produce a set of E-mode CCOSEBIs that, in a systematic-free dataset, are only sensitive to cosmological parameters (\Eqt\ref{eqn:EcBc}).   We compress the B-mode COSEBIs using the same compression matrix.   Cosmic shear is mainly sensitive to a combination of $\sigma_8$ and $\Om$ {\citep[see for example][]{Jain/Seljak:1997}}, hence we choose {only} these two parameters to form the CCOSEBIs modes {\footnote{When survey sizes grow and they become sensitive to other cosmological parameters such as the Hubble parameter, $H_0$ and the scalar spectral index, $n_{\rm s}$, this CCOSEBIs analysis should be extended to include these additional cosmological parameters.}}. The CCOSEBIs modes are highly correlated as $\sigma_8$ and $\Om$ are degenerate in cosmic shear data, and we hence caution the reader, again, against a `$\chi$-by-eye' analysis.  

\fig\ref{fig:CCOSEBIsdata} shows the results for a single-bin analysis (left) and a tomographic analysis (right) for the three sets of angular ranges indicated in the right panels.  The E-modes are shown as open symbols and the B-modes as filled symbols for DES-SV, KiDS-450 and CFHTLenS.  Overlaid is the theoretical expectation for the E-mode signal, shown as curves for visual aid even though the CCOSEBIs modes are discrete. The horizontal axis shows which parameter (for the first-order modes) or two parameters (for the second-order modes) the CCOSEBIs mode is sensitive to.  We highlight that CCOSEBIs represent a significant data compression, particularly in the tomographic case where, for example, we compress the 3-bin 120 data-point DES-SV analysis, and the 6-bin 420 data point CFHTLenS analysis, down to the same 5 CCOSEBIs modes. 
 
Comparing the measured E-modes with the level of B-modes in  \fig\ref{fig:CCOSEBIsdata}  we find that, aside from the largest angular range that also has the lowest signal-to-noise ratio, the E-modes are about an order of magnitude larger than the B-modes. In all panels we see that the KiDS-450 E-mode signal is lower than DES-SV and CFHTLenS, resulting from a smaller upper photometric redshift cut of $z_{\rm phot}<0.9$ in this dataset.

\tab\ref{tab:pvalueCCOSEBIs} shows the $p$-values for CCOSEBIs B-modes. {We do not show $p$-values for the E-modes since for this analysis we have not included the super sample covariance term and our E-mode errors are therefore underestimated. Readers concerned by the apparent offset between the highly correlated $[40',100']$ E-mode measurements and expectation values, should note that the similarly highly correlated B-modes, for the non-tomographic case, are all consistent with zero, even for the cases where they look inconsistent.} The significance of the B-modes is different from the values shown in \tab\ref{tab:pvalue}, where we have used the first 20 COSEBIs modes to measure the $p$-values. This apparent inconsistency is not unexpected, as the bulk of the CCOSEBIs signal comes from the first few COSEBIs modes, which contain the cosmological signal and different levels of systematics in comparison to the full set of 20 COSEBIs modes.  A good example of this difference comes in the tomographic analysis of DES-SV where we find a significant $\sim 5.5 \sigma$ non-zero B-mode signal for COSEBIs, but the CCOSEBI B-mode is not significant at $2.2 \sigma$.  This shows that the first few COSEBIs modes have a smaller contribution to the total DES-SV B-mode signal compared to the higher order modes, which can also be seen in \fig\ref{fig:COSEBIsDES-SVtomo}. KiDS-450, however, shows an insignificant B-mode signal when we consider both high and low COSEBIs modes, in contrast to a $2.7 \sigma$ B-mode detection when only the low $n$-modes are used to construct the CCOSEBIs.  As can be seen in \tab\ref{tab:pvalueCCOSEBIs}, and the upper right panel of \fig\ref{fig:COSEBIsdata1bin}, the low-$n$ B-modes for KiDS-450 data only become significant when the small angular scales are included. 

If the origin of the B-modes detected in the COSEBIs analysis was known to impact the E and B modes equally, then the CCOSEBIs result would be the most relevant for cosmic shear studies.   If the systematics impact the E and B modes differently, however, then the compressed CCOSEBIs result, focused on only low-$n$ modes, could lead to a false null-test for the survey.   It is therefore important to study how different systematics impact the full range of E and B COSEBIs, which we carry out in \sect\ref{sec:ResultsMock}, and discuss this matter further in \sect\ref{sec:Discussion}.  In \App\ref{app:Model Distinction} we also discuss how analysis choices, for example in this case tomographic or non-tomographic, COSEBIs or CCOSEBIs, can optimise or dilute the power of a B-mode null-test.

 \begin{figure}
     \includegraphics[width=\hsize]{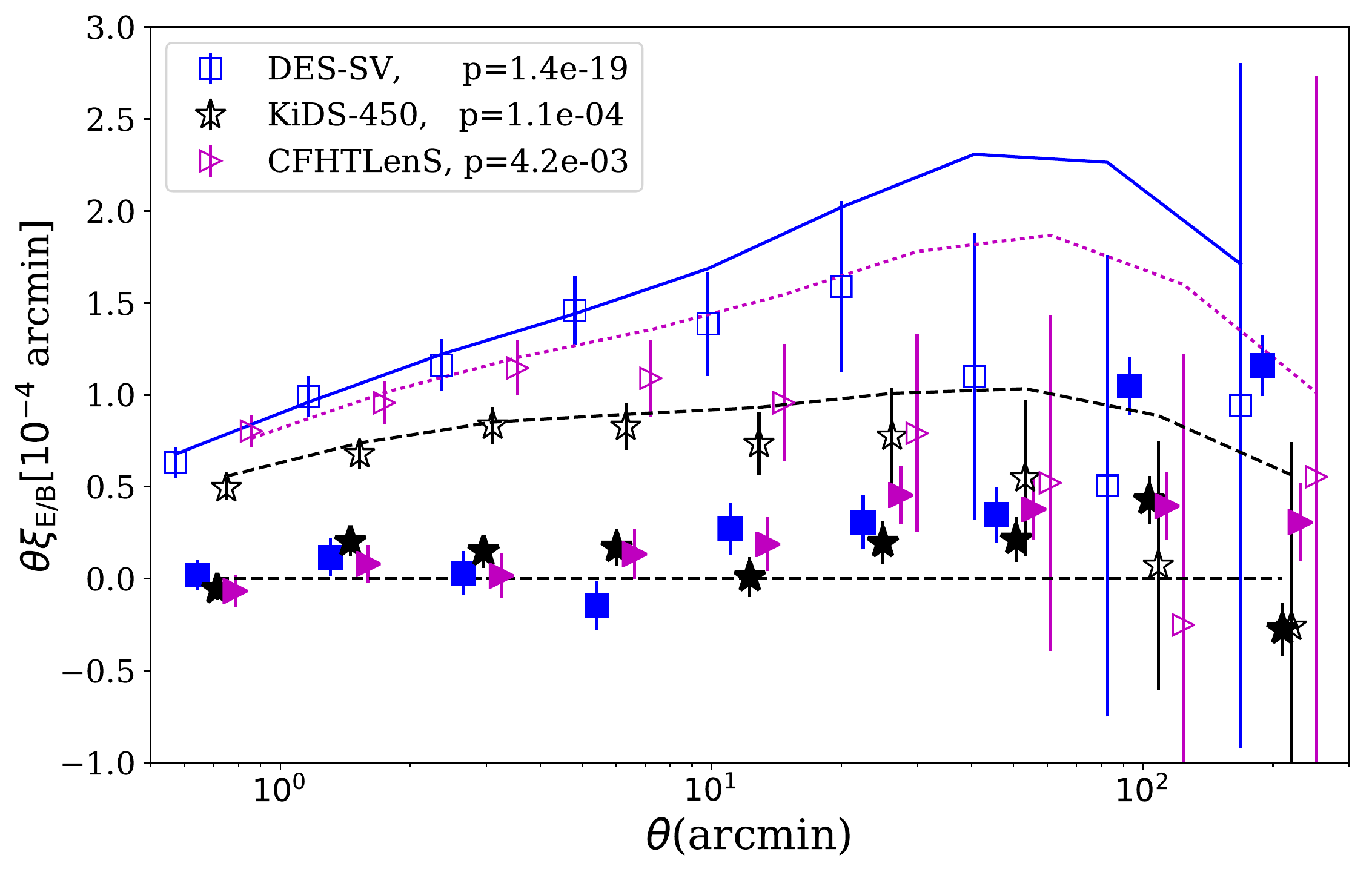}
      \caption{\small{$\xi_{\rm E}$ and $\xi_{\rm B}$ E/B-modes for a single broad redshift bin.  The E-modes are shown as empty symbols, with the B-modes shown as filled symbols, for DES-SV (blue squares), KiDS-450 (black stars) and CFHTLenS (magenta triangles). The DES-SV and CFHTLenS results are horizontally offset relative to KiDS-450 to aid visualisation.  E-mode predictions for $\xi_{\rm E}$ are calculated using the best fitting cosmological parameter values given in \tab\ref{tab:CosmoParam} for DES-SV (solid), KiDS-450 (dashed) and CFHTLenS (dotted). A zero-line is also shown for reference.  We detect significant B-modes in all cases as shown by the $p$-values, in the legend, which determine the probability of the data B-modes given a null B-mode model. }}
     \label{fig:XiData}
 \end{figure}

Finally we turn to \fig\ref{fig:XiData} which shows the measured $\xi_{\rm E/B}$ statistic across the full redshift range for each survey.  Overlaid are the best fitting theory curves for each dataset derived from the published cosmological parameters in \tab\ref{tab:CosmoParam}. 
The $p$-values corresponding to the zero B-mode model are low in all cases, as given in the legend of the figure, {with all surveys showing a tendency for increasing B-mode power and decreasing E-mode power at large scales, which,
we discuss further in \sect\ref{sec:Discussion}}.  B-modes are detected at greater than $\sim 2.6 \sigma$ for all surveys. For DES-SV the significance of the B-modes is particularly high at $\sim 9 \sigma$, but this reduces to $2.3 \sigma$, or $p=0.012$, when we select the angular scales $[4.2',72']$ which roughly correspond to the angular cuts applied to $\xi_+$ in the DES-SV cosmic shear analysis\footnote{The angular cuts used in DES-SV is variable for different redshift bins and are also different for $\xi_+$ and $\xi_-$. Since $\xi_{\rm E/B}$ are estimated using both $\xi_+$ and $\xi_-$ the decision for corresponding angular cuts is ambiguous.}.  In \App\ref{sec:AppData} we present the $\xi_{\rm E/B}$ tomographic analysis for each survey where we find a significant B-mode detection for DES-SV ($p \sim 4\times 10^{-19}$), but no detection of B-modes for KiDS or CFHTLenS ($p \sim 0.7$).

Given the required extrapolation of the data in order to calculate the $\xi_{\rm E/B}$ statistic (see \Eqt\ref{eq:xiPrime}) we emphasize that these results are, by nature, a biased measurement of $\xi_{\rm E/B}$, which may not represent the data accurately.  For this statistic, the errors on $\xi_{\rm B}$ are uncorrelated (see \Eqt\ref{eq:xiBCov}) but also biased as the integral truncation when estimating $\xi_{\rm E/B}$ also affects its noise properties, which we have not taken into account.  We therefore do not place too strong an emphasis on the high significance of the measured B-modes, or the lack of E-mode power on large-scales for all surveys, particularly as these are the scales that are most impacted by the choices made when extrapolating the data. That said, if surveys continue to use 2PCFs as a standard cosmic shear statistic, then it is still relevant to measure $\xi_{\rm E/B}$ as it is the B-mode measurement that is most closely related to the 2PCFs.

%% file: Systematics.tex
\label{sec:syssect}
In \sect\ref{sec:ResultsData} we detected significant B-modes in the DES-SV data as well as in certain tomographic combinations of CFHTLenS and KiDS-450 data.   In this section we introduce models for a range of data-related systematic effects that are appropriate for the datasets described in \sect\ref{sec:Data}.   We consider three models of systematics that affect the shear measurement of all galaxies independently of their redshift.  In addition we model one photometric redshift-dependent systematic, demonstrating how catastrophic errors in photometric redshifts can lead to shape selection bias.   We add these systematic models to mock data to explore their effect on the 2 point statistics introduced in \sect\ref{sec:Method}.  We are particularly interested in measuring the B-modes associated with each systematic (see \sect\ref{sec:ResultsMock}).  By comparing simulated results with and without these systematic effects, the B-mode signatures can then be used as a tool for diagnosing the source of the B-modes in the surveys analysed in \sect\ref{sec:ResultsData}. 

In this analysis, we do not model masking effects, since all of the methods we use rely on measuring 2PCFs, which are insensitive to masking, {provided the mask is uncorrelated with the shear field. If such correlations exist, all statistics will be affected by them.} This is in contrast to methods that rely on Fourier transforms of the shear field, where masks can cause significant systematic effects \citep[see for example][]{Asgari16}.  

\subsection{Shear measurement errors}
\label{sec:Systematics}
For the case of weak shear with $|\gamma | \ll 1$, we can model the observed ellipticity as 
\begin{equation}
\epsilon^{\rm obs}=(1+m)[\epsilon^{\rm int}+\gamma]+\eta+\alpha \epsilon^*+ \beta\, \delta \epsilon^* + c\;, 
\label{eqn:eobs_sys}
\end{equation} 
where $\epsilon^{\rm int}$ is the intrinsic galaxy ellipticity, $\gamma$ is the shear, $\eta$ is random noise on the ellipticity
measurement, $\epsilon^*$ is the PSF model ellipticity, $\delta \epsilon^* = \epsilon^* -  \epsilon^*_{\rm true}$ is the residual ellipticity between the model and true PSF, and $c$ is an additive shear that is uncorrelated with the PSF.    For all these quantities we use complex notation where, for example, $\gamma = \gamma_1 + {\rm i}\, \gamma_2$.    For the two PSF-dependent terms, $\alpha \epsilon^*$ quantifies the fraction of the model PSF ellipticity that leaks into the shape measurement, and $\beta\, \delta \epsilon^*$ quantifies the fraction of the residual PSF arising from PSF modelling errors, that leaks into the shape measurement.  The term $m$ is a multiplicative shear bias that is traditionally calibrated using image simulations.   

We simulate each of the systematic terms in \Eqt\eqref{eqn:eobs_sys} in isolation, in order to characterise their B-mode signature.  One exception is the shear calibration correction, $m$, which we set to zero, as an isotropic shear bias cannot introduce a B-mode signal, only scale an E-mode signal.

\subsubsection{Point spread function (PSF) leakage: $\alpha \epsilon^*$}
\label{sec:PSFLeakage}
\begin{figure*}
    \includegraphics[width=\hsize]{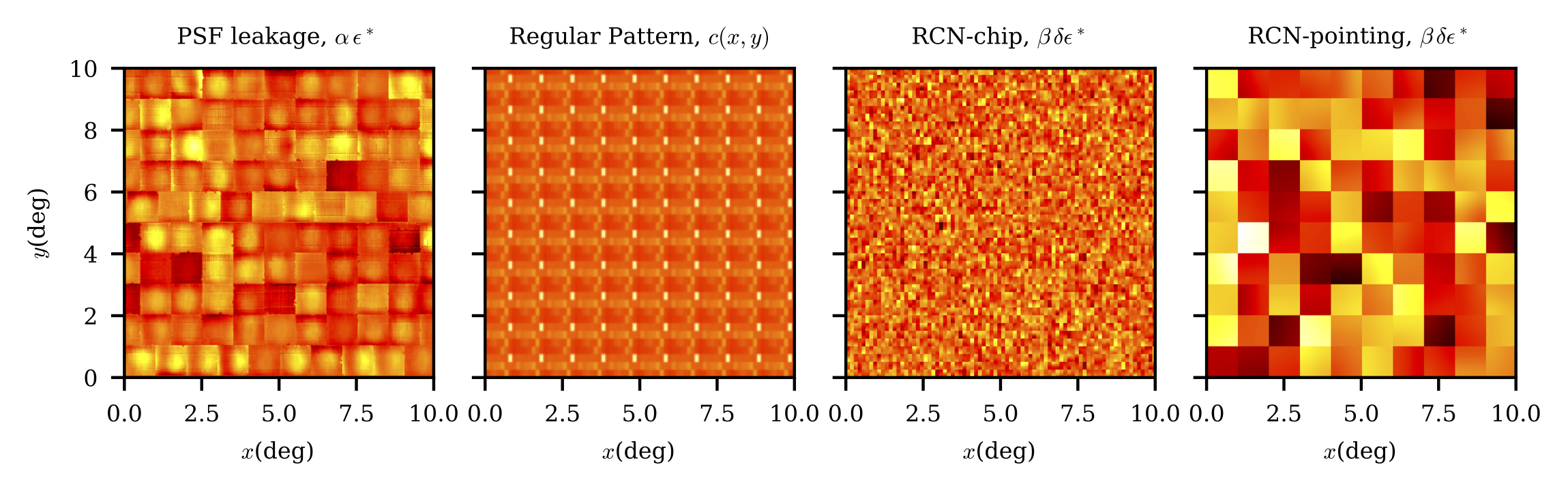}     
    \caption{\small{The first ellipticity component of the spatially varying systematic effects, simulated over a $10^\circ\times 10^\circ$ field. Here the effects are normalized to their maximum value for a better visual comparison.  From the left, the first panel shows the point spread function pattern used to model PSF leakage ($-0.01<\alpha \epsilon^*_{1}<0.01$). The second panel shows a regular pattern using the detector chip bias model from OmegaCam multiplied by a factor of 5 ($0.001<c_1<0.025$). The third panel shows the random correlated noise PSF residual model with a smoothing length similar to the chip size ($-0.006<\beta \delta \epsilon^*_{1}<0.006$), while the last panel shows the same systematic for a roughly pointing size smoothing length ($-0.003<\beta \delta \epsilon^*_{1}<0.003$). }}
     \label{fig:Sys}
 \end{figure*}

In order to mimic the effect of the PSF leakage on cosmic shear measurements we use PSF models from KiDS to make a realistic spatially varying PSF model spanning 100 square degrees.   
We construct this large-scale PSF pattern on a 1 arcmin$^2$ resolution grid, mapping the KiDS PSF measurements onto a $10^\circ\times 10^\circ$ field by stitching together two $5^\circ\times 10^\circ$ sections from the G12 and G15 regions in KiDS-450 data \citep[see][for details]{hildebrandt/etal:2017}. This provides us with a model for $\epsilon^*_i(x,y)$, where $\epsilon^*_1(x,y)$ is shown in the left panel of \fig\ref{fig:Sys}. In KiDS the PSF  is modelled with polynomials of third order within each pointing, where the lowest order is allowed to vary between CCDs to allow for discontinuities between CCD chips \citep[see][for more details]{Kuijken15}.  Similar modelling approaches are taken by CFHTLenS and DES-SV.  The mean of the PSF ellipticity and its one sigma deviation is $\epsilon^*_i=0.006\pm 0.016$ and its full range is covered by $-0.1<\epsilon^*_i<0.1$ for both components. \fig\ref{fig:Sys} shows how the PSF pattern changes within and across each $\sim$ square degree pointing. In areas where the KiDS data are masked and the PSF model unconstrained, we linearly interpolate the value of the PSF ellipticity to accommodate all galaxy positions in our unmasked mock data analysis.

We choose to apply a $10\%$ PSF leakage by setting $\alpha=0.1$.   This level of leakage corresponds to the $\alpha$ measured in the poorer-seeing KiDS $i$-band data \citep[see][]{Amon_iband18}.  For the high-quality KiDS $r$-band data that are used for the main cosmic shear analysis, $\alpha$ was found to be consistent with zero \citep{hildebrandt/etal:2017}. 

\subsubsection{Regular repeating additive pattern: $c(x,y)$}
\label{sec:RegularPattern}
In the absence of PSF-related errors, the amplitude of any remaining additive bias that is uncorrelated with the PSF, $c$, can be estimated directly from the data.  Since we expect $\langle \epsilon^{\rm int} + \gamma_i + \eta \rangle=0$ over a large area, $\langle \epsilon_i^{\rm obs} \rangle = c$ when $\alpha = \beta = 0$.  Surveys typically correct for any significant measurement of $c$ before any analysis, but this empirical correction usually takes an average over all galaxies and is therefore insensitive to small scale spatial variations $c(x,y)$ \citep{vanUitertSchneider16}.  Systematic effects that are stable and associated with the camera or telescope would result in a repeating pattern in the survey which is built from a series of different pointings.  To determine the impact of such a systematic, we model a spatially varying, but repeating additive term, which remains constant between pointings.

We use the data from \cite{Hoekstra2018}, who present a detailed analysis of imaging from the KiDS OmegaCam camera.  \cite{Hoekstra2018} report low-level but significant detector and electronic defects that introduce an additive shear contribution per CCD that is uncorrelated with the PSF and spans the range $0.0002<\epsilon^*_1<0.005$ and $-0.0004<\epsilon^*_2<0.00015$, shown in the second panel in \fig\ref{fig:Sys}.  The white "chip 15" of OmegaCam shows the largest bias at the level of $\epsilon_1=0.005$.   For the purposes of this analysis we multiply the \cite{Hoekstra2018} detector-bias model by a factor of five to amplify its effect, as we find that the original level of this effect is too small to show any significant E/B-modes for the current datasets.     

\subsubsection{Random but correlated noise: $\beta\, \delta \epsilon^*$}
\label{sec:RCNP}
Errors in PSF modelling, $\delta \epsilon^*$, can be systematic if the stars used in the modelling are unrepresentative of the PSF experienced by the galaxies \citep[][]{Antilogus2014,Guyonnet2015, Gruen2015}.   In this case the resulting systematic behaves similarly to the PSF leakage model outlined in \sect\ref{sec:PSFLeakage}, and we therefore do not consider this type of PSF modelling error.  

Instead we consider the random errors in the PSF modelling that arise from noise in the PSF measurement.  The impact of measurement noise on the PSF model increases as the number of stars available to characterise the model at each position in the field decreases.   
%In KiDS the PSF is modelled with polynomials of third order within each pointing, where the lowest order is allowed to vary between CCDs, to allow for discontinuities between CCD chips \citep[see][for more details]{Kuijken15}.  Similar modelling approaches are taken by CFHTLenS and DES-SV.   
The PSF modelling strategy (see \sect\ref{sec:PSFLeakage}) means that any local random errors from the sparse PSF measurement will propagate as random but correlated errors across the PSF model for the full field of view.

We mimic the impact of random but correlated PSF residual errors by assigning a randomly generated number from a Gaussian distribution with zero mean and unit dispersion to each $5 \times 5$ arcsecond pixel within a $10^\circ\times 10^\circ$ field. We first verify that the uncorrelated version of this systematic does not produce any coherent signal, as expected from a random error.  We then correlate the random PSF measurement noise over each pointing using a Gaussian filter convolution defined within the boundaries of the pointing. These convolved fields are then renormalized to produce an overall dispersion equal to $10\%$ of the shear dispersion in the mock data, $\sigma_{\rm RCN}=0.1 \sigma_\gamma$.
We investigate two kernel sizes with a correlation length of roughly the CCD chip level ($\sim$1.6 arcmin) and the pointing scale level ($\sim$43 arcmin).  
The resulting systematic patterns are shown in the two right panels of \fig\ref{fig:Sys}, where the systematic ranges are $-0.006<\beta \delta\epsilon^*_i<0.006$ (chip-level correlation) and $-0.003<\beta \delta\epsilon^*_i<0.003$ (pointing-level correlation).   For this systematic we chose both components of the contaminating ellipticity to be equal.

\subsection{Photometric redshift selection bias}
\label{sec:sysZPSF}
Cosmic shear surveys rely on photometric measurements to estimate the redshifts of galaxies.  The photometric redshift (photo-$z$) of a galaxy can be estimated by comparing {the magnitude of the galaxy in several colour-bands} to template catalogues of galaxy spectral energy distributions (SEDs) or to spectroscopic training samples \citep[see][and references therein]{Salvato/etal:2018}.   The most probable value for the redshift of each galaxy, given the measured photometric colours, $z_{\rm phot}$, is then used to divide the galaxy sample into tomographic redshift bins.  The true redshifts of these galaxies may not all lie within the boundaries of their appointed photometric redshift bins but provided the true underlying redshift distribution of the galaxies is known, this can be accounted for in the theoretical predictions of the cosmic shear signal (\Eqt\ref{eqn:gchi}).   The dispersion in true redshift within these tomographic bins will however depend on the precision of each galaxy's photo-$z$ estimate, which in turn depends on the error on the measured flux of the galaxy in each photometric band.   As a galaxy imaged with different noise realisations can therefore appear in different photometric redshift bins, in cases where the flux error is correlated with the shape or orientation of the galaxy, selecting a galaxy sample based on photometric redshifts could therefore lead to an ellipticity-redshift selection bias and hence a systematic error in a cosmic shear analysis.
{In this section we explore and introduce this new concept of photometric redshift selection bias as a systematic for lensing surveys.}

Consider two identical elliptical galaxies of fixed size and flux.  The first galaxy is convolved with an elliptical PSF aligned with its major axis.  The second is convolved with an elliptical PSF aligned perpendicular to its major axis.   In the resulting image our second galaxy will appear to cover a larger area than our first galaxy and with a lower surface brightness and lower significance.   It will therefore have larger photometric errors compared to the first galaxy.  \cite{Kaiser:2000} recognised that this effect implied that any cuts on observed significance would introduce a PSF-dependent selection bias in the ellipticity of the galaxies \cite[see also][]{Bernstein/Jarvis:2002}.   The introduction of tomographic photo-$z$ selection in a cosmic shear analysis, which implicitly depends on the significance of each detection, can therefore also lead to an ellipticity-dependent selection bias.

In addition to this core effect, flux errors that are correlated with the relative orientation of the galaxy and PSF can also arise simply from the methodology used to measure the photometry in each band.  DES-SV use {\sc SExtractor} automated aperture magnitudes where the aperture is fixed by the galaxy shape in the detection image \citep{Bonnett/etal:2016, Rykoff/etal:2016}.   Whilst this method ensures that the physical apertures are matched between the bands, it does not take into account the differing PSFs.  \citet{Hildebrandt12} show that this approach leads to an overall degradation in the photometric redshifts.  For example, if the PSF in the $i$-band is perpendicular to the PSF in the detection $r$-band, the resulting $i$-band flux, assuming a fixed-detection aperture, will be underestimated.  This approach therefore results in flux errors that are correlated with the relative orientation of the galaxy and PSF in each band.  \citet{Hildebrandt12} demonstrate the importance of homogenising the PSFs between bands before determining the matched-aperture photometry.  Both CFHTLenS and KiDS Gaussianise the PSFs before measuring the photometry using the methodology proposed by \cite{Kuijken/2008} and \cite{Kuijken15}.  These surveys should therefore be fairly immune to this additional error and we note that the DES photometry methodology has been significantly improved since the release of the DES-SV data that we analyse in this paper \citep{DrlicaWagner/etal:2018}.

In this analysis we make the first step to examine photometric-redshift selection bias, by simulating a mock galaxy catalogue where we introduce an anti-correlation between the signal-to-noise ratio of the measured flux and the ellipticity of the galaxies relative to the local PSF ellipticity, $|\epsilon-\epsilon^*_{\rm x}|$, in four bands ${\rm x}=u,g,r,i$. We use the following linear relation for the anti-correlation, 
\begin{equation}
\frac{\rm{Flux}}{\rm Flux \, error} = a_{\rm x} | \epsilon-\epsilon^*_{\rm x} | + b_{\rm x}
\label{eqn:fluxpsfcorr}
\end{equation}
where the value for $a_{\rm x}$ and $b_{\rm x}$ are determined by fitting to KiDS-450 multi-band data (see \tab\ref{tab:FluxErr}). Given that KiDS can only measure the noisy observed ellipticity, we recognise that the majority of the anti-correlation that we find in the KiDS-450 data, derive from taking the mean of the absolute value of the observed ellipticity where the measurement noise, $\eta$ in \Eqt\eqref{eqn:eobs_sys}, increases with decreasing signal-to-noise. Using \Eqt\ref{eqn:fluxpsfcorr} to apply a correlation between the signal-to-noise of a galaxy and its relative ellipticity to the local PSF therefore provides an upper limit for this effect. Future work will use multi-band image simulations to determine values for $a_{\rm x}$ and $b_{\rm x}$ where the true ellipticity is known.  Our current approach is however sufficient for the purposes of examining the B-mode signature that is introduced by such an effect.

We produce mock ellipticity catalogues by randomly associating ellipticities to galaxies, using a fit to the observed KiDS-450 galaxy ellipticity distribution, such that $\langle \epsilon_{\rm mock} \rangle = 0$.   We simulate 15 fields of 100 deg$^2$ each with a total galaxy number density of 5.5 arcmin$^{-2}$.   We choose a simple model of constant PSF per 1 deg$^2$ pointing taken randomly from a uniform distribution between $[-0.1, 0.1]$ for each component of the PSF ellipticity\footnote{{We chose this strong PSF over the very low ellipticity KiDS-450 PSF model, $\epsilon^*=-0.006\pm 0.018$, shown in the left-hand panel of \fig\ref{fig:Sys}, as initial KiDS-like studies did not result in a significant photometric redshift selection bias.}}.

\begin{figure}
  \includegraphics[width=\hsize]{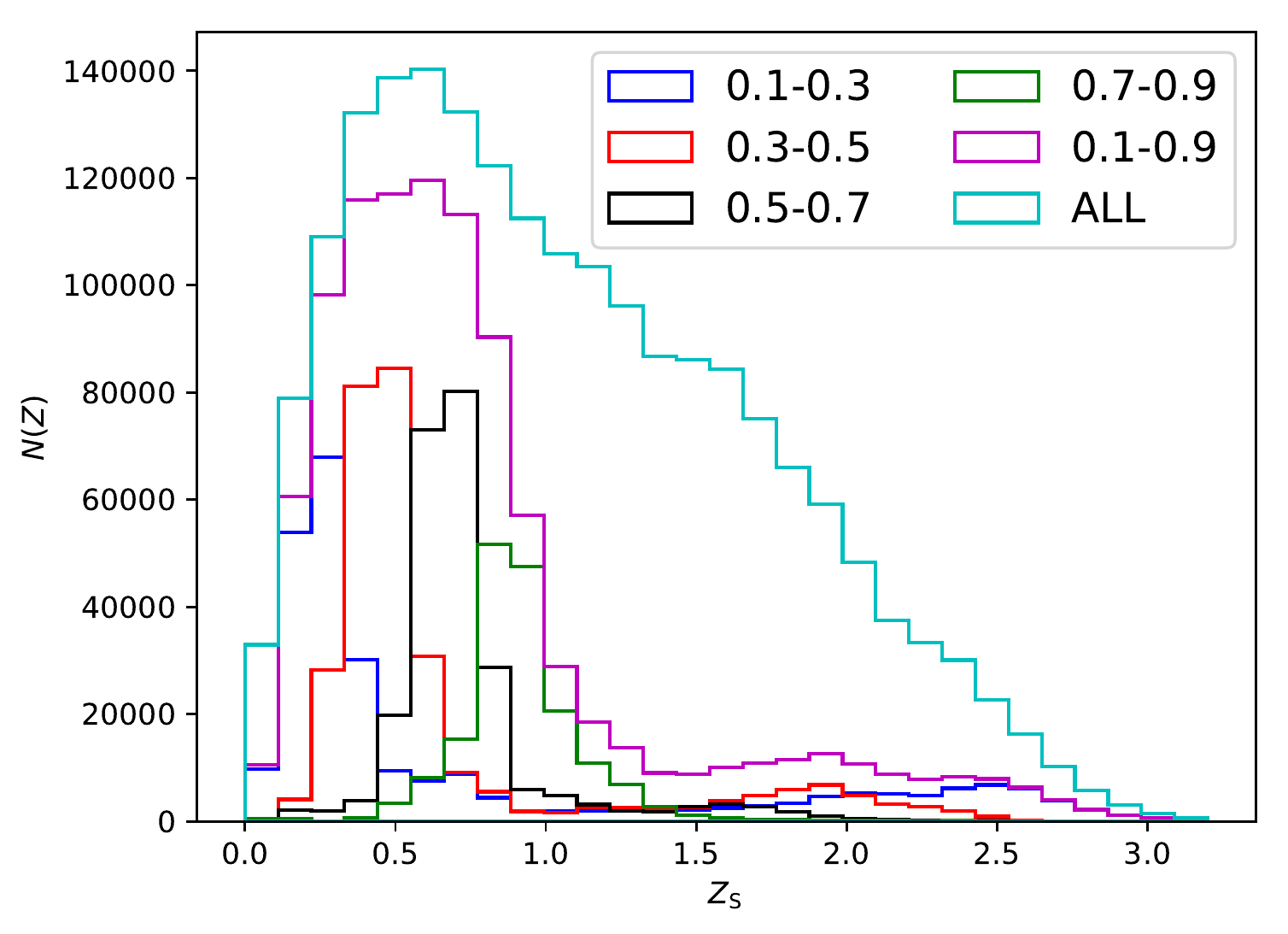}\\
    \caption{\small{The true redshift distribution of the mock galaxies, separated into photometric redshift bins.  The $z_{\rm B}$ selection is shown in the legend. The cyan histogram shows the true redshift distribution of the galaxies in the parent noise-free sample.  In order to determine the photometric redshifts we introduce flux errors that mimic a KiDS-like survey and depend on the relative ellipticity of the mock galaxy to the mock PSF. }}
   \label{fig:Zdists}
\end{figure}

We associate noise-free multi-band fluxes to the mock galaxies using simulations similar to the ones presented in Sect. 3.1  of \cite{Hildebrandt2009} but adapted to KiDS. These simulations were created with the {\sc HyperZ} package \citep[][]{Bolzonella2000} and are based on SED templates from the library of \cite{BruzualCharlot1993}, $i$-band number counts from COSMOS \citep{Capak2007}, and redshift distributions from {\sc bpz} \citep{BPZ2000}. 
These magnitude simulations contain half a million galaxies with magnitudes given in each of the four bands, selected to recover the KiDS redshift distributions given in \fig\ref{fig:Zdists}.  

For each galaxy, we assign an error on the flux in each band using \Eqt\eqref{eqn:fluxpsfcorr}.  Noise is then added to the mock galaxy flux, sampling from a Gaussian distribution.  This approach correlates high values of observed galaxy ellipticity with high flux errors, as expected from the ellipticity measurement noise in the data.   In addition, the flux error may also depend on the relative-orientation of the galaxy and the PSF, in each band, as expected from some photometry measurement methods in addition to the \citet{Kaiser:2000} effect.   As this is the first investigation into photometric redshift selection bias, we have not tried to separate these effects in our analysis.  We also note that this method of assigning noise to our mock galaxy sample ignores the additional dependence of the signal-to-noise ratio on galaxy size and magnitude.  Future work will need to investigate this in more detail, using multi-colour image simulations.

We use the Bayesian Photometric Redshifts software {\sc bpz} to estimate photo-$z$'s for each of our mock galaxies using a template fitting method \citep{BPZ2000,BPZ2004,BPZ2006}. The inputs are the noisy flux measurements and their associated errors.  The output is the best fitting photometric redshift, $z_{\rm B}$, which we use to then bin the galaxies into the four redshift bins that were used in the KiDS-450 cosmic shear analysis, $z_i\leq z_{\rm B}<z_{i+1}$, with $z_i=\{0.1,0.3,0.5,0.7,0.9\}$ as well as a broad single bin encompassing the full redshift range of KiDS-450, $0.1\leq z_{\rm B}<0.9$.    DES-SV and CFHTLenS use a similar number of tomographic bins, spanning similar ranges in photometric redshifts. Note that the SED templates that we use in {\sc bpz} to estimate the redshift of the mock galaxies is independent from the ones used to make the mocks, which shows the robustness of this method to the choice of templates.

\fig\ref{fig:Zdists} shows the true redshift distribution of the mock galaxies for each tomographic redshift bin in $z_{\rm B}$.   The distributions are broad due to the noise with extended high and low redshift tails which we label as catastrophic outliers in the distribution.    The mean and median of each tomographic bin is similar to those in the KiDS-450 data, demonstrating that our method to assign noise to our mock galaxy sample is sufficient for this analysis.

%% file: Results2.tex
\label{sec:ResultsMock}
In this section we present the two-point statistic signatures of the systematics introduced in \sect\ref{sec:Systematics} using the statistics explained in \sect\ref{sec:Method}. {We show the effect of the three redshift-independent shear systematics in \sect\ref{sec:shearsysres} considering only a single redshift distribution for the galaxies. The effect of the remaining systematic, photometric redshift selection bias, is explored in \sect\ref{sec:resultZdep} where redshift binning is applied.}
With these signatures identified, our goal is to use B-mode measurements as a diagnostic tool to uncover the origin of the systematic signals identified in DES-SV, CFHTLenS and KiDS in \sect\ref{sec:ResultsData}.   We can also determine the impact of these systematics on the measured E-mode signals. Our approach is complementary to previous studies by \citet{AmaraRefregier08,Kitching16,TaylorKitching18} who propagated cosmic shear systematics through to cosmological parameters in order to set requirements on their (in)significance.

\subsection{SLICS shear simulations}
The basis of our systematics analysis makes use of the ensemble of mock KiDS-450 catalogues constructed from the SLICS\footnote{Available here: http://slics.roe.ac.uk/} simulations suite described in \cite{Harnois15} and \cite{Harnois-Deraps/etal:2018}. Each SLICS line of sight corresponds to a  $10^\circ\times10^\circ$ field that includes galaxy positions, shear and their true and photometric redshifts. On average the redshift distribution and number density of galaxies in these 
mocks correspond to the KiDS-450 data, which is not so dissimilar from the properties of both DES-SV and CFHTLenS. 

\begin{figure*}
   \begin{center}
     \begin{tabular}{c}
     \resizebox{100mm}{!}{\includegraphics{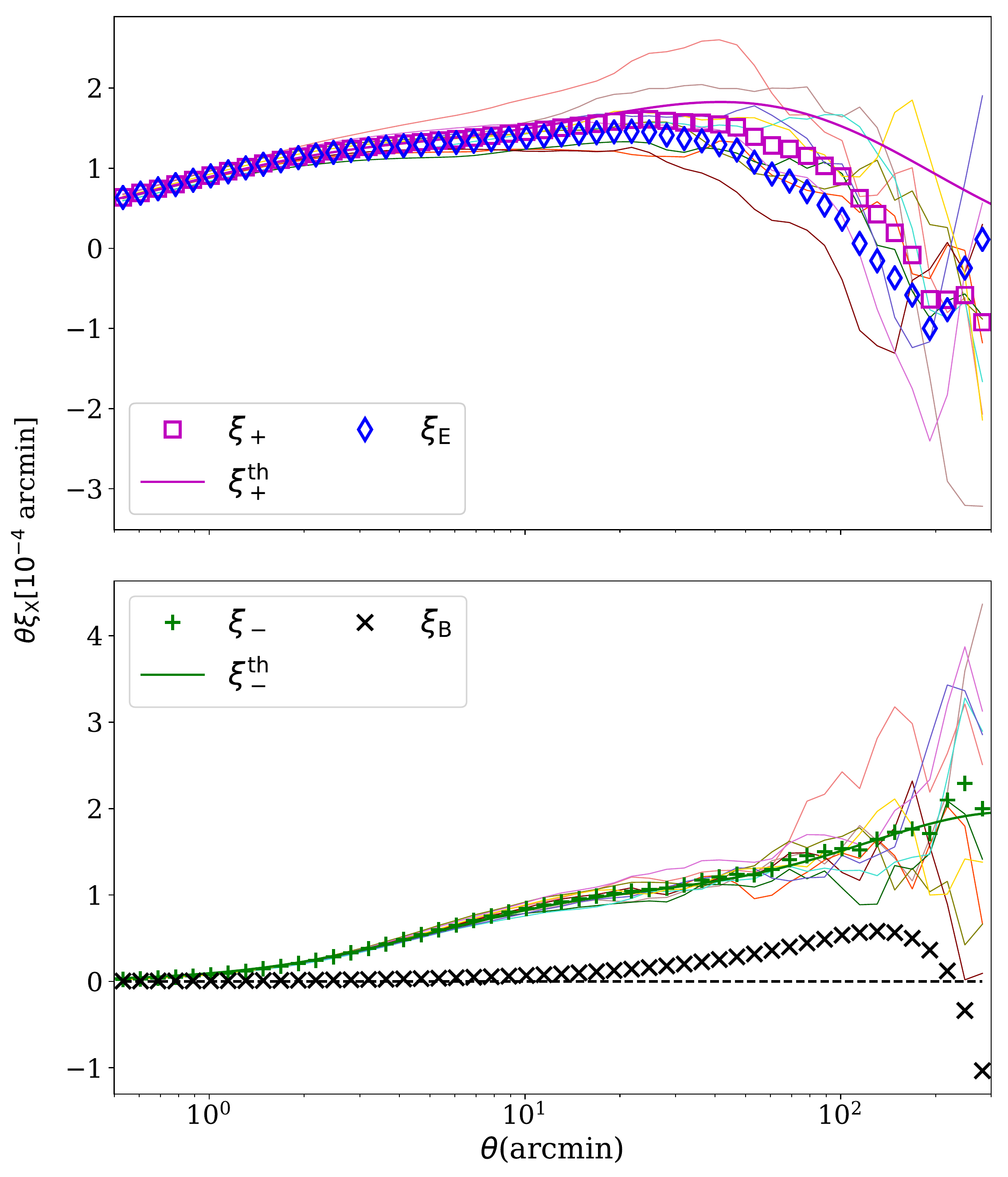}}
     \hspace{0.5cm}
      \resizebox{65mm}{!}{\includegraphics{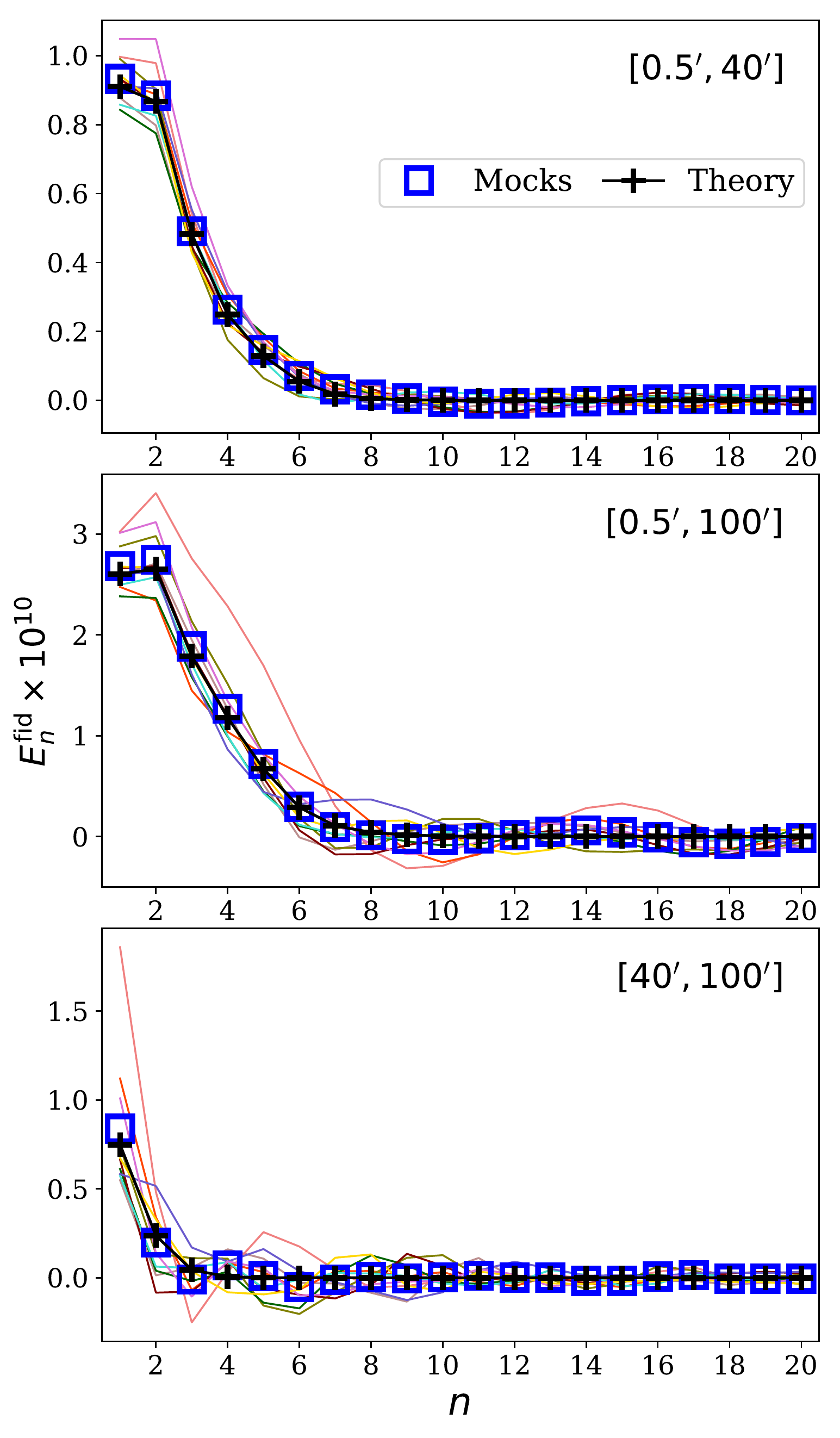}}
%       \vspace{0.1cm}
     \end{tabular}
   \end{center}
     \caption{\small{SLICS 2-point statistics, $\xi_\pm$ and $\xi_{\rm E/B}$ (left) and E-mode COSEBIs (right), averaged over 10 noise-free lines-of-sight, which serves as our fiducial `systematics-free' measurement.  The mean result can be compared to the theoretical expectation (smooth solid curves).  For $\xi_\pm$ and COSEBIs we also show the measurements for each individual line-of-sight with thin solid curves with matching colors between different panels.  The upper left panel shows the measured $\xi_+$ (magenta squares) and $\xi_{\rm E}$ (blue diamonds), with the lower left panel showing $\xi_-$ (green pluses) and $\xi_{\rm B}$ (black crosses). The expectation value for $\xi_{\rm B}$ is zero, shown with the dashed black line. The COSEBIs E-modes (right panels) are shown for the three angular ranges indicated in each row. {The COSEBIs B-modes in SLICS are 4 orders of magnitude smaller than the E-modes and are therefore not shown.} The measurements are shown as squares and their expected theory value as plus symbols.  Note that COSEBIs modes are discrete and the points are only connected together as a visual aid. }}
     \label{fig:MockvsTheory}
 \end{figure*}

In \fig\ref{fig:MockvsTheory} we present the SLICS cosmic shear measurements, $\xi_\pm$, $\xi_{\rm E/B}$ and COSEBIs, averaged over 10 shape noise-free lines of sight (i.e. $\epsilon^{\rm int} = 0$).  On the left we show $\xi_\pm$ and $\xi_{\rm E/B}$, for $\theta\in[0.5',300']$ in 50 logarithmic bins. The top left panel shows $\xi_+$ and $\xi_{\rm E}$ and the lower left panel shows $\xi_-$ and $\xi_{\rm B}$. The right panels belong to the E-mode COSEBIs for a range of angular scales.   The measurements from SLICS can be compared to the theoretical prediction (Eqs.\,\ref{eqn:xipm}, \ref{eq:xiEBPower}, \ref{eq:EnBnFourier}), shown as thick solid curves  for $\xi$ and pluses for COSEBIs.  Here we adopt a flat $\Lambda$CDM model given by the input cosmology of the SLICS simulations in \tab\ref{tab:CosmoParam}.  We use a \cite{BondEfstathiou84} transfer function to estimate the linear matter power spectrum and the \cite{smith03} halofit model for the non-linear scales.  This combination, although dated, was chosen as the resulting theoretical predictions fit the mocks better than the more modern alternatives \citep{Harnois15}.  The thin coloured lines in \fig\ref{fig:MockvsTheory} show the measured values for each line-of-sight (LOS), which show a considerable scatter, especially for larger angular scales. Even with the inclusion of a larger number of LOS, however, we do not expect the theory to match the mean of the mocks perfectly, as the finite box-size of the N-body simulations, where $L_{\rm box}=505 h^{-1}\, {\rm Mpc}$, results in a loss of power on large scales \citep{Harnois-Deraps/etal:2018}.

For a B-mode free dataset, $\xi_+=\xi_{\rm E}$. In the upper left panel of \fig\ref{fig:MockvsTheory}, we see that this is not the case for SLICS as at large angular scales $\xi_{\rm E}$ is smaller than $\xi_+$.  Looking at the lower panel we see that  $\xi_{\rm B}$ is non-zero for the same angular ranges.  This leakage from E to B in the $\xi_{\rm E/B}$ statistic is a result of using the theoretical $\xi_-$ at large scales for calculating the integrals in \Eqt\eqref{eq:xiPrime}, which differs 
from the $\xi_-$ of SLICS due to its finite box-size.  We find that COSEBIs do not suffer from either of these effects, with the COSEBIs B-mode signal in the mocks found to be $\sim 4$ orders of magnitude smaller than the E-modes (not shown). The reason for the robustness of COSEBIs to the finite box bias comes from the weight functions that convert the shear power spectrum to these statistics. The low $\ell$ behaviour of the weight functions for COSEBIs have a leading-order term proportional to $\ell^4$ such that the function reaches zero at small $\ell$-values in contrast to the $\xi_+$ kernel, $\rm{J}_0$, which has power at small arguments (see \Eqt\ref{eqn:xipm}).   At high $\ell$ the COSEBI weights also diminish rapidly in contrast to the 2PCFs which include some degree of power from all scales \citep[see figure 4 in][for a comparison between the kernels corresponding to COSEBIs and 2PCFs]{Kilbinger17}. 

\subsection{The B-mode signature of shear measurement systematics}
\label{sec:shearsysres}

\begin{figure*}
	\centering
	\includegraphics[width=17cm]{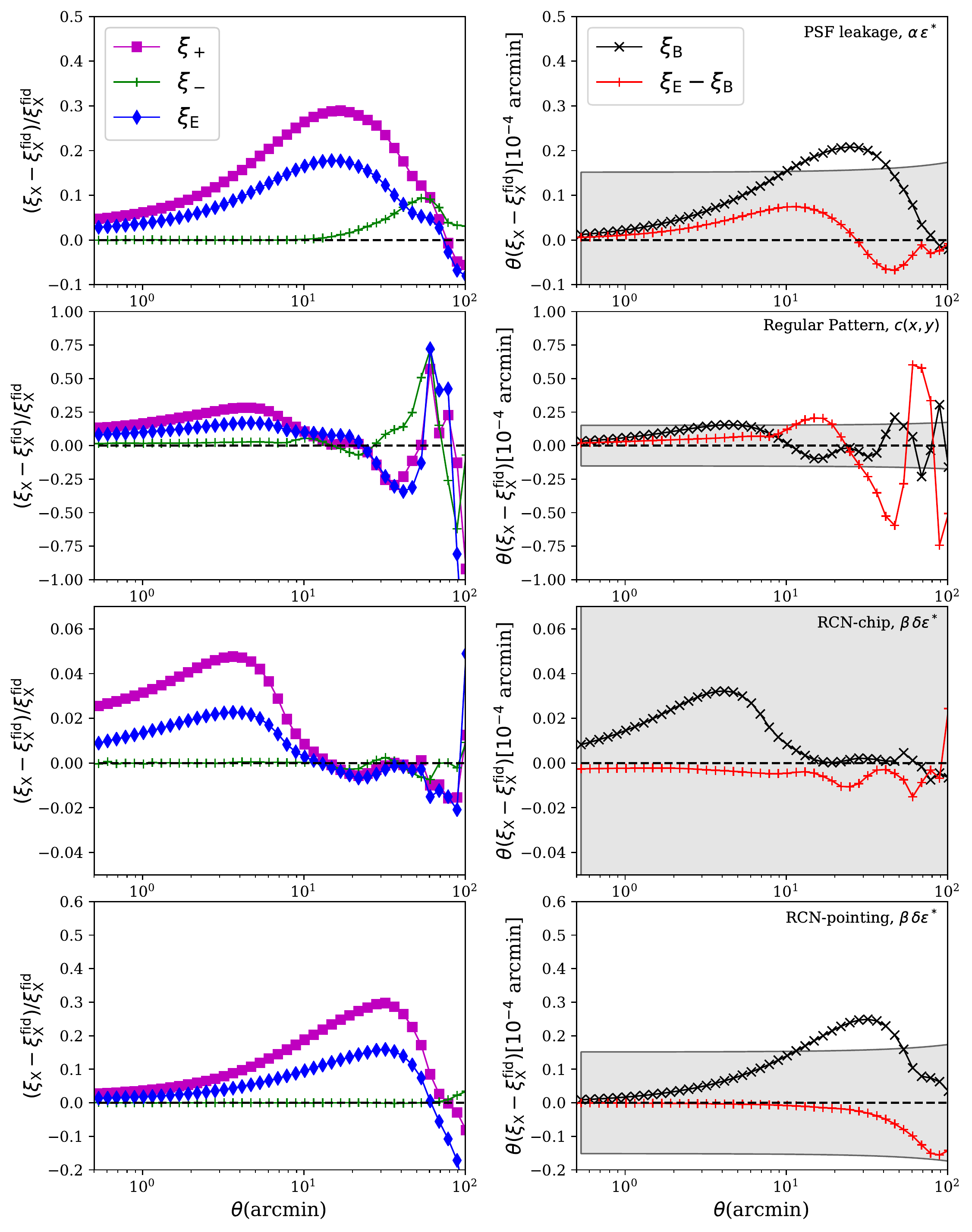}
     \caption{\small{The impact of shear measurement systematics on  $\xi_\pm$ and $\xi_{\rm E/B}$ for four different types of shear measurement systematics; {From the top down:} PSF leakage, a repeating additive pattern, and random but correlated noise, correlated on chip and pointing scales (see \fig\ref{fig:Sys}).   For $\xi_+$ (magenta squares), $\xi_-$ (green pluses) and $\xi_{\rm E}$ (blue diamonds) we present, in the left panels, the fractional difference between the measured signal in the systematic-induced KiDS-like SLICS mocks and the fiducial systematic-free case.   As $\xi_{\rm B}$ (black crosses) and the E/B difference $\xi_{\rm E} - \xi_{\rm B}$ (red pluses) tends to zero, we present, in the right panels, the difference between these measurements and the fiducial case, multiplied by the angular distance in arcminutes and scaled by $10^4$.  The measured B-modes can be compared to the expected shape-noise error for KiDS-450 (shaded area).}}
     \label{fig:2PCFSys}
 \end{figure*}

We add the shear measurement systematic effects, developed in \sect\ref{sec:Systematics}, in turn to the SLICS simulations.  We follow the standard approach of applying an empirical systematics correction to each mock by subtracting the average observed ellipticity from each line of sight before commencing our statistical analysis.  In \fig\ref{fig:2PCFSys} we compare the resulting 2PCFs with the signal measured in the systematic-free fiducial data {(see \fig\ref{fig:MockvsTheory})}.  The 2PCFs are $\xi_+$ (squares), $\xi_-$ (pluses), $\xi_{\rm E}$ (blue diamonds) and $\xi_{\rm B}$ (crosses). The left panels show {the fractional difference of $\xi_\pm$ and $\xi_{\rm E}$ to their fiducial values}, calculated from systematic free mocks and shown with a `fid' superscript, for each systematic, while the right panels show the difference between $\xi_{\rm B}$ and its fiducial value as well as $\xi_{\rm E}-\xi_{\rm B}$ (pluses) for each case. Here we use 50 logarithmic $\theta$-bins, to show the angular dependence of each systematic in detail.

\begin{figure*}
	\centering
	\includegraphics[width=17cm]{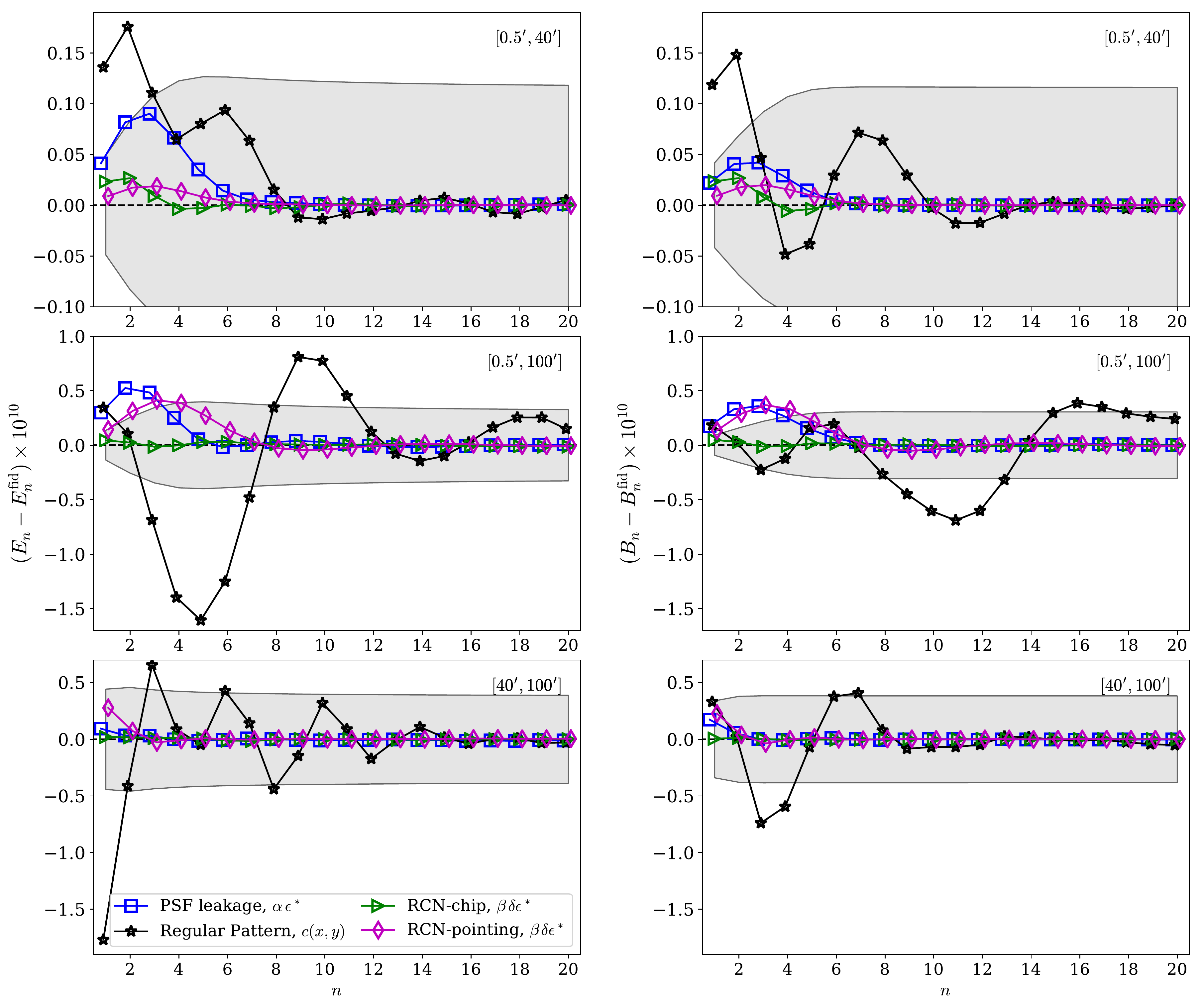}
     \caption{\small{The impact of shear measurement systematics on E-mode (left) and B-mode (right) COSEBIs for four different types of shear measurement systematics; PSF leakage (blue squares), a repeating additive pattern (black stars), and random but correlated noise on chip (green triangles) and pointing (magenta diamonds) scales (see \fig\ref{fig:Sys}).  The analysis is conducted for three different angular ranges spanning $[0.5',40']$ (upper panels), $[0.5',100']$ (middle), and $[40',100']$ (lower panels).  We present the difference between the measured signal in the systematic-induced KiDS-like SLICS mocks and the fiducial systematic-free case scaled by $10^{10}$.  The measured B-modes and the resulting change to the E-mode can be compared to the expected shape-noise error for KiDS-450 (shaded area).}}
     \label{fig:COSEBIsSys}
 \end{figure*}

Each row in \fig\ref{fig:2PCFSys} shows the impact of the systematic which from top-down cover PSF leakage, a repeating additive pattern, and random but correlated noise (RCN), similar to PSF residuals, correlated over chip-scales and then pointing scales. The grey regions in the right panels show the level of noise expected for KiDS-450 data, which is similar to the noise in the DES-SV and CFHTLenS analyses. {As the simulations are free of shape-noise, the error associated with them is negligible compared to the expected errors from either of the three surveys, therefore we have excluded the simulation error from this figure (\fig\ref{fig:2PCFSys}) and the next one (\fig\ref{fig:COSEBIsSys})}. We note that all these systematics also produce parity violating signals {$\xi_{\times}=\langle \epsilon_t \epsilon_{\times} \rangle$, which is expected to be zero for a shear only field}. We find that their amplitude is about an order of magnitude smaller than the B-modes, however, and are therefore harder to detect in the data. As a result, we limit our systematics study to the effect of systematics on E/B-modes.

{One interesting result from \fig\ref{fig:2PCFSys} comes from the non-zero signal in the $(\xi_{\rm E}-\xi_{\rm E}^{\rm fid}) -(\xi_{\rm B}-\xi_{\rm B}^{\rm fid})$ curves.  If a systematic adds equal power to both observed E and B-modes, $P_{\rm sys}$, from \Eqt\eqref{eq:xipmPower} and {\Eqt\eqref{eq:xiEB} we find that the observed $\xi_{\rm B}$ is equal to the excess signal in the observed $\xi_{\rm E}$ due to this systematic,
\begin{equation}
\xi_{\rm E}^{\rm obs}- \xi_{\rm E}^{\rm true} = \xi_{\rm B}^{\rm obs}\;,
\end{equation}
where  $\xi_{\rm E/B}^{\rm obs}$ are the observed E/B-mode 2PCFs and $\xi_{\rm E}^{\rm true}$ is the E-mode signal produced from a shear only field.} Furthermore as $\xi_-$ is proportional to $P_{\rm E}-P_{\rm B}$, an equal $P_{\rm sys}$ contribution to the E and B-modes will cancel such that $\xi_-^{\rm obs}  - \xi_-^{\rm true}  =0$.   In this case there is a clear route to correct the measured E-mode by the measured B-mode or to select `clean' angular scales for the E-mode analysis which are B-mode free \citep[see for example][]{hildebrandt/etal:2017}.  For the sample of systematics that we have simulated, however,  we see that this common assumption of the equal contribution of systematic power to the E and B modes is far from reality, especially for larger angular scales}\footnote{Note that the finite box-size bias has already been accounted for by subtracting the fiducial values for each statistic. Any remaining correlation originates from the systematic effects that we have introduced.}.  This implies that if a $\xi_{\rm B}$ signal is detected at any angular scale,  its origin should be identified and mitigated at the catalogue or image level.  Without understanding the origin it is unclear how that systematic will contaminate the $\xi_{\pm}$ signal.

In \fig\ref{fig:COSEBIsSys} we present the COSEBIs analysis of the mocks.  We show the relative effect of systematics on the E-mode (left) and B-mode (right) COSEBIs, as the difference between their fiducial values and those estimated from the systematic induced mocks including PSF leakage, $\alpha \epsilon^*$ {(blue squares), repeating additive pattern, $c(x,y)$ (black stars), and random but correlated noise (RCN), $\beta \delta \epsilon^*$, correlated on chip scales (green triangles) and pointing scales (magenta diamonds)}.  All values are shown for the mean of the 10 SLICS lines-of-sight. The grey regions show the one sigma errors corresponding to a KiDS-450-like survey. Random but correlated noise at the chip level shows small deviations from the fiducial values in agreement with \fig\ref{fig:2PCFSys}. Within a single pointing, this systematic has a similar form to the additive pattern, {where we find} similar low-$n$ behaviour between these two systematics with similar peaks at $n=2$ and a dip at $n=4,5$, albeit at different amplitudes. 
PSF leakage and the random but correlated noise at the pointing level are more significant, exhibiting a similar signal from the lower COSEBIs modes. The repeating additive pattern has the most chaotic effect on COSEBIs, in comparison to the other systematics that we have simulated. The erratic high frequency changes that can be seen in the 2PCFs in \fig\ref{fig:2PCFSys} are reflected in the significant power seen in the higher COSEBIs modes. As these systematics produce varying correlations for different angular scales, COSEBIs modes are affected by them in differing amplitudes, which also depend on the angular range they probe. Comparing \fig\ref{fig:2PCFSys} and \fig\ref{fig:COSEBIsSys} gives insight into the sensitivity of COSEBIs modes to correlation at various angular scales.

For all four systematics, we find that their effects on the COSEBIs are more prominent when the full angular range is used (middle panels). Comparing the right and left panels we see that all systematics affect both E and B-modes, but not equally.   In general a significant B-mode signal translates to a significant contamination to the E-modes on the same scales. The repeating additive pattern forms a clear exception to this rule though. This draws us to the same conclusion as the $\xi_{\rm B}$ analysis, 
the origin of any COSEBIs B-mode signal should be traced back to its source and corrected for at the pixel-data product level where phase information is still available, since it is unclear how these systematics will impact the E-mode at another angular scale.

The characteristic patterns that we have identified should be used in any future approach to diagnose and correct shear-measurement systematics.  As an example, if COSEBI B-modes are found to be oscillatory and extend to high-$n$, the survey should investigate additive biases that repeat on a fixed angular scale across the survey, for example detector-level effects.  If the B-modes are localised at low-$n$ with little high-$n$ power, the survey should investigate the PSF modelling.  With COSEBIs alone we cannot distinguish between PSF leakage, $\alpha \epsilon^*$ or correlated noise in the PSF model, $\beta \delta \epsilon^*$, but these two effects can be separated by measuring the correlation between galaxy shape and PSF ellipticity, which will be significant if $\alpha$ is non-zero \citep{Bacon/etal:2003}.    If our ellipticity model in \Eqt\eqref{eqn:eobs_sys} is reasonable in its approach to add systematic terms linearly, we would expect the B-modes from each individual effect to also add linearly.   When we see both significant power at $n<7$ and high-$n$ oscillatory power, as we do for DES-SV for example, we can conclude that there is a likely superposition of systematics from both the PSF modelling errors and repeating additive biases.  

The simulation approach that we use here should be specialised to the survey in question in future work.  This would allow for a more precise exploration of how survey-specific issues flow through to cosmological biases.  In this analysis we have presented results that use KiDS to motivate the angular dependence of the systematics that we have simulated.  In addition, we have also tested a variety of alternative schemes in the development of this work such that we are confident that the global behaviour of the B-mode signatures, presented in \fig\ref{fig:COSEBIsSys}, are broadly representative of how these systematics would feature in any weak lensing survey. {These tests included modelling different possible patterns and fixed amplitudes for each of the systematics, with the pointing size fixed to 1 deg$^2$ in all cases in accordance to both KiDS and CFHTLenS. DES pointings are however hexagons of width $2.2$ degrees.}.

\subsection{The B-mode signature of photometric redshift selection bias}
\label{sec:resultZdep}

\begin{figure*}
   \begin{center}
     \begin{tabular}{c}
     \resizebox{160mm}{!}{\includegraphics{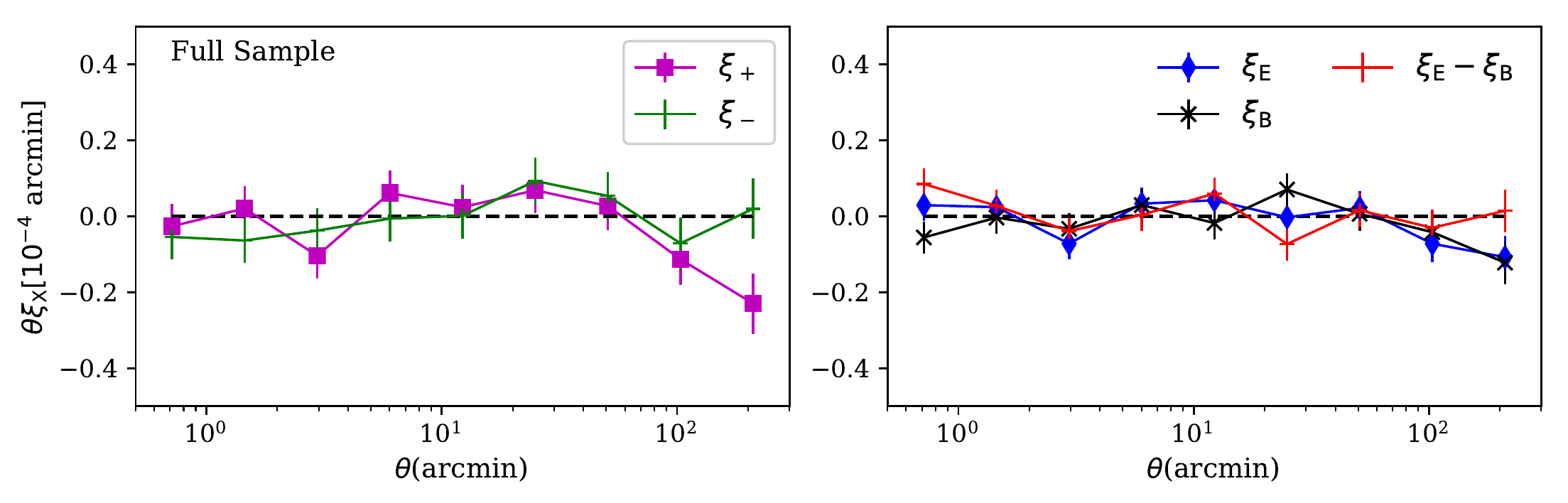}}\\
     \resizebox{160mm}{!}{\includegraphics{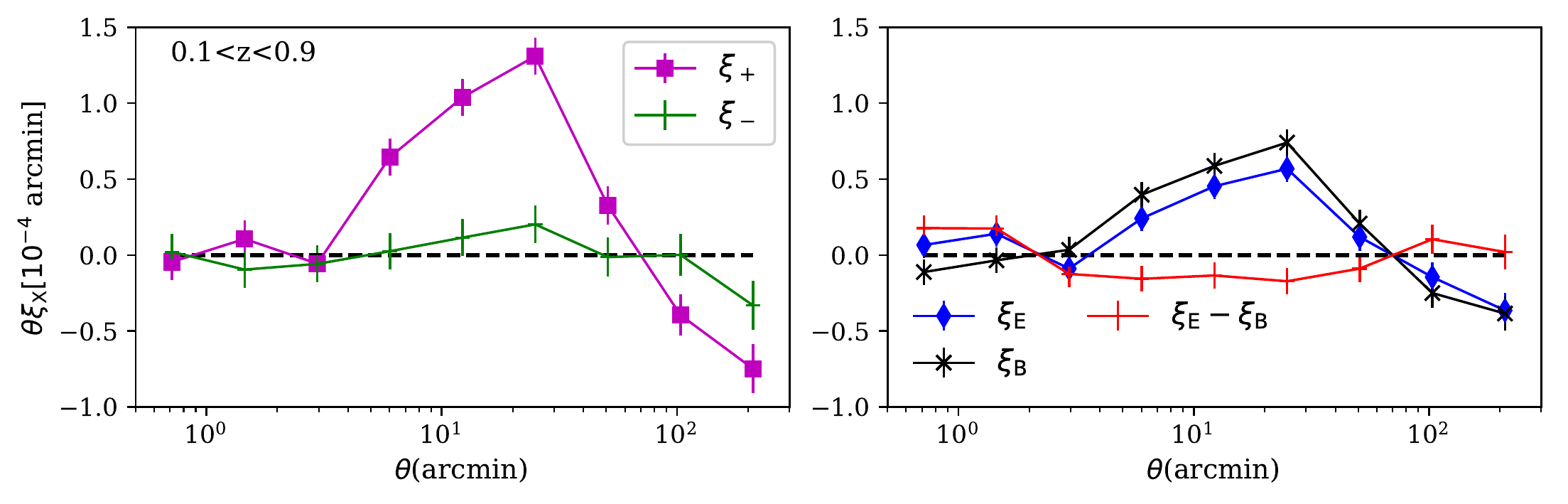}}
     \end{tabular}
   \end{center}
     \caption{\small{Photometric redshift selection bias:  the upper panels show the correlations measured from a random intrinsic ellipticity field using the full galaxy sample.  The lower panels show the correlations measured from the same random intrinsic ellipticity field after a $0.1\leq z_{\rm phot}\leq 0.9$ photometric redshift selection has been applied.  The 2PCFs shown on the left are $\xi_+$ (squares) and $\xi_-$ (pluses).  On the right, $\xi_{\rm E}$ (diamonds),  $\xi_{\rm B}$ (crosses) and the difference between the two (pluses) are shown.  The one-sigma error bars correspond to the level of ellipticity noise in the mock data. The signals are shown multiplied by $\theta$ in arcminutes and scaled by $10^4$. }}
     \label{fig:PSFz2PCFs}
 \end{figure*}

We simulate photometric redshift selection bias following the approach developed in \sect\ref{sec:sysZPSF}.  As this effect is subtle, we choose to analyse the correlations in a random intrinsic ellipticity field alone, in contrast to the analysis in \sect\ref{sec:shearsysres}. where we also include the correlated SLICS cosmic shear field.  Starting with the traditional statistics in \fig\ref{fig:PSFz2PCFs}, we compare two cases. The top panels show the 2PCFs using the full galaxy sample which are consistent with zero by construction.   The lower panels show the same 2PCF analysis, including a photometric redshift selection with $0.1\leq z_{\rm phot}\leq 0.9$.  Comparing these two results we immediately see that the photometric redshift selection has produced a significant signal in all but the $\xi_-$ statistic.  As such we find similar levels of E/B-modes in the lower right panel where the difference between $\xi_{\rm E}-\xi_{\rm B}\approx 0$. {Comparing the bottom panel of \fig\ref{fig:PSFz2PCFs} with the top panel of \fig\ref{fig:2PCFSys} we see that the location of the peak in  $\xi_+$ is similar. This is because the PSF-leakage systematic (\fig\ref{fig:2PCFSys}) is related to photometric redshift selection bias, although, here we have used a simplified PSF model without the small scale variations that affect the PSF-leakage systematic (\fig\ref{fig:Sys}).}

\begin{figure*}
	 \centering
 	\includegraphics[width=17cm]{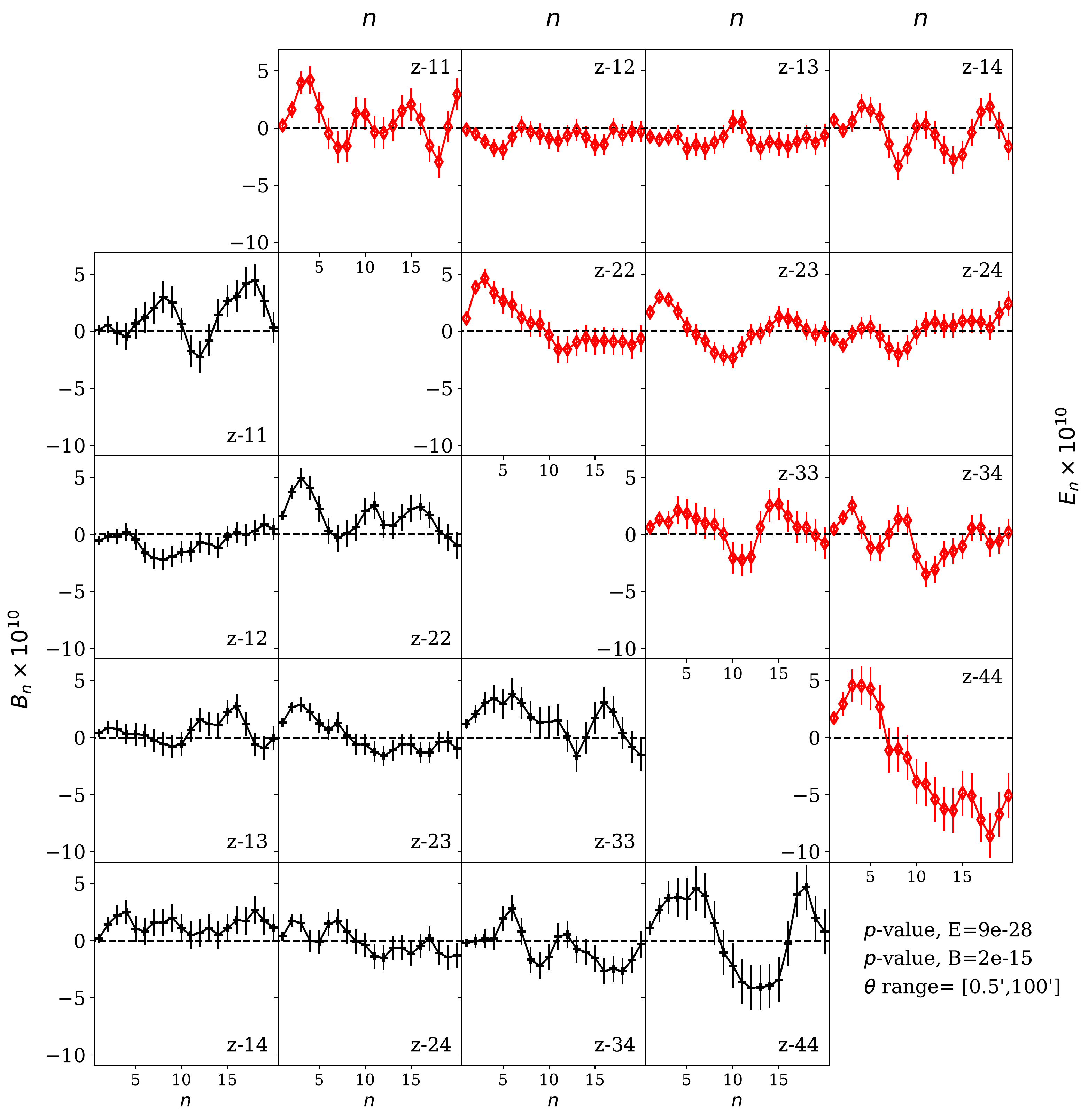}
     \caption{\small{The impact of photometric redshift selection bias for a COSEBIs 4-bin tomographic analysis of a random intrinsic ellipticity field. The upper triangle shows the E-modes and the lower triangle shows the B-modes, where the error bars in both cases correspond to the level of ellipticity noise in the mocks. Each panel shows COSEBIs for a tomographic redshift bin pair, z-${ij}$, corresponding to the correlation between photometric redshift bins $i$ and $j$.   As our photometric redshift mocks are devoid of any cosmological correlations, in the absence of any selection bias, we would expect both the E and B modes to be consistent with zero.}}
     \label{fig:PSFzCOSEBIs4bins}
 \end{figure*}

 To see the full effect of the photometric redshift selection bias on the measured correlation function, we use COSEBIs to analyse four photometric redshift bins, corresponding to those chosen in the KiDS-450 analysis. \fig\ref{fig:PSFzCOSEBIs4bins} shows the results revealing significant signal in the different tomographic slices, with the strongest effect in the highest redshift bins.   To quantify the significance of this effect we use a theoretical covariance to estimate a $\chi^2$ value for COSEBIs relative to the null hypothesis of zero signal.  We then calculate the $p$-value corresponding to that $\chi^2$ finding vanishingly small values of $p \sim 10^{-28}$ and $p \sim 10^{-15}$ for the E and B modes respectively, {clearly confirming the significance of the bias.}
 
As with the analysis of the shear measurement systematics in \sect\ref{sec:shearsysres} we see {similarities (generally low-$n$) and differences (generally high-$n$)} in the measured E and B-modes, again leading us to the conclusion that $B$-modes can be used as a diagnostic but cannot blindly be used to correct the E-modes.

\subsection{Cosmological parameter inference}

Although we can see the signature of each systematic in Figures~\ref{fig:2PCFSys} to~\ref{fig:PSFzCOSEBIs4bins}, it is not immediately clear how they would affect cosmological parameter inference.   One could carry out a likelihood analysis to find any biases introduced by these systematics \citep[see for example][]{AmaraRefregier08}, but our preference is to use compressed COSEBIs (CCOSEBIs, see \sect\ref{sec:CCOSEBIs}) as a faster alternative approach.  CCOSEBIs are formed of linear combinations of COSEBIs that are sensitive to cosmological parameters. If the systematics that we identify in the COSEBIs analyses are null in both the E and B mode CCOSEBIs case, then we can conclude that the systematics are unlikely to be detrimental to the cosmological inference.

In this analysis we focus on the CCOSEBIs that are sensitive to $\Sigma_8=\sigma_8(\Om/0.3)^\alpha$, as this is the combination of parameters that cosmic shear data are mostly sensitive to. We find that for a KiDS-450 redshift distribution, $\alpha=0.65$ best describes the COSEBIs degeneracy direction (see \App\ref{app:CCOSEBIs}).  Because we are interested in $\Sigma_8$ we only consider the 5 first and second-order CCOSEBIs for $\sigma_8$ and $\Om$; $E^c_{\sigma_8}$, $E^c_{\Omega_{\rm m}}$, $E^c_{\sigma_8 \sigma_8}$, $E^c_{\sigma_8\Om}$ and $E^c_{\Om\Om}$, where $E^c$ is defined in \Eqt\eqref{eqn:Ec}. Although it is possible to construct a CCOSEBIs mode that is sensitive to $\Sigma_8$ directly, we choose to look at these 5 modes instead  in order to also provide an internal consistency check.   For each of the five modes we calculate a single compressed value, $E^c_{\mu(\nu)}$, that can be compared to its expectation value, given a set of cosmological parameters, noise covariance and source redshift distribution.

\begin{figure*}
   \begin{center}
     %\begin{tabular}{c}
     \resizebox{170mm}{!}{\includegraphics{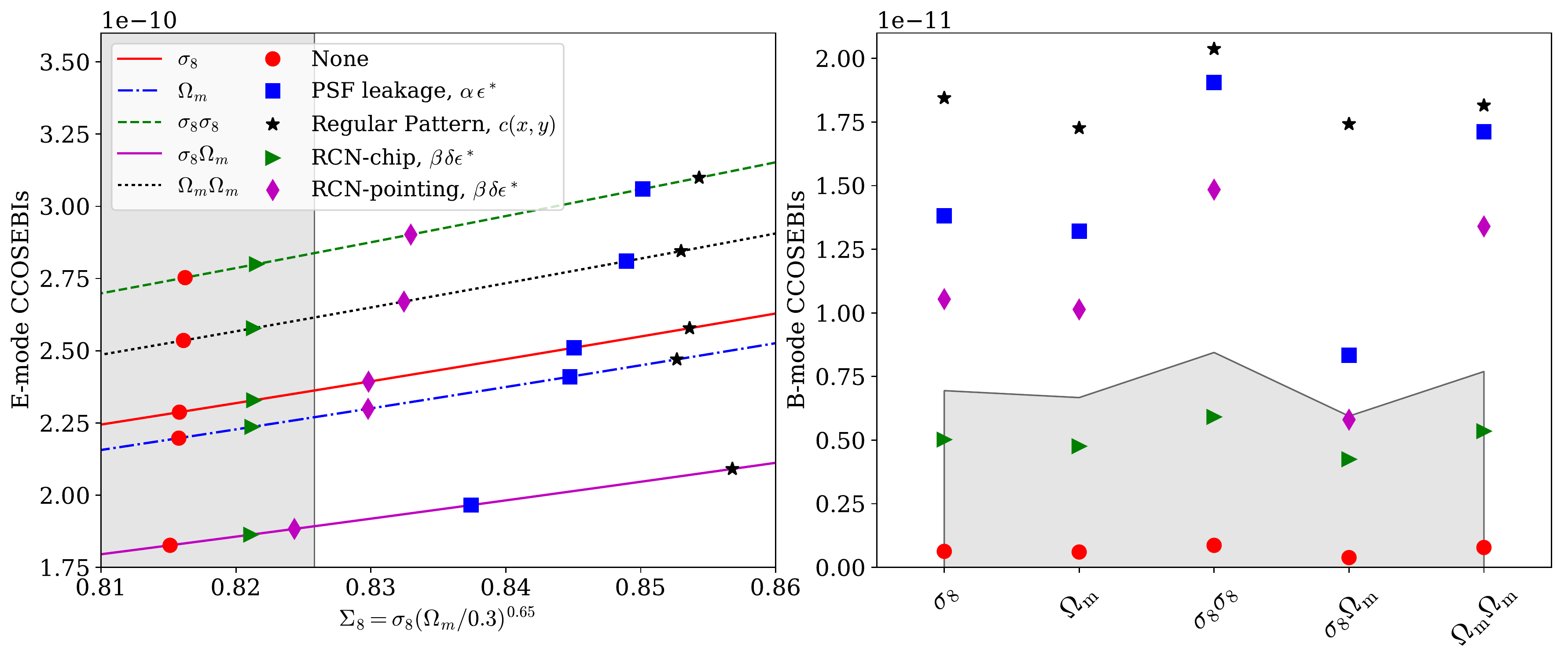}}
     %\end{tabular}
   \end{center}
     \caption{\small{{\it Left panel}: the inferred values of $\Sigma_8=\sigma_8(\Omega_{\rm m}/0.3)^{0.65}$ from an E-mode CCOSEBIs analysis of four mock cosmic shear surveys that suffer from PSF leakage (blue squares), a repeating additive pattern (black stars), and random but correlated noise on chip (green triangles) and pointing (magenta diamond) scales.   The curves show the theoretical values of the 5 E-mode CCOSEBIs when varying $\Sigma_8$, calculated using the KiDS-450 noise only covariance matrix and redshift distribution; $E_{\sigma_8\sigma_8}$(dashed), $E_{\Omega_{\rm m}\Omega_{\rm m}}$ (dotted), $E_{\sigma_8}$ (middle solid), $E_{\Omega_{\rm m}}$ (dot-dashed) and $E_{\sigma_8\Omega_m}$ (lower solid).   The inferred cosmology for each mock systematic survey can be compared between the 5 different modes.  The higher the recovered E-mode is relative to the fiducial `no-systematics case' (circles), the stronger the bias is on the inferred value of $\Sigma_8$ and the more discrepant the inferred cosmology is between the different CCOSEBIs modes.  The bias in $\Sigma_8$ can be compared to the grey region which shows the one-sigma error for $\Sigma_8$ from the KiDS-450 cosmic shear analysis, centred on the fiducial case.   {\it Right panel:} B-mode CCOSEBIs from the four mock cosmic shear surveys. The measured B-mode signal, {which does not depend on $\Sigma_8$}, can be compared to the shape noise on a KiDS-450-like survey (shown in grey).}}
\label{fig:CCOSEBIsSys}
 \end{figure*}

\fig\ref{fig:CCOSEBIsSys} shows the measured E-mode (left panel) and B-mode (right panel) CCOSEBIs for the full angular range of $[0.5', 100']$.   The symbols correspond to the range of shear measurement systematics\footnote{Note that as we do not include a cosmic shear signal in the photometric redshift selection bias mocks developed in \sect\ref{sec:sysZPSF}, we do not present a CCOSEBIs analysis of this systematic.} simulated using the SLICS cosmic shear simulations in \sect\ref{sec:Systematics}.  The lines connecting the E-mode points indicate which of the 5 CCOSEBIs modes are shown and also show their theoretical value. Each measured E-mode can be compared to the value of $\Sigma_8=\sigma_8(\Om/0.3)^{0.65}$ that would be inferred from the measurement.  In the absence of systematic errors, we would expect to find the inferred parameters to be consistent with each other and the input SLICS cosmology with $\Sigma_8 = 0.808$.   We would also expect to find the B-mode signal consistent with zero, but looking at the fiducial `no-systematics' mocks (circles), we do recover a very small residual B-mode and a slightly high best-fit $\Sigma_8 = 0.815$.  This result is expected, however, given the imperfect match between the two-point statistics measured from SLICS and the theoretical expectation shown in \fig\ref{fig:MockvsTheory}. Here no errors are associated to the E-mode CCOSEBIs, because the mock data used to produce them are free of shape-noise.

We find that the introduction of the random, but correlated noise {(RCN)} increases the recovered $\Sigma_8$ value, but within the statistical tolerance of KiDS (shown as a grey bar) in the case of the chip-scale correlation.  The PSF leakage and repeating additive pattern result in the largest bias in cosmological parameters with a $\sim 5\%$ deviation from the true input cosmology.  Applying this level of bias to either $\sigma_8$ or $\Om$ can produce excess correlations of only up to $13\%$ which is significantly less than the up to $40\%$ biases seen in the two-point correlation function analysis in \fig\ref{fig:2PCFSys}, from which we can conclude that the impact of these systematics on the data can be only weakly correlated with the impact of varying cosmological parameters.  

We find that the stronger the bias in the recovered cosmology, the larger the inconsistency between the 5 CCOSEBIs modes, providing another important diagnostic tool.   We also note that all the shear measurement systematics tested in this analysis serve to increase the inferred value of  $\Sigma_8$.   If these types of systematics were present in the weak lensing data, {correcting for them would decrease the recovered} $\Sigma_8$, exacerbating the current hints of cosmological parameter tension between weak lensing surveys and Planck \citep[see for example][]{Troxel18}. 

Comparing the power in the measured E-modes (left panel) and B-modes (right panel) reveals a close connection, where a larger bias in the E-modes corresponds to larger B-modes. We note that although the magnitude of the E/B-modes are connected, they can take opposite signs.  For example, in \fig\ref{fig:COSEBIsSys} we see that for the large angular scale analysis with the repeating additive pattern systematic, the sign of the first four E-modes differs from the sign of the first four B-modes. {Although we do not show the large-scale CCOSEBIs result, we can confirm that, this difference in sign is also reflected there, as the CCOSEBIs  are sensitive to the first few modes that contain a large proportion of the cosmological information.} This example demonstrates that although the large angular scales have the lowest signal to noise, they can and should be used as an investigative tool for hunting systematics that could also impact small angular scales. 

%% file: Discussion.tex
\label{sec:Discussion}
In this section we discuss how we can use the measured COSEBIs B-mode signatures from our systematics mocks in \sect\ref{sec:ResultsMock} to diagnose the origin of the B-modes recovered in the CFHTLenS, DES and KiDS surveys in \sect\ref{sec:ResultsData}.  Firstly, we focus on the non-tomographic COSEBIs B-mode measurements in \fig\ref{fig:COSEBIsdata1bin}.  One feature that stands out for all three surveys is the high $n$-mode oscillatory pattern in the full angular range, shown in the middle right panel.    This oscillatory pattern is the signature of a repeating additive systematic, shown in \fig\ref{fig:COSEBIsSys}, which we find to be the systematic that was most detrimental to cosmological parameter estimation in \fig\ref{fig:CCOSEBIsSys}.

We find the level of B-modes for KiDS-450 and CFHTLenS for higher $n$-modes to be small and hence the repeating additive signature is not highly significant in these cases.  The similarity between the B-modes in the data and this systematic signature does however warrant further exploration, particularly as we also see similarities in the E-modes for KiDS and the repeating pattern E-mode signature.  Here the unexpected E-mode `secondary peak' seen at $n \sim 6$ in the small angular scales of KiDS-450 E-modes (upper left panel of \fig\ref{fig:COSEBIsdata1bin}) is replicated at $n \sim 6$ in the E-mode analysis of the same angular scales of the repeating additive bias mocks (upper left panel of \fig\ref{fig:COSEBIsSys}).  If a repeating additive systematic persists it would likely become significant in future releases of KiDS.  It could also be responsible for the power seen in the low-$n$ modes that lead to the significant detection of the KiDS CCOSEBIs B-modes.

For DES-SV we find a significant detection of B-modes, noting that in addition to the high-$n$ oscillatory pattern, DES-SV presents significant additional signal for modes around $n=8$ and $n=4$.  For an instrument-based repeating additive pattern, the resulting B-mode signature will depend on the dithering strategy and camera field-of-view.  Both KiDS-450 and CFHTLenS have a field-of-view of $\sim 1$ deg$^2$ with small dithers.  DES-SV, however, has a hexagonal field-of-view, 2.2 degrees across, and uses half-field dithers.   This means the frequency of any repeating additive pattern will differ for DES-SV in comparison to the KiDS-like imaging strategy that we have simulated in our mocks.   Looking at only the first few modes for DES-SV data, however, we find that the signal resembles the signature of both PSF leakage, and random but correlated noise on the pointing level.  This result is consistent with the findings of \cite{Zuntz/etal:2018} who report and correct for a small but significant PSF residual  in their analysis of the first year of DES observations.  We therefore conclude that the B-mode signature recovered for DES-SV is likely a superposition of different shear measurement systematics.  

By comparing the $p$-values in \tab\ref{tab:pvalue} we can see that for DES-SV the significance of the COSEBIs B-modes substantially increases when the data are separated into tomographic bins. This could be understood by considering the photometric redshift selection bias explained in \sect\ref{sec:sysZPSF}. This systematic correlates the PSF ellipticity with the redshift estimation for a galaxy, and can produce significant B-modes when the data are binned into smaller photometric redshift bins.   It is likely that all surveys will suffer from this systematic to some degree, but the level will depend on how the multi-band photometry is measured in each survey and how the PSF ellipticity varies in each optical band.   We cannot directly compare our mock analysis with the B-modes in the DES-SV tomographic analysis, but our first-look at this effect certainly motivates further exploration with more detailed simulations that fully mimic the photometric redshift measurement in each survey.   

Interestingly, comparing the DES-SV tomographic and non-tomographic $p$-values in \tab\ref{tab:pvalueCCOSEBIs},  we find that for the analyses that include small-scale information, the significance of the B-modes, measured using the cosmological-parameter-sensitive CCOSEBIs, decreases when the data are separated into tomographic bins. 
This promising result means that if the systematic that was introduced when the DES-SV tomographic selection is applied adds equal power to the E and B modes, that systematic would not introduce modifications to the E-mode signal that would bias the inferred cosmological parameters.   Unfortunately, however, the photometric redshift selection bias systematic was found to exhibit different E and B mode signals in \fig\ref{fig:PSFzCOSEBIs4bins}.   Passing the CCOSEBIs B-mode null-test therefore cannot validate the CCOSEBIs E-mode measurement.  In addition, this CCOSEBIs B-mode result does not hold for the large angular scales, $[40',100']$, where again we see a substantial increase in the measured B-mode when the data are separated into tomographic bins.   

Our findings for DES-SV contradict \cite{Becker15} who conclude that the B-modes in DES-SV are insignificant using two Fourier space methods.   We argue that as power spectra pick up features of the data at different scales compared to $\xi_\pm$, they are not suitable statistics for verifying the absence of B-modes in a $\xi_\pm$ cosmic shear analysis. In addition, power spectra measurements are binned in a range of Fourier modes, such that any high-frequency variations in Fourier space will average out. COSEBIs are sensitive to these variations and can therefore be used to diagnose the origin of the B-modes in the data.  

At first sight our findings for KiDS also contradict \cite{hildebrandt/etal:2017} who report a low-level but significant detection of $\xi_{\rm B}$.  This is in contrast to our tomographic $\xi_{\rm B}$ analysis which concludes that $\xi_{\rm B}$ is consistent with zero {(see \fig\ref{fig:XiKiDStomo})}.    We find that the $\xi_{\rm B}$ statistic is sensitive to the choice of the maximum $\theta$-scale measured from the data and the maximum $\theta$-scale used for completing the integral to infinity using a theoretical prediction (in this analysis we use $1000'$ instead of $3000'$ used in \citealt{hildebrandt/etal:2017}).   We also find that $\xi_{\rm B}$ is sensitive to the method used to bin $\xi_\pm$ as explained in \sect\ref{app:binning} (see \Eqt\ref{eq:binningSimple}).  This sensitivity to data analysis choices provides another reason to archive the traditional $\xi_{\rm E/B}$ approach.   In this paper we promote COSEBIs as the optimal statistic for both E and B mode measurements as it can be estimated accurately and free of any biases connected to binning and extrapolating the data.    Analysing all 20 COSEBIs modes, we find no significant evidence for B-modes in KiDS.  In our compressed CCOSEBIs analysis, however, we arrive at the same conclusion of both \cite{hildebrandt/etal:2017} and \cite{vanuitert/etal:2017}, that low-level but significant B-modes are present in KiDS-450.  In our $[0.5',100']$ CCOSEBIs analysis, we find a $\sim 2.7\sigma$ detection of a B-mode signal that is less than 10\% of the amplitude of the E-mode.  This difference between the significance of the COSEBIs and CCOSEBIs B-mode analysis might seem confusing or even contradictory.   We therefore refer the reader to \App\ref{app:Model Distinction} where we explore how choices over the number of modes used in a null-test can dilute or optimise the detection of systematics.

%% file: Conclusions.tex
%conclusions
\label{sec:Conclusions}

Two-point shear correlation functions (2PCFs) have been the primary observables in cosmic shear analysis to date, but they are not immune to systematics. These statistics mix E and B-modes in the data, giving rise to a mixed lensing and non-lensing signal in the presence of systematic errors.  In order to test for systematics most surveys turn to alternative statistics to separate E/B-modes, using $\xi_{\rm E,B}$ or power spectrum measurements.  We argue that these alternative statistics are biased as they depend on infinite integrals over 2PCFs and are sensitive to binning choices. In addition, treating the E/B-mode decomposition with a statistic that has a different scale-dependence to the statistic used in the cosmological parameter inference, causes a disparity in the analysis.  For future cosmic shear analyses, we therefore advocate the use of COSEBIs for both parameter inference and systematic analyses (see \sect\ref{sec:COSEBIs}). COSEBIs cleanly and completely separate E/B-modes over a finite angular range, without loss of information. They have discrete modes and therefore are insensitive to binning choices. The first few modes of COSEBIs contain almost all of the cosmological information and as such a COSEBIs analysis is also an efficient approach to data compression. {For a B-mode analysis, however, a larger number of modes need to be considered, as systematics can affect the E and B-mode at different scales. }

In this paper we analysed the E and B-mode signals in three public cosmic shear surveys, CFHTLenS, DES-SV and KiDS-450.  We compared the $\xi_{\rm E,B}$ statistic with COSEBIs and CCOSEBIs, using $p$-values to quantify the level of B-modes in the data. To determine COSEBIs filter functions we need to first define an angular range of interest. For this study we chose three sets of angular separation ranges: small separations, $[0.5', 40']$, large separations, $[40',100']$, and the overall separation range, $[0.5',100']$.
We measure COSEBIs up to mode $n=20$.  We considered two cases for each survey; one using the same redshift bins as used in each survey's primary cosmic shear analysis, and another combining those bins into a single redshift bin.  We see that for DES-SV data the tomographic cases show significant B-modes at a level between $4\sigma$ and $5.5\sigma$.  For the non-tomographic DES-SV analyses, B-modes are detected at the level of $2.8 \sigma$.  For KiDS-450 and CFHTLenS, we find no significant detection of B-modes for the majority of our analyses.  There is however some exceptions in each case.  The CCOSEBIs analysis of the small separations (non-tomographic case only) and the analysis of the full angular range show B-modes at up to $2.7 \sigma$ for KiDS-450.  The tomographic COSEBIs analysis over the small angular range $[0.5', 40']$ detects a B-mode signal at $2.8 \sigma$ for CFHTLenS.

In order to diagnose the origin of the B-modes detected in each survey,  we modelled several non-astrophysical systematic effects relevant to current data in order to determine their E/B mode signature and assess their impact on cosmological parameter inference.  We modelled four shear measurement systematics. PSF-leakage, was modelled using the mosaic PSF pattern from KiDS-450 assuming a $10\%$ leakage with $\alpha = 0.1$. An instrument-based additive bias term resulting in a repeating pattern from pointing to pointing.  Here we used the low-level CCD bias of OmegaCam \citep{Hoekstra2018}, multiplied by a factor of five to amplify and model this effect.  To model biases arising from random PSF modelling errors, we correlated low levels of random noise using two kernel sizes, corresponding to roughly KiDS CCD and pointing scales. In addition to these shear measurement systematics, we { have introduced a new effect by modelling} the impact of photometric redshift selection bias that arises from the correlation between the relative orientation of PSF ellipticity and galaxy ellipticity, and the measured signal-to-noise of the galaxy.

All of the systematics simulated were detected in our B-mode analysis.   The PSF-leakage and random but correlated noise systematics introduced low $n$-mode COSEBIs signal.  This was in contrast to the repeating additive bias which introduced high frequency variations in the shear field which are picked up as oscillatory behaviour in the high $n$-mode COSEBIs measurements.    Photometric redshift selection bias also resulted in high $n$-mode power in the high photometric redshift bins.   Comparing the B-mode signatures recovered by our mocks to the B-modes measured in each survey we conclude that DES-SV is likely to suffer from a combination of all the systematics that we have simulated.  The significant increase in DES-SV B-modes when the tomographic redshift selection is applied is particularly striking, motivating future work to enhance the realism of the first-look photometric redshift simulations that we have analysed in this paper.   KiDS-450 and CFHTLenS show oscillatory behaviour in the high $n$-mode indicating a repeating additive bias in the data, although this result is not significant. 

The simulated systematics produce E-modes that would bias cosmological parameter inference.  For the analysed shear measurement systematics we found that $\Sigma_8=\sigma_8(\Om/0.3)^{0.65}$ is biased high in all cases. As a result, we conclude that these types of systematics, if present, cannot explain the mild tension between some current cosmic shear and Planck results {and could in principle exacerbate the tension as they bias $\Sigma_8$ to even higher values (see \fig\ref{fig:CCOSEBIsSys})}.   It is interesting to note that the DES-SV cosmological parameter constraints on $S_8=\sigma_8(\Om/0.3)^{0.5}$ are higher than those from KiDS-450, CFHTLenS and the first year DES results, which include a number of improvements over the DES-SV analysis.   Given the significant DES-SV B-modes detected in our analysis, the direction of the difference in $S_8$ between the surveys is expected.  The published cosmological parameter constraints from all three surveys are, however, in good agreement. 

For the analysis of KiDS-450, we find an interesting case where the survey formally passes the COSEBIs B-mode analysis, but a flag is raised with a $2.7 \sigma$ B-mode detection in the compressed CCOSEBIs analysis. Here the COSEBIs B-modes that are insignificant overall are weighted in such a way that the resulting CCOSEBIs signal becomes significant.  We therefore recommend measuring both CCOSEBIs and COSEBIs B-modes in future analyses, ensuring that both are consistent with zero. The CCOSEBIs B-modes will robustly identify systematics that will lead to a bias in the cosmological parameter inference, if the systematic impacts the E/B-modes in the same way.  In contrast the COSEBIs B-modes detect systematics that can affect E and B-modes differently.  {For this reason we have shown that it is not sufficient to correct data by simply subtracting the B-modes from the E-modes.} 
As we have seen from our repeating additive bias systematic, where the E and B mode behaviour is very different, it will be crucial to look
{at both a COSEBIs and CCOSEBIs analysis and return to correct the input catalogues if a significant B-mode is detected in either case.}

%% file: binning.tex
\label{app:binning}
When using the traditional $\xi_\pm$ statistic an issue arises from the choices that can be made when binning the data. $\xi_\pm(\theta)$ is usually binned in broad $\theta$-bins as a form of data compression. As we expect the number of galaxy pairs to roughly scale with $\theta$, see \Eqt\eqref{eqn:Npairapprox}, the sampling of $\xi_\pm(\theta)$ within the bin is non-uniform.  If we bin these functions into broad angular bins, their value will therefore be biased towards larger $\theta$ scales in each bin. This is not an issue provided the theory is treated in the same way, but this is not the standard approach that is taken, as it is computationally more expensive.  \cite{Troxel18} compare the differences in KiDS-450 cosmological parameter inference if one takes the logarithmic mid-point of the bin or the weighted mean value of $\theta$ in each bin and evaluate the theoretical 2PCFs at each $\theta$ value.   They argue that the latter approach is correct, supported by \citet{Krause2017} who conclude that this approach is sufficiently accurate for the first year DES analysis.  Here we provide more detail on the question of binning bias,  quantifying how inexact each treatment of the theory is, where we find up to 10\% biases in both approaches.  A full integration of the theory within the bin is the correct approach to this problem.  If future surveys wish to use an approximation, however, we demonstrate that using the linear mid-point of the bin provides the closest match to the binned data, with less than $2.5\%$ bias.

\subsection{Binning theory}
Consider making measurements of a function $f(x)$ from noisy data, with samples drawn non-uniformly in $x$.  We denote the sampled data points by $f_{\rm data}(x)$ and the distribution of measured $x$ by $\D(x)$. 
Given that the sampled data points are noisy, we want to combine them to find an estimate for the function for a given binning in $x$. 
One way to bin the data is to write it as
\begin{equation}
\label{eq:binningDef}
\hat{f}_{\rm binned}(x_{\rm b})=\frac{1}{N_{\rm bin}}\sum_x f_{\rm data}(x)\, \Delta(x-\xw) \;,
\end{equation}
where $N_{\rm bin}$ is the number of data points in the given bin and $\Delta(x-\xw)$ is the binning function defined as 
\begin{equation}
\Delta(x-\xw) =
  \begin{cases}
      1    & \quad {\rm if\; } x\; {\rm\; is \; in \; bin}\; \xw\;,\\
      0    & \quad {\rm if\; } x\; {\rm\; is \; not\; in\; bin\;} \xw.
  \end{cases}
\end{equation}
This estimate for the binned function corresponds to a weighted binning, with more weight given to the values where there are more sampled points. The expectation value of $\hat{f}_{\rm binned}(x_{\rm b})$ is in general not equal to $f(\xw)$,
\begin{equation}
\label{eq:binningExpect}
\langle\hat{f}_{\rm binned}(x_{\rm b})\rangle=\frac{\int_{x_{\rm min}(\xw)}^{x_{\rm max}(\xw)} \d x\, f(x) \D(x)}{\int_{x_{\rm min}(\xw)}^{x_{\rm max}(\xw)}\d x\, \D(x)}\neq f(\xw)\;,
\end{equation}
where $x_{\rm min}(\xw)$ and $x_{\rm max}(\xw)$ are the edges of bin $\xw$. Even if we define $\xw$ as the weighted mean of the $x$-values in the bin, as advocated by \cite{Troxel18},
\begin{equation}
\xw=\frac{\int_{x_{\rm min}(\xw)}^{x_{\rm max}(\xw)} \d x\, x\, \D(x)}{\int_{x_{\rm min}(\xw)}^{x_{\rm max}(\xw)}\d x\, \D(x)}\;,
\end{equation}
we would only recover the true value of the binned data, if $f(x)$ is either a constant or has a linear relation to $x$.
 
Let's now take very fine $x$-bins, such that the variation in the sampling of $f(x)$ is negligible, i.e. $\D(x)\approx$ constant within each bin. The expectation value of the finely binned function $\langle\hat{f}_{\rm f}(\xf) \rangle$, where the subscript f represents "fine" and $\xf$ is the mid-point of the fine bin, is given by
\begin{align}
\label{eq:binningFine}
\langle\hat{f}_{\rm f}(\xf)\rangle & = \frac{\int_{x_{\rm min}(\xf)}^{x_{\rm max}(\xf)} \d x\, f(x)\, \D(x)}{\int_{x_{\rm min}(\xf)}^{x_{\rm max}(\xf)} \d x\, \D(x)} \\ \nonumber 
&\cong \frac{1}{2\,\delta x}\int_{\xf-\delta x}^{\xf+\delta x} \d x\, f(x)\cong f(\xf)\;,
\end{align}
where $2\,\delta x$ is the width of the fine bin. If we first measure finely binned $\hat{f}_{\rm f}(\xf)$ from the data, then we have the flexibility to re-bin the measurements as desired,
\begin{equation}
\label{eq:binningWideFine}
\hat{f}_{\rm w}(x_{\rm b})=\frac{\sum_\xf w(\xf) \hat{f}_{\rm f}(\xf)\Delta(\xf-\xw)}{\sum_\xf w(\xf)\Delta(\xf-\xw)} \;,
\end{equation}
where $w(\xf)$ is a weight function assigned to each fine bin. If we chose to set $w(\xf)$ to $\D(\xf)$ we would recover the weighted binning defined in \Eqt\eqref{eq:binningDef}. 

If we choose $w(\xf)$ such that it does not vary between different realisations of the data, the expectation value of $\hat{f}_{\rm w}(x_{\rm b})$ is given by
\begin{equation}
\label{eq:binningWideExpenctation}
\langle\hat{f}_{\rm w}(x_{\rm b})\rangle=\frac{\sum_\xf w(\xf) f(\xf)\Delta(\xf-\xw)}{\sum_\xf w(\xf)\Delta(\xf-\xw)}\;,
\end{equation}
and the covariance of the binned data for bins $x_{\rm b}$ and $y_{\rm b}$  is given by
\begin{align}
\label{eq:binningCov}
& \Cov_{\rm w}(x_{\rm b},y_{\rm b})  =\langle\hat{f}_{\rm w}(x_{\rm b})\hat{f}_{\rm w}(y_{\rm b})\rangle-\langle\hat{f}_{\rm w}(x_{\rm b})\rangle \langle\hat{f}_{\rm w}(y_{\rm b})\rangle \\ \nonumber
&=\frac{\sum_\xf \sum_\yf w(\xf) w(\yf)\Cov_{\rm f}(\xf,\yf) \Delta(\xf-\xw)\Delta(\yf-\yw)}{\sum_\xf \sum_\yf w(\xf)w(\yf)\Delta(\xf-\xw) \Delta(\yf-\yw)}\;,
\end{align}
where $\Cov_{\rm f}(\xf,\yf)$ is the covariance of the finely binned measurements,  $f_{\rm f}(\xf)$ and $f_{\rm f}(\yf)$. If we assume no cross-correlation between the bins, which is the case for a shape-noise only covariance, then \Eqt\eqref{eq:binningCov} simplifies and the variance of $f_{\rm w}(x_{\rm b})$ can be written as,
\begin{eqnarray}
\label{eq:binningVariance}
\sigma_{\rm w}^2(x_{\rm b})  &=&\langle\hat{f}_{\rm w}^2(x_{\rm b})\rangle-\langle\hat{f}_{\rm w}(x_{\rm b})\rangle^2 \\ \nonumber
&=&\frac{\sum_\xf w^2(\xf)\sigma^2_{\rm f}(\xf)\Delta(\xf-\xw)}{\sum_\xf w^2(\xf)\Delta(\xf-\xw)}\;.
\end{eqnarray}
From this equation we can see that the variance of the binned data is also not equal to the variance of the function at $\xw$,
 $\sigma_{\rm w}^2(x_{\rm b}) \neq \sigma^2(x_{\rm b}) $, which complicates the calculation of covariance matrices for binned data.

To simplify covariance calculation we can set the weights in \Eqt\eqref{eq:binningWideFine} equal to unity and obtain an unweighted rebinned estimate,
\begin{equation}
\label{eq:binningSimple}
\hat{f}_{\rm unweighted}(x_{\rm b})=\frac{\sum_\xf \hat{f}(\xf)\Delta(\xf-\xw)}{\sum_\xf \Delta(\xf-\xw)}\; .
\end{equation}
In this case the expectation value of the estimator is,
\begin{equation}
\label{eq:binningSimpleExpectation}
\langle\hat{f}_{\rm unweighted}(x_{\rm b})\rangle= \frac{1}{2\, \Delta x}\int_{x_{\rm min}(x_{\rm b})}^{x_{\rm max}(x_{\rm b})} \d x\, f(x)\;,
\end{equation}
where $2\, \Delta x$ is the width of the bin. If the relative variation of the sampled points within a broad bin is large, then this estimator may not be optimal and can produce larger errors compared to the estimator in \Eqt\eqref{eq:binningWideFine}. 

Equation\thinspace\eqref{eq:binningWideExpenctation} is useful for predicting the theoretical value of the binned function, especially when the sampling frequency of the data points, $\D(x)$, is derived from the data itself. In the case of cosmic shear $\xi_\pm(\theta)$ the sampling of the data points roughly scales with $\theta$, however, survey geometry and masking effects together with variations in the depth of the images complicates the analytical estimation for the distribution of data in angular scale. Hence we suggest measuring $\D(\theta)$ from the data and use \Eqt\eqref{eq:binningWideExpenctation} to predict the binned $\xi_\pm$ values.

\subsection{Application to cosmic shear}
To demonstrate the level of bias introduced by partial treatment of the theory in a $\xi_\pm$ cosmic shear analysis we use a theoretical prediction for $\xi_\pm$ as our function, $f(x)$, assuming a single KiDS-450 redshift bin. To sample $\xi_\pm(\theta)$ in a non-uniform way, we randomly pick $\theta$ values from a $\D(\theta)=\theta/{\rm arcmin} \times 2000$ distribution in the angular range of  $[0.5',300']$. We then add a constant Gaussian random noise with $\sigma=0.01$ to each sampled point to produce the noisy sampled data points, $\xi_{\pm {\rm data}}(\theta)$ and then bin $\xi_{\pm {\rm data}}(\theta)$ into 1000 fine logarithmic bins to produce $\hat{\xi}_{\pm {\rm f}}(\theta_{\rm f})$ and 9 broad logarithmic bins to get $\hat{\xi}_{\pm {\rm binned}}(\theta_{\rm b})$ (see \Eqt\ref{eq:binningDef}). 

\begin{figure}
    \includegraphics[width=\hsize]{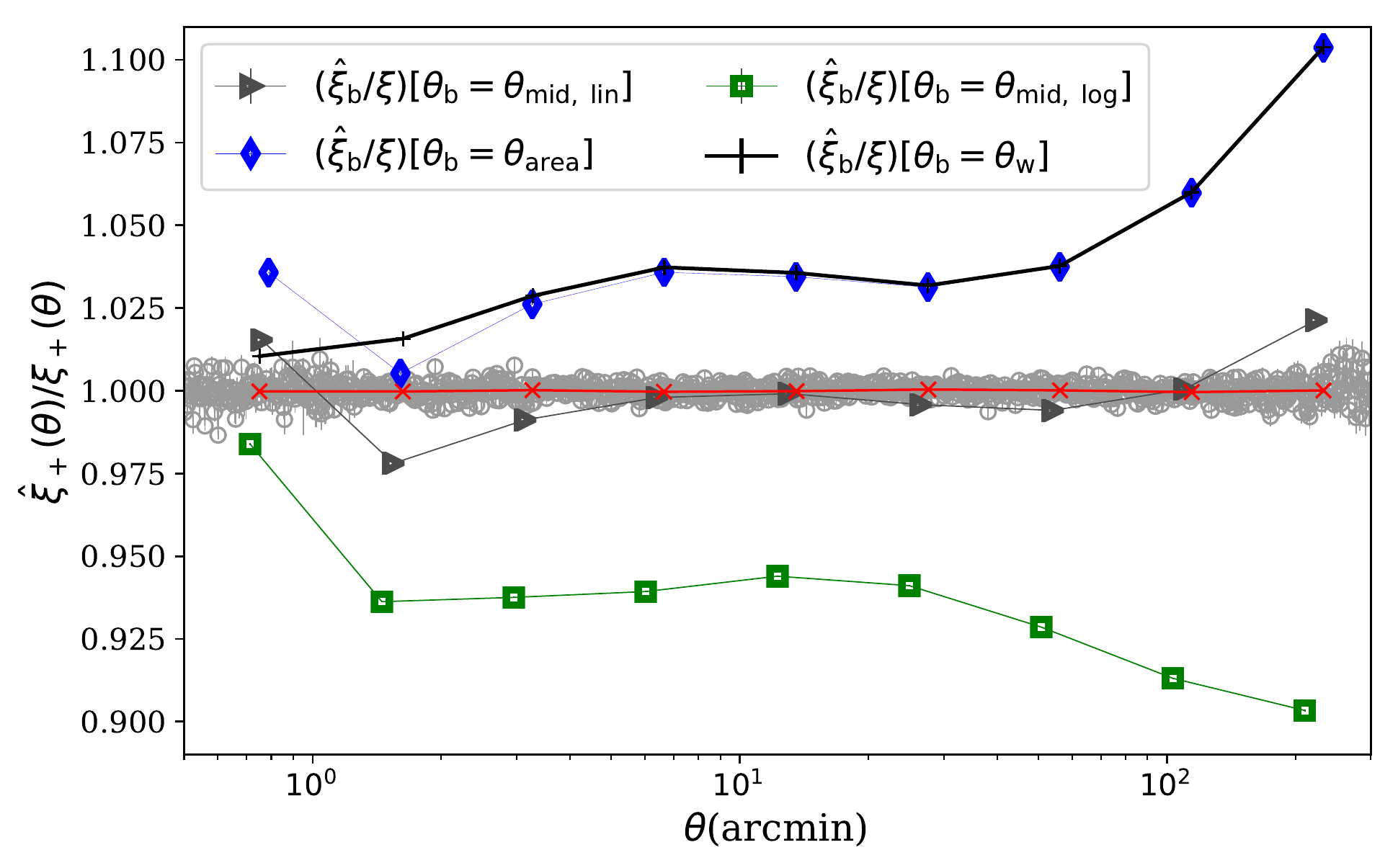}\\
    \includegraphics[width=\hsize]{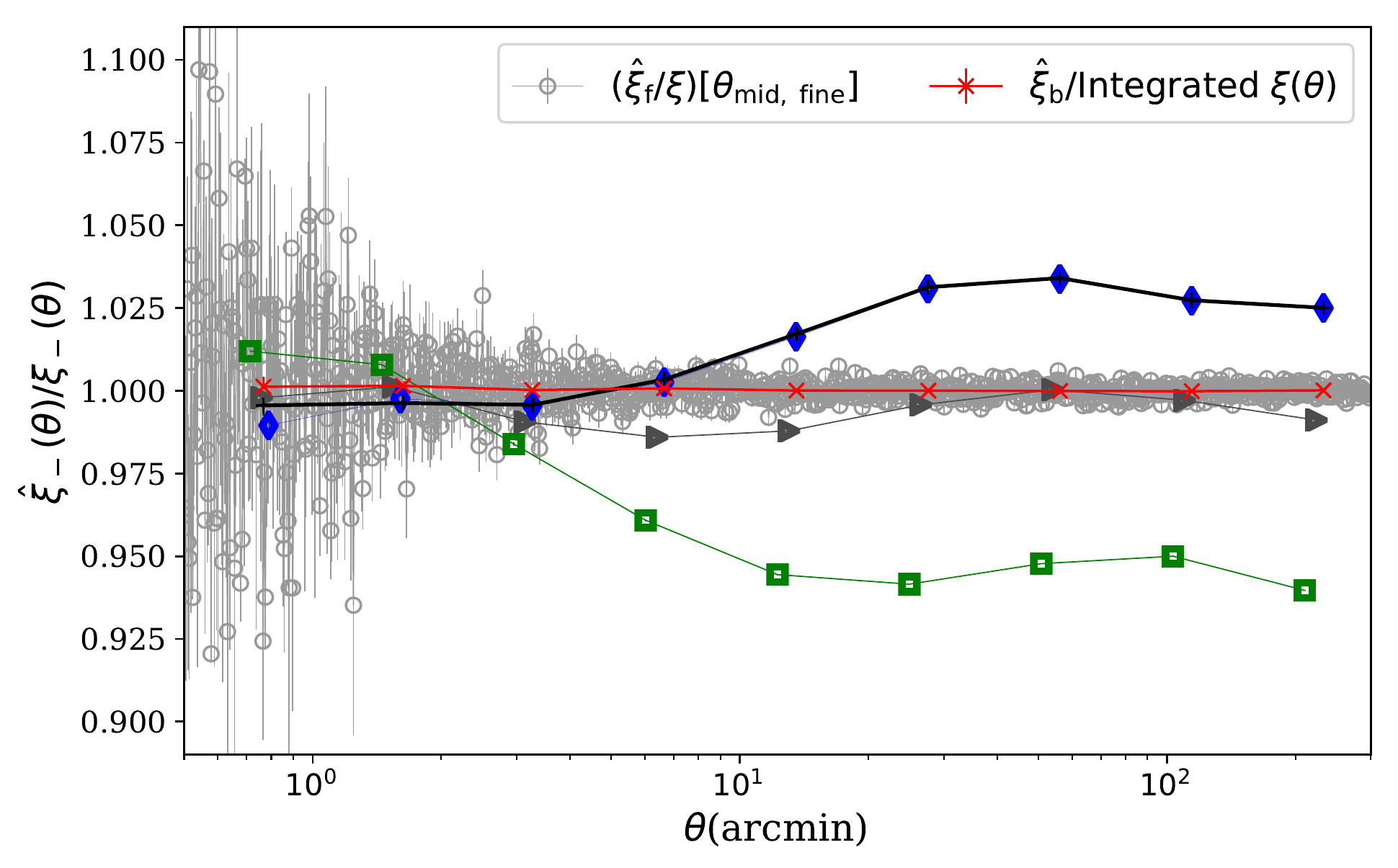}
        \caption{\small{2PCF binning bias introduced for a range of analysis choices, shown as the ratio between the measured $\hat{\xi}_\pm$ and their proposed theoretical value as a function of angular scale, $\theta$. The legends in this figure are shared between the two panels. For the weighted broad bin estimator, $\hat{\xi}_{\pm {\rm binned}}(\theta_{\rm b})$, the bias is calculated assuming $\theta_{\rm b}$ is given by the logarithmic mid-point of the bin, $\theta_{\rm mid, log}$ (green squares), the weighted mean of the bin, $\theta_{\rm w}$ (black pluses, under the blue diamonds),  the geometric mean or linear mid-point of the bin, $\theta_{\rm mid, lin}$ (dark grey triangles), or an area-weighted bin centre, $\theta_{\rm area}$ (blue diamonds). These estimators can be compared to the fine binning case, $\hat{\xi}_{\pm {\rm fine}}(\theta_{\rm mid,\; fine})$, where the theory is estimated at the weighted mean of each bin (light grey circles), and the exact case (red crosses) where the theoretical value is calculated as a weighted integral over the signal within the bin (\Eqt\ref{eq:binningWideExpenctation}). All points are plotted with errorbars, but in the case of broad binning the errors are too small to be visible.  }}
     \label{fig:RatioFunctionBinning}
 \end{figure} 

In \fig\ref{fig:RatioFunctionBinning} we show the binning bias introduced for a range of cases as a ratio between the measured and the proposed theoretical value of $\xi_+$ (top panel) and $\xi_-$ (bottom panel), as a function of angular scale, $\theta$.   The noisy finely binned data, $\hat{\xi}_{\pm {\rm f}}(\theta)$ (light grey circles), shows no significant bias relative to its expectation value (see \Eqt\ref{eq:binningFine}).  As was shown in \Eqt\eqref{eq:binningExpect} and \Eqt\eqref{eq:binningWideExpenctation} the expectation value of the broad binned,  $\hat{\xi}_{\pm {\rm binned}}(\theta_{\rm b})$, should be calculated using an integral over $\xi_\pm$ with the appropriate weights. The red crosses in the figure correspond to this theoretical prediction which is unbiased as expected. The remaining curves show the biases introduced when broad binning is applied to the measurements and the theory is evaluated at a single point in the bin denoted as $\theta_{\rm b}$. The green squares assume that $\theta_{\rm b}$ is given by the logarithmic mid-point of the bin \citep[as used in][]{hildebrandt/etal:2017}, for the black pluses $\theta_{\rm b}$ is the weighted mean of the bin \citep[as used in][]{Heymans13,troxel/etal:2017,Troxel18}, the grey triangles use  the geometric mean or linear mid-point of the bin (not used to date) and finally blue diamonds assume that $\theta_{\rm b}$ is the area-weighted bin centre \citep[as advocated by][]{Krause2017}, where
\begin{equation}
\label{eq:binningArea}
\theta_{\rm area}= \frac{\int_{\theta_{\rm min}}^{\theta_{\rm max}}\d\theta\, \theta 2\,\pi \theta}{\int_{\theta_{\rm min}}^{\theta_{\rm max}}\d\theta\, 2\pi \theta}=\frac{2(\theta^3_{\rm max}-\theta^3_{\rm min})}{3(\theta^2_{\rm max}-\theta^2_{\rm min})}\;.
\end{equation}
Here $\theta_{\rm min}$ and $\theta_{\rm max}$ are the minimum and maximum values of the bin.  

We find that the weighted mid-point and the area weighted values are similarly biased, boosting the signal at the $\sim 3 \%$ level at 10 arcmin, rising to $\sim 10 \%$ bias at large scales for $\xi_+$.  Taking the logarithmic mid-point of the bin has the opposite effect, decreasing the signal at $\sim 7 \%$ level at $10'$ and $\sim 10 \%$ bias at large scales for $\xi_+$.   That the biases work in the opposite sense here increases the inferred impact of binning bias when comparing the two KiDS analyses in \cite{Troxel18}. 

In all cases we see that the choice of binning affects $\xi_+$ more than $\xi_-$, since $\xi_+$ has more curvature than  $\xi_-$. We note that these biases will be smaller for narrower angular bins and as such their effect will not be as significant for the first year DES analysis \citep{troxel/etal:2017}  which uses the weighted mean for  $\theta_{\rm b}$ with roughly twice as many bins in the same angular range as shown here.

If future surveys conclude that it is too computationally expensive to calculate the impact of binning theoretically, especially in the case of the covariance matrix, our proposed solution is to use the linear mid point of the $\theta$-bin in the binned $\xi_\pm$ analysis.  We find that this approximation presents the weakest bias with at most $2.5\%$ bias at large and small scales and below percent level bias between $0.5' < \theta < 300'$.     Another alternative is to move to a COSEBIs analysis.  As COSEBIs are discrete they are not subject to any of the binning biases presented in this Appendix.

%% file: Appendix_CCOSEBIs.tex
\label{app:CCOSEBIs}
Cosmic shear is most sensitive to a combination of $\sigma_8$ and $\Omega_m$ \citep{Jain/Seljak:1997}, where the degeneracy can be written as 
\begin{equation}
\Sigma_8=\sigma_8\left(\frac{\Omega_{\rm m}}{\Omega^{\rm fid}_{\rm m}}\right)^\alpha\;.
\end{equation}
Here $\Omega^{\rm fid}_{\rm m}$ is arbitrary but is usually taken to be $0.3$.  In the majority of cosmic shear analyses $\alpha$ has been taken to be $\alpha = 0.5$,  even though the optimal value of $\alpha$ will depend on the statistic used, the redshift distributions and the angular ranges used in the analysis.  As an example, \cite{hildebrandt/etal:2017} present joint $\Sigma_8-\Omega_m$ constraints with $\alpha=0.5$.  The tilt seen in their Figure 6 of these constraints demonstrates that $\alpha=0.5$ does not best represent the degeneracy direction of $\Om$ and $\sigma_8$ for the KiDS-450 2PCF tomographic analysis.

\begin{figure}
   \begin{center}
     \begin{tabular}{c}
     \includegraphics[width=\hsize]{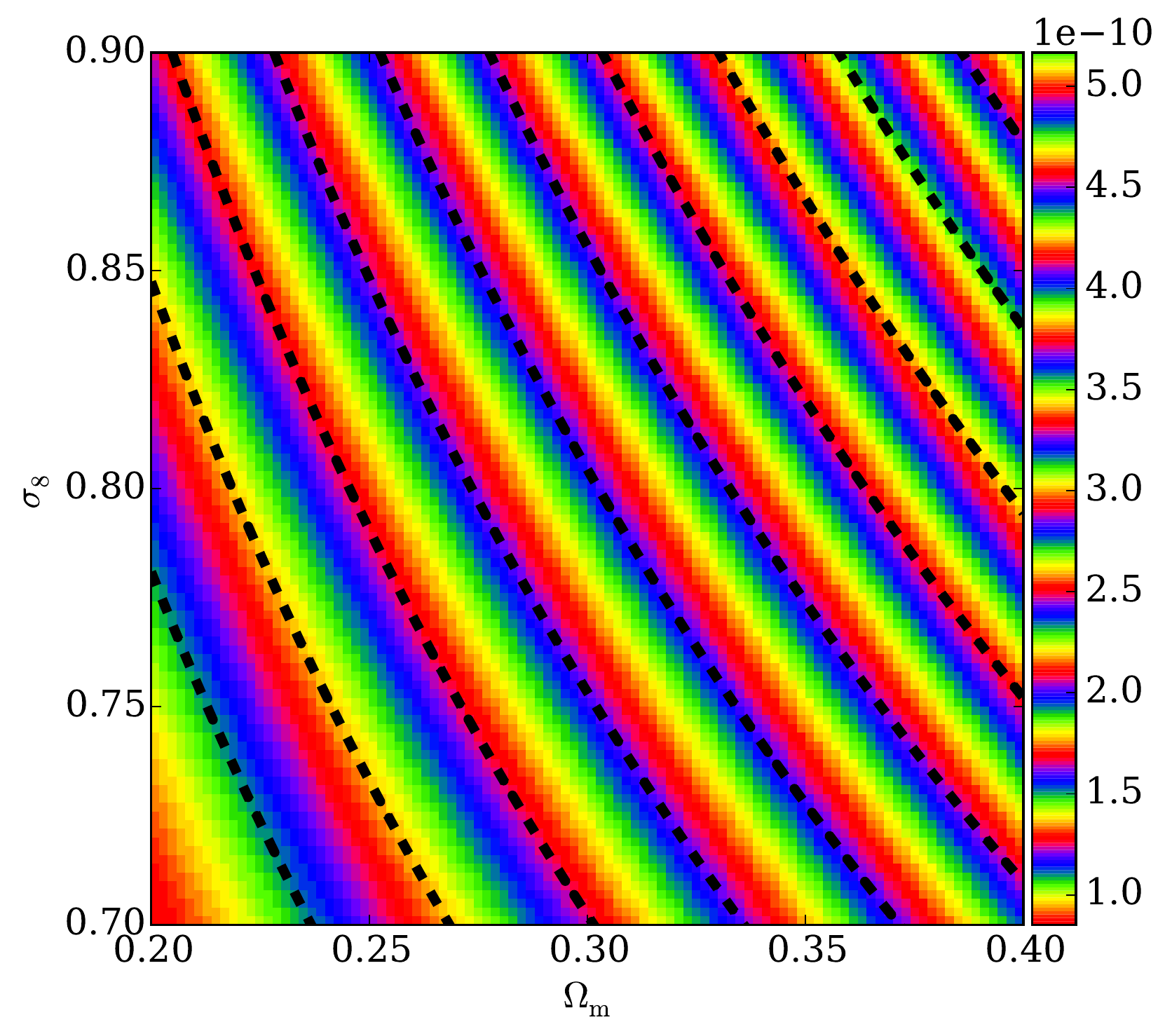}
     \end{tabular}
   \end{center}
     \caption{\small{The degeneracy direction of $\sigma_8$ and $\Omega_{\rm m}$ for a CCOSEBIs analysis of the KiDS-like data. The colours in the image show the value of the CCOSEBIs $E_{\sigma_8}$ mode, in comparison to dashed lines of constant $\Sigma_8=\sigma_8(\Omega_{\rm m}/0.3)^{\alpha}$ with $\alpha=0.65$. The repeating color scheme was chosen to capture the variations in the values of $E_{\sigma_8}$. The lower left corner has the smallest value of $E_{\sigma_8}$ which gradually increases, perpendicular to the dashed curves, towards the upper right corner.}}
     \label{fig:DegeneracyDirection}
 \end{figure}
 
In \fig\ref{fig:DegeneracyDirection} we show the value of the CCOSEBIs mode $E_{\sigma_8}$ (see colour bar) for a range of $\sigma_8$ and $\Omega_m$ values assuming a KiDS-like survey.   The degeneracy shown in $E_{\sigma_8}$  can be compared to the dashed lines of constant $\Sigma_8=\sigma_8(\Omega_{\rm m}/0.3)^{\alpha}$ where $\alpha = 0.65$.   We have carried out this test for all the CCOSEBIs modes in our analysis;  $E_{\Omega_{\rm m}}$,$E_{\sigma_8,\sigma_8}$, $E_{\sigma_8,\Omega_{\rm m}}$ and $E_{\Omega_{\rm m},\Omega_{\rm m}}$ to confirm that $\alpha = 0.65$ is an optimal choice for our CCOSEBIs analysis.

%% file: ModelDistinction.tex
\label{app:Model Distinction}
All null tests are subject to the choices we make in our data analysis.  As an example, if we limit our B-mode analysis of CFHTLenS to the first 7 COSEBIs modes,  following \citet{asgari/etal:2017}, we conclude there are no significant small scale B-modes in CFHTLenS.  In contrast, our analysis of the first 20 COSEBIs modes, in \sect\ref{sec:ResultsData}, finds a significant B-mode detection for CFHTLenS on the same scales.   In this Appendix we explore the question of how many COSEBIs modes should be used to determine the overall significance of the B-modes in a dataset.  

\begin{figure}
   \begin{center}
     \begin{tabular}{c}
      \includegraphics[width=\hsize]{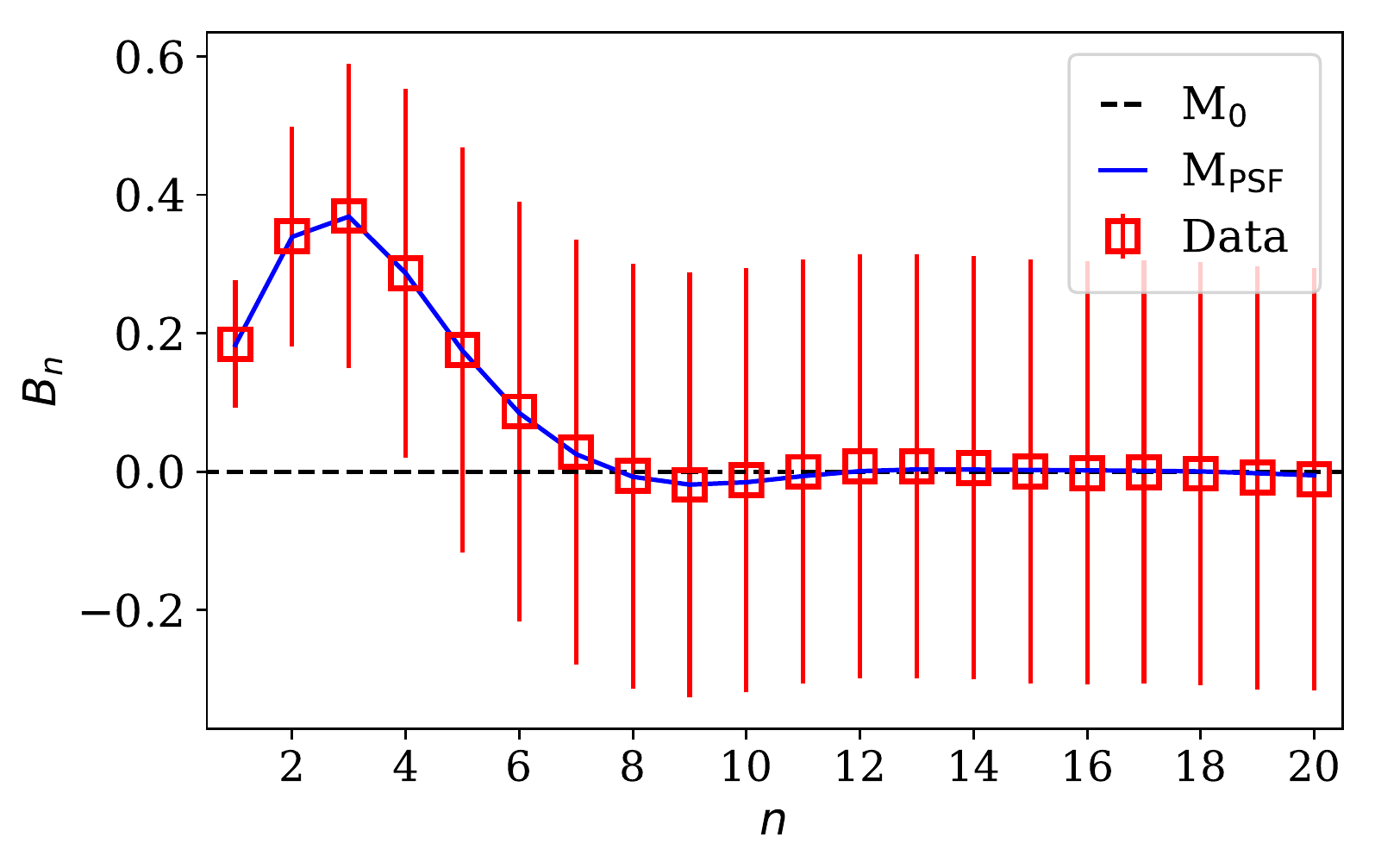}
     \end{tabular}
   \end{center}
     \caption{\small{Model comparison using $10,000$ random samples of the PSF-leakage systematic model, $\tens{M}_{\rm PSF}$, (solid) defined in \sect\ref{sec:PSFLeakage}.   The data points show the mean of the random samples, with the errors reflecting the noise on a single realisation.   The data samples are analysed to determine how often the input model $\tens{M}_{\rm PSF}$ can be distinguished from the no-systematics zero B-mode model, $\tens{M}_0$, (dashed).}}
     \label{fig:BmodeTest}
 \end{figure} 

As an illustrative example, we take two parameter-free models for COSEBIs B-modes, shown in \fig\ref{fig:BmodeTest}: $\MM_0$ where $B_n=0$ for all $n$, and $\MM_{\rm PSF}$ where $B_n$ corresponds to the measured PSF-leakage systematic defined in \sect\ref{sec:PSFLeakage}.  The difference between these two models is captured by the first few modes, with almost zero power for $n\gtrsim 10$. We create $10,000$ random samples of $B_n$ for the full angular range of $[0.5', 100']$ given the model $\MM_{\rm PSF}$ and the KiDS noise-only covariance for the non-tomographic case. \fig\ref{fig:BmodeTest} shows the mean of these samples (red squares) with errors corresponding to a single sample as well as the input model (blue curve).

We can determine which of the two models best represents the data using a Bayesian evidence analysis. If we give the same weight to both models then the ratio of the Bayesian evidences for these models is given by the Bayes factor,
\begin{equation}
\label{eq:BayesFactor}
{\rm Bayes \; Factor}=\frac{P(D|\MM_1)}{P(D|\MM_2)}=\frac{\int P(D|\MM_1,\Phi_1)\,P(\Phi_1 |\MM_1)\, \d\Phi_1}{\int P(D|\MM_2,\Phi_2)\,P(\Phi_2 |\MM_2)\, \d\Phi_2}\;,
\end{equation}
where $D$ is the data, $\MM_i$ is model $i$ and $\Phi_i$ represents the set of parameters for model $i$. For the simplified case of parameter free models that we consider here,  \Eqt\ref{eq:BayesFactor} simplifies to,
\begin{equation}
\label{eq:SimpleBayesFactor}
{\rm Bayes\; Factor}=\frac{P(D|\MM_0)}{P(D|\MM_{\rm PSF})}\;.
\end{equation}
The resulting Bayes factor will however depend on the number of $n$-modes that are included in the analysis. The Bayesian evidence can only be used when an alternative model exists, but in the case of null tests, such as a B-mode test, the only available model is the null hypothesis and therefore we need to use classical methods to identify the significance of the B-modes. Here we use $\chi^2$ and $p$-values to test the null hypothesis. The $p$-value for the $\chi^2$ is defined as the probability of calculating a $\chi^2$ value larger than the measured one, $\chi^2_{\rm m}$, given the model $\MM$,
\begin{equation}
p{\rm{\text - value}}={\rm Pr}(\chi^2>\chi^2_{\rm m}|\MM)= \int_{\chi^2_{\rm m}}^\infty\d\chi^2\; {\rm Pr}(\chi^2|\MM)\;.
\end{equation}

\begin{figure*}
   \begin{center}
     \begin{tabular}{c}
     \includegraphics[width=\hsize]{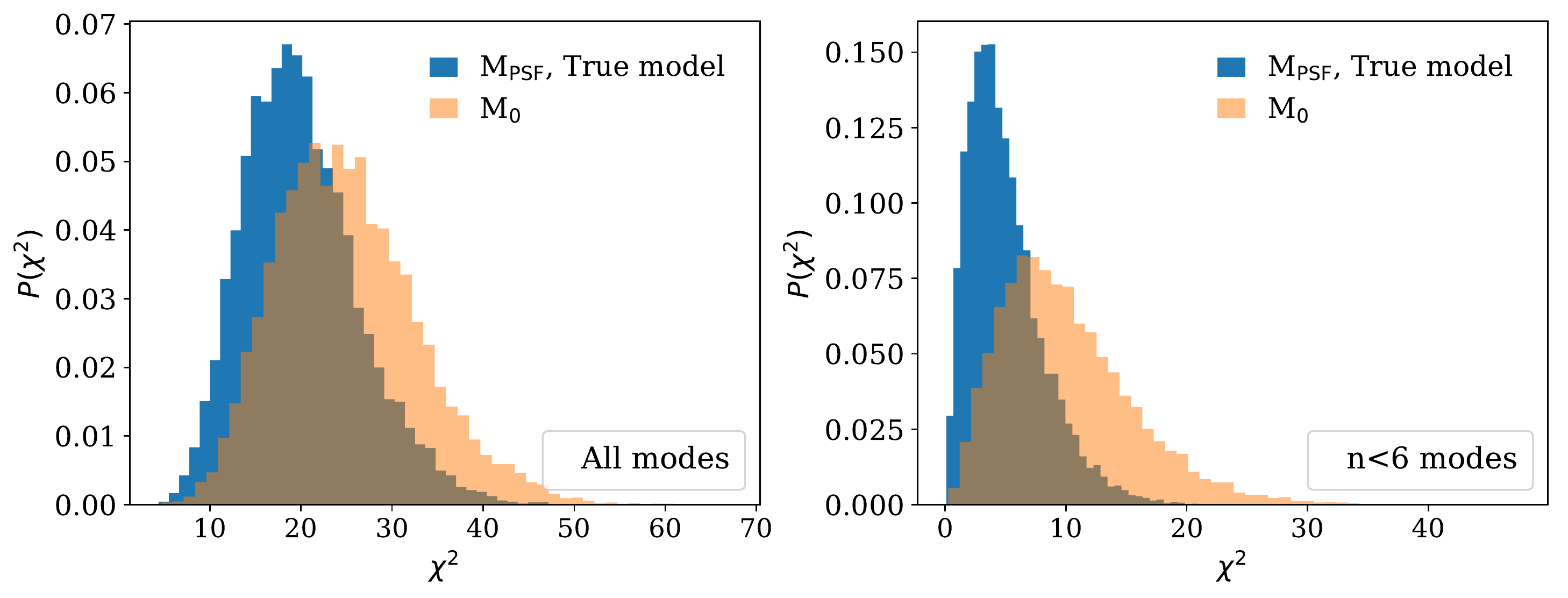}
     \end{tabular}
   \end{center}
     \caption{\small{$\chi^2$ distribution of the mock data given the true PSF leakage model, $\tens{M}_{\rm PSF}$, (blue histogram) or given the null model, $\tens{M}_0$, (orange histogram). The left panel shows the analysis of the $n \le 20$ COSEBIs modes.  In the right panel only the $n \le 5$ modes are considered. }}
     \label{fig:chiSBmodeTest}
 \end{figure*} 

\fig\ref{fig:chiSBmodeTest} shows the distribution of the measured $\chi^2$ across our $10,000$ random samples when the data are fit using the input $\MM_{\rm PSF}$ model (blue histogram) and the $\tens{M}_0$ no systematics model (orange histogram).  In the left panel we take the null-test case where all modes up to $n=20$ are included in the analysis (All-modes).  In the right panel, only the first 5 modes ($n<6$) are analysed.   As $\MM_{\rm PSF}$ is the correct model, we naturally find better fits to the data, i.e. lower $\chi^2$ values, for this model.  The difference between the two distributions for the $\chi^2$ values is however enhanced when the modes analysed are limited to the range where the two models differ significantly.   This means that the power of the null test is optimised over this reduced, $n \le 5$, range.

 \begin{figure*}
   \begin{center}
     \begin{tabular}{c}
     \includegraphics[width=\hsize]{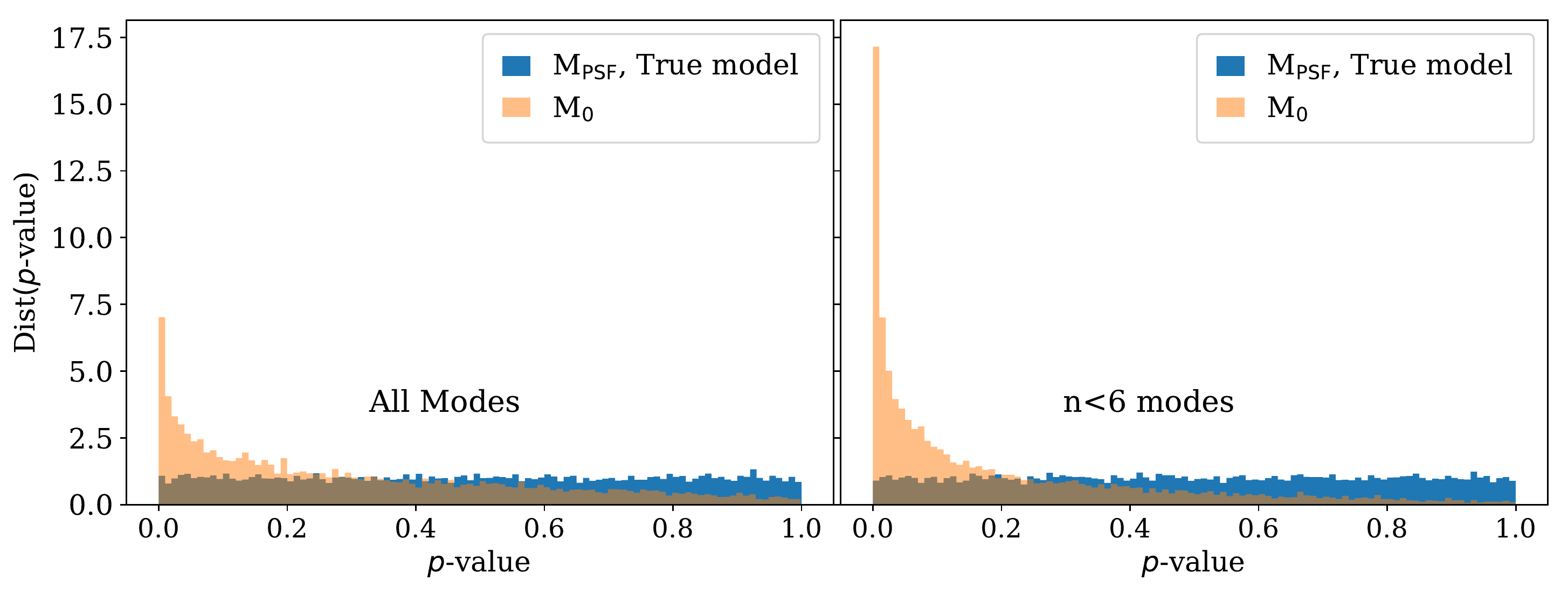}
     \end{tabular}
   \end{center}
     \caption{\small{The distribution of $p$-values for the 10,000 data samples, showing the probability that the $\MM_{\rm PSF}$ model (blue histogram) or the $\tens{M}_0$ no systematics model (orange histogram) is the true underlying model, given each data sample.  The left panel shows the $p$-values from an analysis of the $n \le 20$ COSEBIs modes.  In the right panel only the $n \le 5$ modes are considered.  }}
     \label{fig:pvalueBmodeTest}
 \end{figure*} 
 
\fig\ref{fig:pvalueBmodeTest} shows the distribution of $p$-values for the $\chi^2$ values shown in \fig\ref{fig:chiSBmodeTest}.  
If the model used to fit the data is the true underlying model, any particular $p$-value is as likely to be measured as the other.  If the model is not representative of the data, however, then one is more likely to obtain smaller $p$-values from the sample. As expected with $\MM_{\rm PSF}$ as the correct model, we find a uniform distribution of $p$-values and a skewed distribution for the $\MM_0$ model.  When all 20 COSEBIs modes are included this $p$-value distribution is less skewed compared to when we only include the $n \le 5$ modes.   By adding more data points to the analysis, we have diluted the systematic signal of the PSF leakage, making this null-test less effective.

Based on this analysis, we must recognise that finding that the B-modes pass a null-test using a large data vector does not ensure that analysing a smaller dataset will give the same result.  A good example of this is KiDS-450 passing the 20-mode COSEBIs null-test, but failing the CCOSEBIs null-test which is most sensitive to the $n \le 5$ modes.   In contrast DES-SV and CFHTLenS fail the 20-mode COSEBIs null-test, even though they pass the CCOSEBIs null-test.  Their B-modes therefore appear when adding in more data points to the analysis.   As our example shows how increasing the size of your data set serves to reduce the stringency of the null-test,  we can therefore conclude that the significant DES-SV and CFHTLenS B-mode, seen with COSEBIs and not with CCOSEBIs, is present in the high-$n$ data that is not included in the CCOSEBIs analysis.  If we had only performed a COSEBIs null-test, we would have missed the presence of a systematic signal in KiDS.  If we had only performed a low-$n$ CCOSEBIs null-test, we would have missed the presence of a systematic signal in DES-SV and CFHTLenS.

To illustrate our discussion of null-tests we have used COSEBIs, but the concept holds for any statistic or null-test.   If a systematic produces a feature at a particular scale, but is otherwise identical to the standard model, by adding data from other scales we will dilute the power of the statistical test to distinguish between the two cases.   As null B-mode tests are generally performed independently of alternative models, it is not clear which data points should be added to the null-test analysis.   We therefore propose that future null-tests are performed with the B-mode signatures shown in \sect\ref{sec:ResultsMock} in mind.  In this way one can optimise the modes over which to carry out a model comparison.

%% file: Appendix_Data.tex
\begin{figure*}
   \begin{center}
     \begin{tabular}{c}
     \resizebox{150mm}{!}{\includegraphics{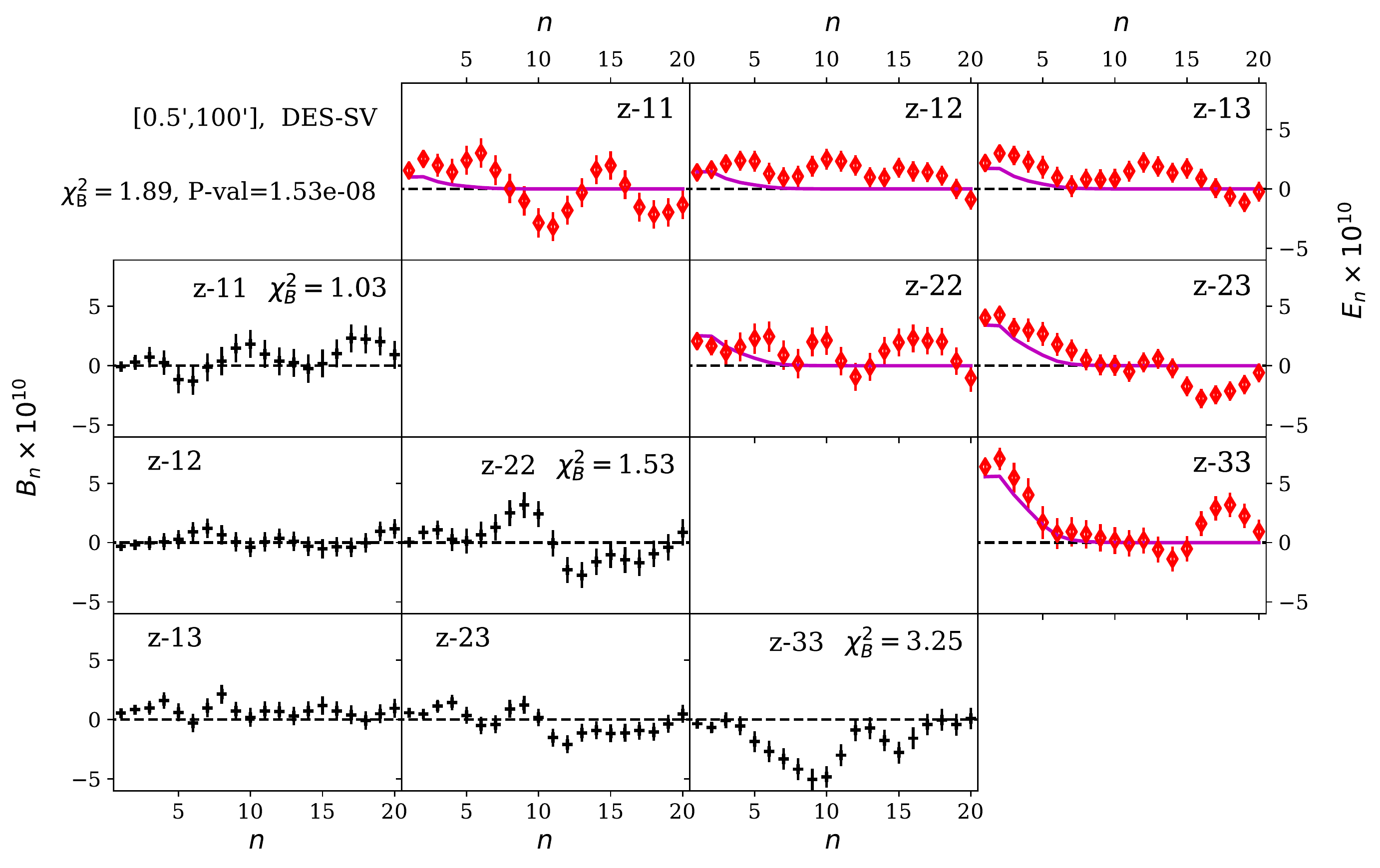}}     \end{tabular}
   \end{center}
     \caption{\small{Tomographic E/B mode COSEBIs analysis of DES-SV, spanning an angular range of $[0.5',100']$. Each panel shows the COSEBIs modes for the tomographic redshift bin pair z$-ij$, corresponding to the correlation between photometric redshift bins $i$ and $j$ (see \sect\ref{sec:Data} for the definition of the redshift bins).  The E-modes (upper right) can be compared to the theoretical expectation given by the cosmological parameters listed in Table~\ref{tab:CosmoParam}.  Note that COSEBIs modes are discrete and we only connect the theoretical model in a curve as a visual aid. The B-modes (lower left) can be compared to the null-case (dashed) where the reduced $\chi^2$ value for the B-modes being equal to zero is given, for the auto-correlation cases, in their corresponding panels. The reduced  $\chi^2$ and $p$-value of the full data vector is listed in the upper left.}}
     \label{fig:COSEBIsDES-SVtomo}
 \end{figure*}
\begin{figure*}
   \begin{center}
     \begin{tabular}{c}
     \resizebox{160mm}{!}{\includegraphics{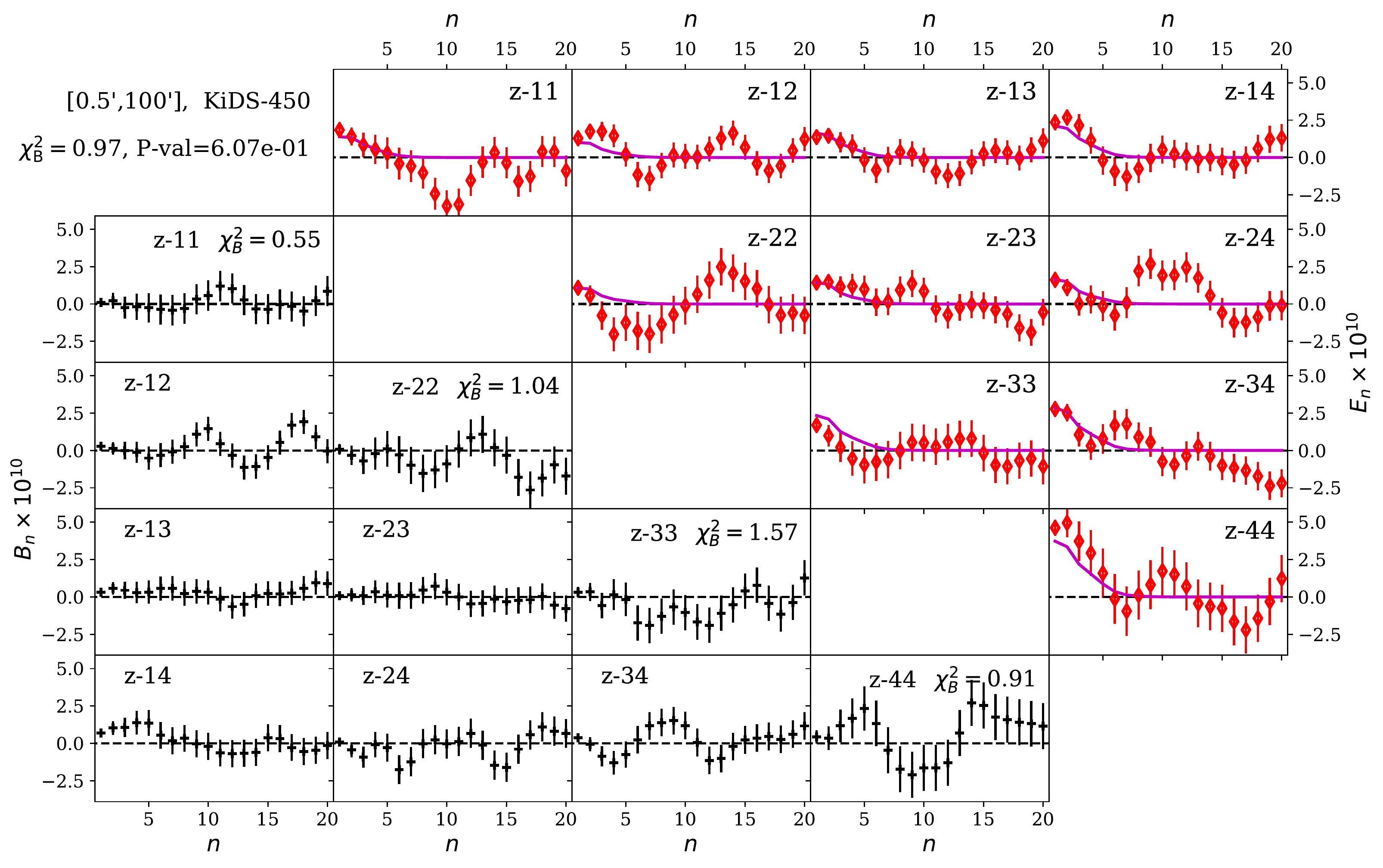}}
          \end{tabular}
   \end{center}
     \caption{\small{Tomographic E/B mode COSEBIs analysis of KiDS-450. See the caption of \fig\ref{fig:COSEBIsDES-SVtomo} for details.}}
     \label{fig:COSEBIsKiDStomo}
 \end{figure*}
\begin{figure*}
   \begin{center}
     \begin{tabular}{c}
     \resizebox{180mm}{!}{\includegraphics{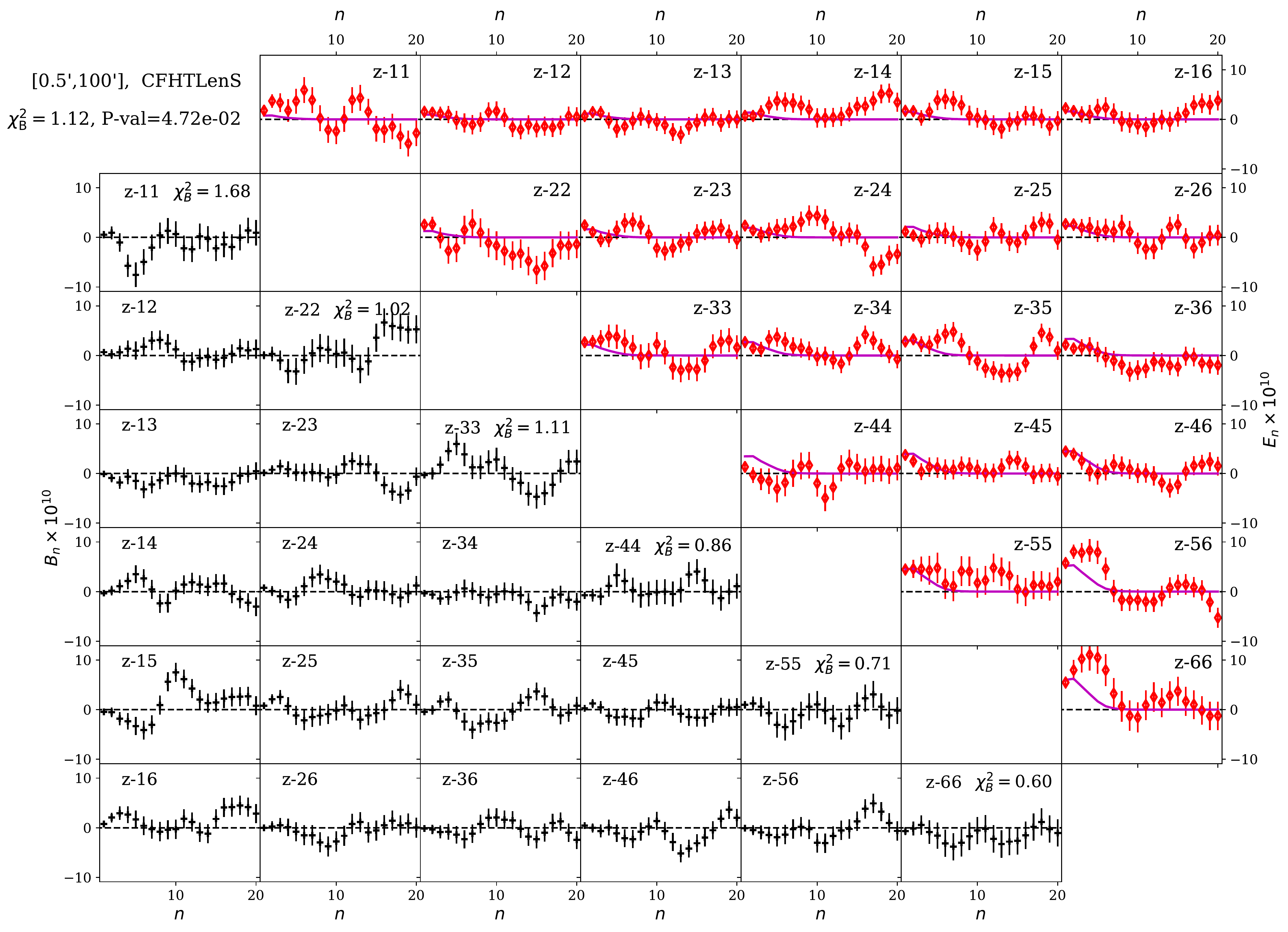}}     \end{tabular}
   \end{center}
     \caption{\small{Tomographic E/B mode COSEBIs analysis of CFHTLenS. See the caption of \fig\ref{fig:COSEBIsDES-SVtomo} for details. }}
     \label{fig:COSEBIsCFHTLenStomo}
 \end{figure*}

\label{sec:AppData}
Figures \ref{fig:COSEBIsDES-SVtomo}, \ref{fig:COSEBIsKiDStomo} and \ref{fig:COSEBIsCFHTLenStomo} show the tomographic COSEBIs measurements, using the angular range of $[0.5',100']$, for DES-SV, KiDS-450 and CFHTLenS respectively.   In each figure, the upper panels present the E-modes, the lower panels present the B-modes, and the significance of the B-modes are indicated with a $p$-value shown in the upper left corner.  The $p$-values for the other two angular ranges analysed are given in \tab\ref{tab:pvalue}. The predicted E-modes, given the best-fitting cosmology parameters listed in Table~\ref{tab:CosmoParam}, are shown as curves.  
 
\begin{figure*}
   \begin{center}
     \begin{tabular}{c}
     \resizebox{150mm}{!}{\includegraphics{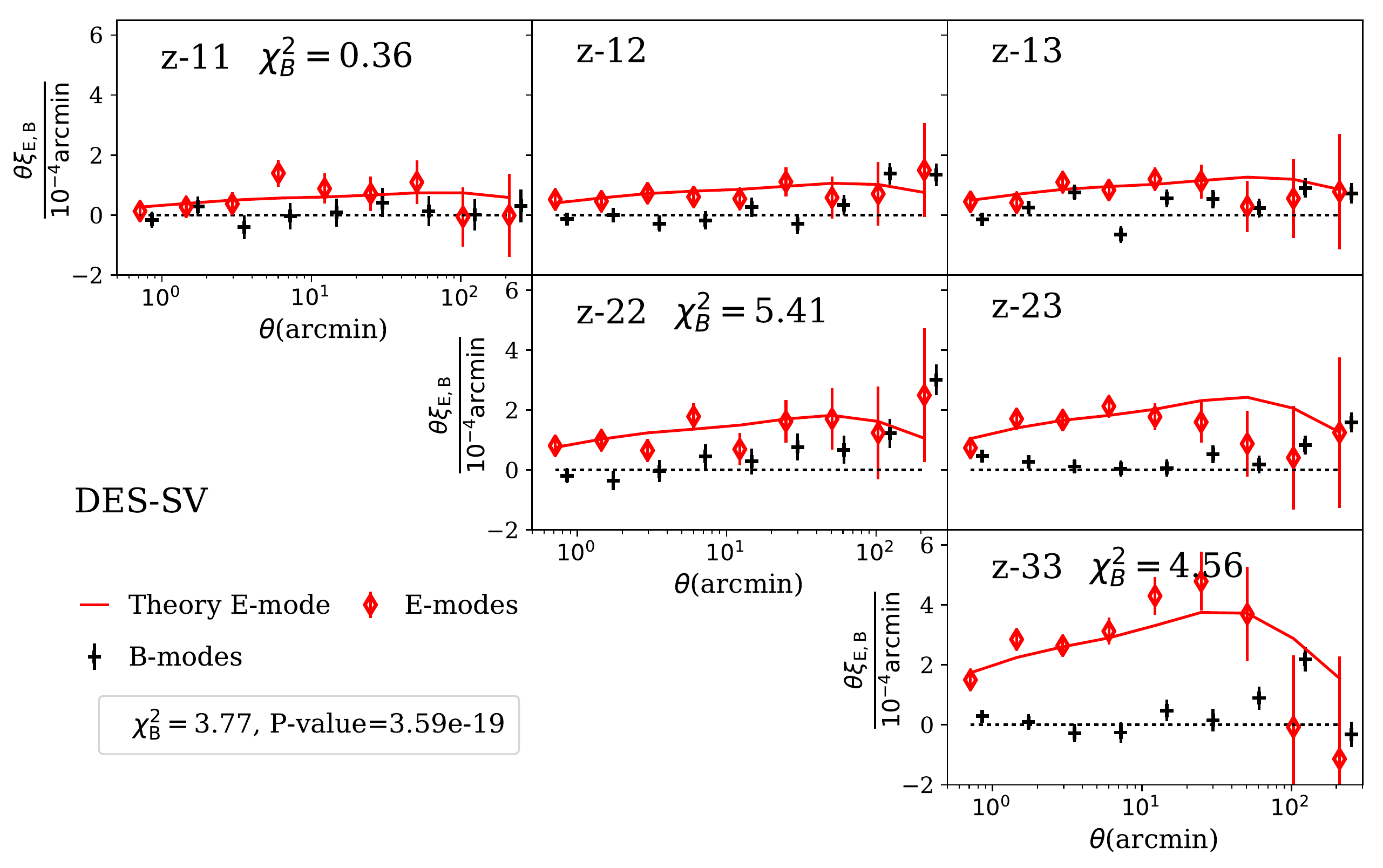}}
     \end{tabular}
   \end{center}
     \caption{\small{Tomographic $\xi_{\rm E/B}$ analysis of DES-SV. 
     Each panel shows $\xi_{\rm E}$ (diamonds) and $\xi_{\rm B}$ (pluses) for the tomographic redshift bin pair z$-ij$, corresponding to the correlation between photometric redshift bins $i$ and $j$ (see \sect\ref{sec:Data} for the definition of the redshift bins).  The E-modes, $\xi_{\rm E}$, can be compared to the theoretical expectation given by the cosmological parameters listed in Table~\ref{tab:CosmoParam}.  The B-modes, $\xi_{\rm B}$, can be compared to the null-case (dashed) where, for the auto-correlation cases, the reduced $\chi^2$ value for the B-modes being equal to zero is listed. The reduced  $\chi^2$ and $p$-value of the full data vector is listed in the lower left. }}
     \label{fig:XiDES-SVtomo}
 \end{figure*}

\begin{figure*}
   \begin{center}
     \begin{tabular}{c}
     \resizebox{160mm}{!}{\includegraphics{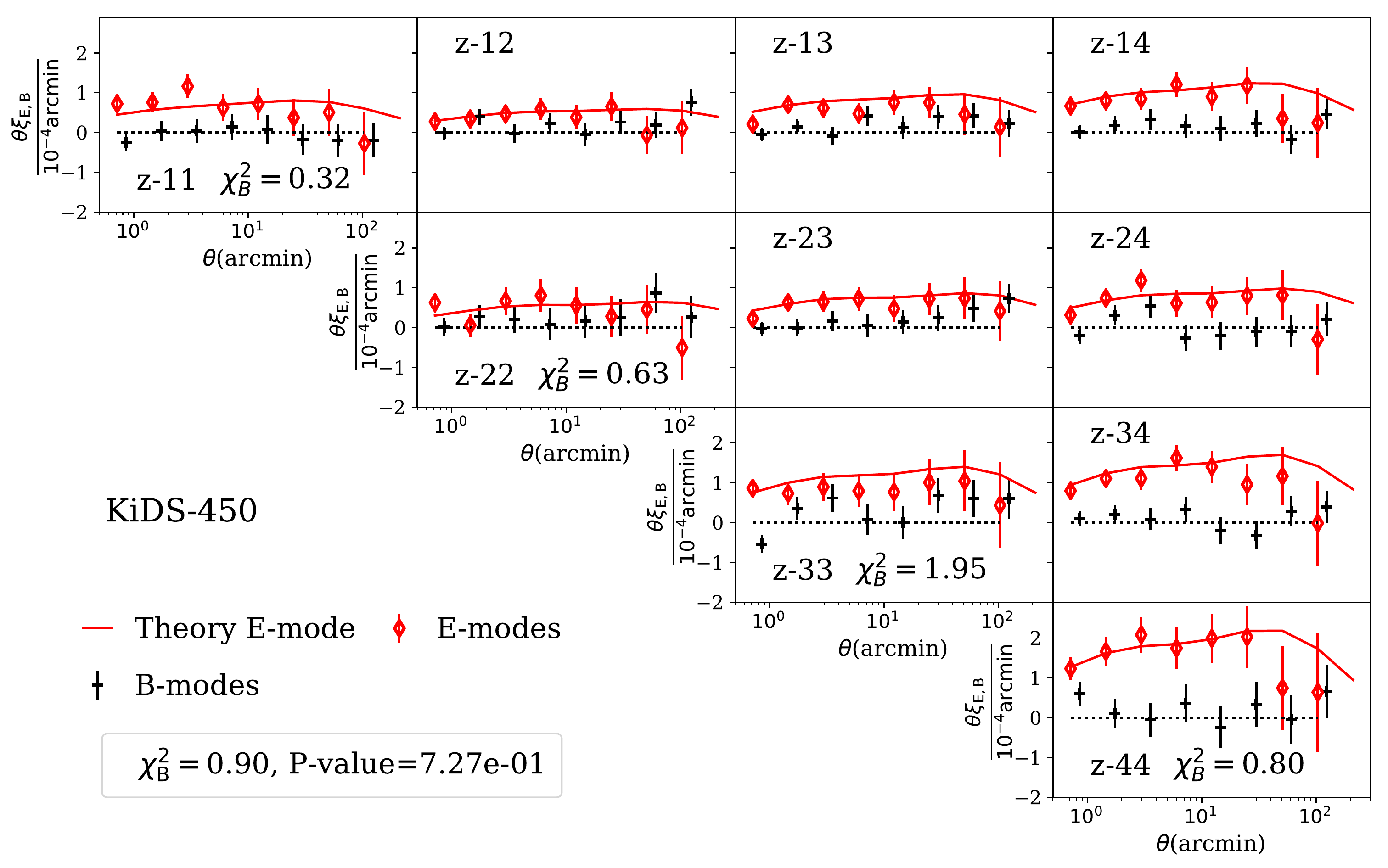}}
          \end{tabular}
   \end{center}
     \caption{\small{Tomographic $\xi_{\rm E/B}$ analysis of KiDS-450. See the caption of \fig\ref{fig:XiDES-SVtomo} for details.}}
     \label{fig:XiKiDStomo}
 \end{figure*}

\begin{figure*}
   \begin{center}
     \begin{tabular}{c}
     \resizebox{180mm}{!}{\includegraphics{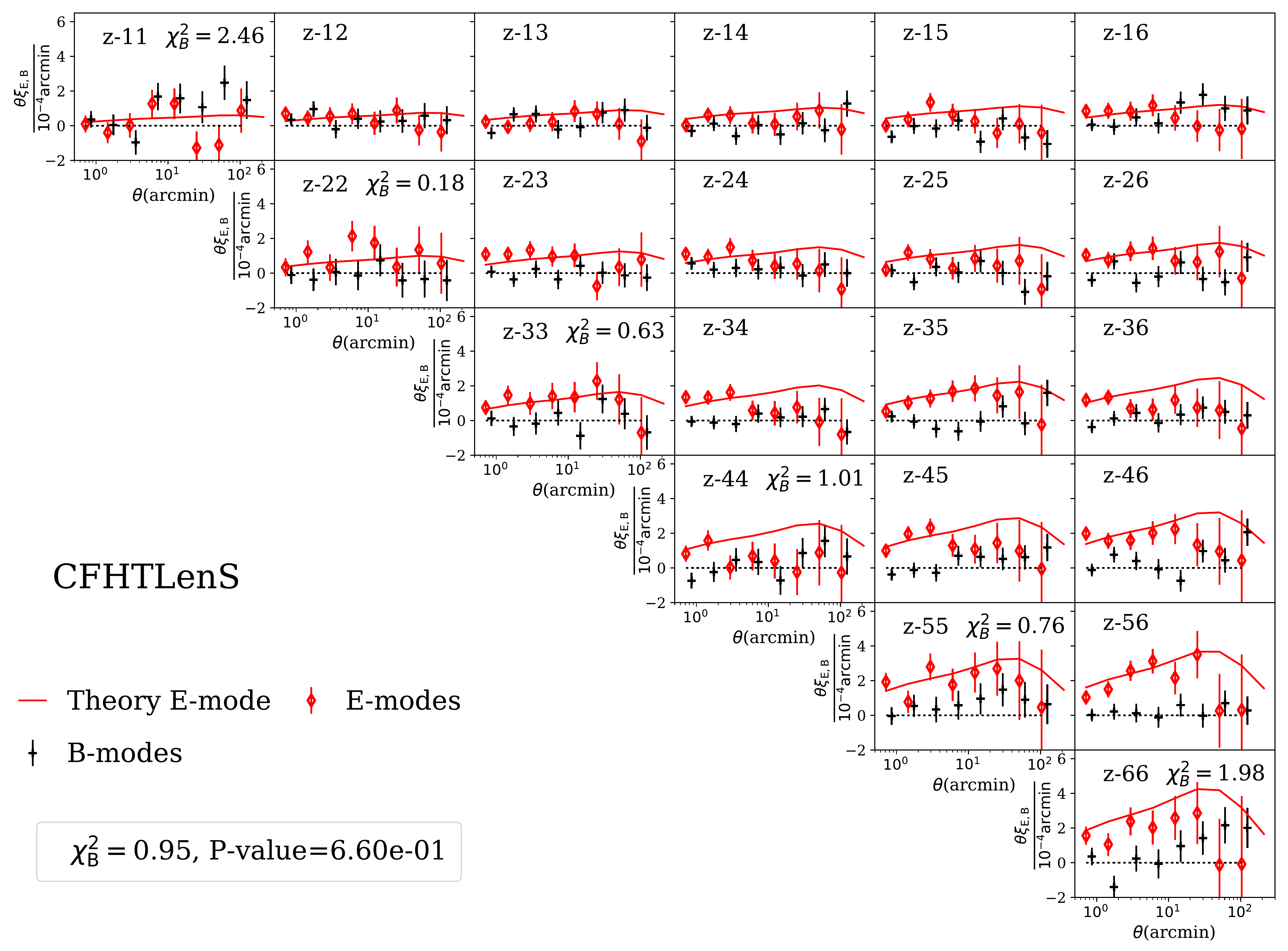}}
          \end{tabular}
   \end{center}
     \caption{\small{Tomographic $\xi_{\rm E/B}$ analysis of CFHTLenS. See the caption of \fig\ref{fig:XiDES-SVtomo} for details. }}
     \label{fig:XiCFHTLenStomo}
 \end{figure*}

Figures \ref{fig:XiDES-SVtomo}, \ref{fig:XiKiDStomo} and \ref{fig:XiCFHTLenStomo} show $\xi_{\rm E/B}$ for the tomographic cases for DES-SV, KiDS-450 and CFHTLenS respectively.  We show $p$-values for the significance of the B-modes in each figure, but caution the reader that due to binning and the truncated integrals discussed in \sect\ref{sec:ResultsData}, this method is not robust. However, as $\xi_{\rm B}$ data points are uncorrelated, they can help with identifying the source of the systematic even though it was seen in \sect\ref{sec:ResultsMock} that systematics do not always affect the same angular ranges for E and B-modes. The prediction for $\xi_{\rm E}$, given the best-fitting cosmology parameters listed in Table~\ref{tab:CosmoParam}, is shown as curves.

For DES-SV, we note that the significance of the tomographic $\xi_{\rm B}$ signal significantly decreases when we restrict the analysis to an angular range of $[4.2',72']$, as adopted by \citet{DES2016}, with the $p$-value increasing from $p = 4 \times 10^{-19}$ to $p=0.012$.   If the systematics that source the B-modes detected in the standard $[0.5',100']$ analysis add equally to the E and B modes, then the chosen DES-SV angular selection would serve to mitigate the impact of these systematics.   As shown in \sect\ref{sec:ResultsMock}, however, we find that the range of tested systematics exhibit different E and B mode responses.   We would therefore caution against concluding that a choice selection of angular scales, based on the B-mode response, is sufficient to remove the systematic contamination to the E-modes within those chosen scales.

 \begin{table}
\centering
\caption{\small{Best fitting shear calibration and photo-z biases for different redshift bins in the DES-SV data (from private communication with Joe Zuntz). }}
\label{tab:calibDES}
\resizebox{\columnwidth}{!}{\begin{tabular}{ c   c  c  c  c  }
                                               &     $z_1$       &  $z_2$         & $z_3$          &  Single bin     \\
  \hline 
  Shear calibration bias               &  0.0163         &  $-0.0051$    &  $-0.0058 $   &   0.0163    \\
  \hline
  Photo-z bias                            &  0.0314        &  $-0.0138$     &   0.0106   &     0.0   \\
  \hline
\end{tabular}}
\end{table}

In \tab\ref{tab:calibDES} we list the best-fitting values of the calibration parameters for DES-SV used to calculate the E-mode predictions for DES-SV  shown in \sect\ref{sec:ResultsData} and this Appendix . The first row shows the value of the multiplicative shear calibration bias and the second row the additive photometric redshift bias for redshift bins one to three. The last column shows the values we adopted for the single bin case, which was not analysed in \cite{DES15_CP}. For this case we use a multiplicative shear calibration equal to the first redshift bin value and a vanishing photometric redshift bias. 

\tab\ref{tab:FluxErr} lists the fitted values for $a_{\rm x}$ and $b_{\rm x}$ to KiDS-450 multi-band data using \Eqt\eqref{eqn:fluxpsfcorr}. This values are used to produce a correlation between the measured ellipticity of galaxies that are binned in photometric redshift bins with their local PSF ellipticity (see \sect\ref{sec:sysZPSF} and \sect\ref{sec:resultZdep}).

\begin{table}
\centering
\caption{\small{Best fitting values for  $a_{\rm x}$ and $b_{\rm x}$ as defined in \Eqt\ref{eqn:fluxpsfcorr}, fitted to KiDS-450 data. In this equation the signal-to-noise of measured fluxes are written as a linear function of $|\epsilon-\epsilon^*_{\rm x}|$, where $\epsilon$ is the observed ellipticity of galaxies and  $\epsilon^*_{\rm x}$ is the PSF ellipticity in photometric band x=$u,\; g\;,r\;,\; i$.  }}
\label{tab:FluxErr}
% \resizebox{\columnwidth}{!}
{\begin{tabular}{ c   c  c  c  c  }
                &   $u$-band  &  $g$-band    & $r$-band       &  $i$-band    \\
  \hline 
  $a_{\rm x}$   &   $-3.5$    &   $-17$      &   $-37$        &   $-19 $   \\
  \hline
  $b_{\rm x}$   &    5.5        &   23           &   47             &   23     \\
  \hline
\end{tabular}}
\end{table}

\begin{figure*}
   \begin{center}
     \begin{tabular}{c}
     \resizebox{180mm}{!}{\includegraphics{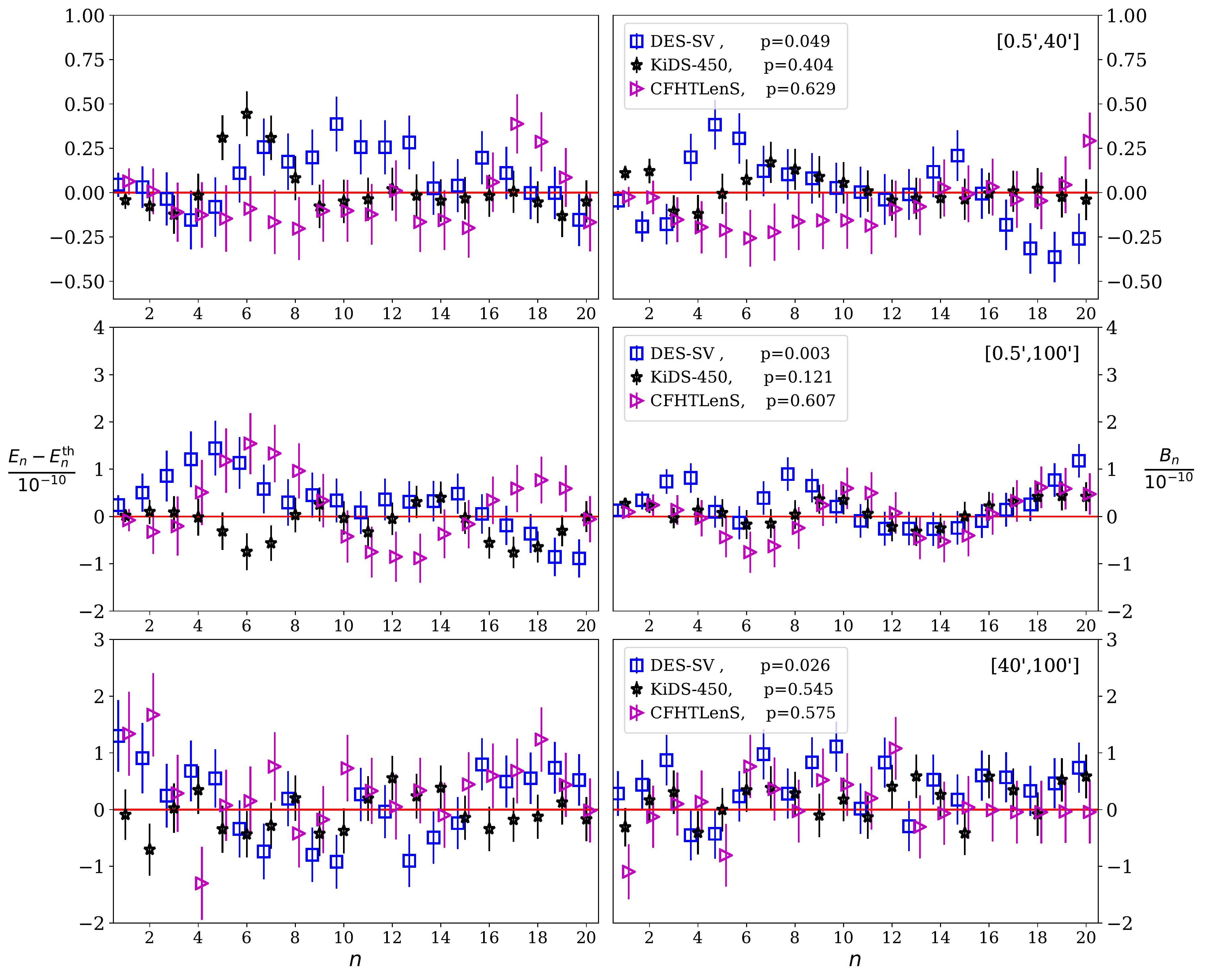}}
     \end{tabular}
   \end{center}
     \caption{\small{{The difference between COSEBIs E-modes and their theory value (left) and COSEBIs B-modes (right) for a single broad redshift bin. The only difference between this figure and \fig\ref{fig:COSEBIsdata1bin}, is the left column. Results for DES-SV are shown with blue squares, KiDS-450 with black stars and CFHTLenS with magenta triangles. The angular ranges are shown for each row in the upper right corner. In addition, the significance of the B-modes is shown as $p$-values for each survey and angular range. E-mode predictions are calculated using the best fitting cosmological parameter values given in \tab\ref{tab:CosmoParam} for DES-SV (solid), KiDS-450 (dashed) and CFHTLenS (dotted). Note that COSEBIs modes are discrete and the theory values are connected to each other only as a visual aid. A zero-line is also shown for reference.}}}
\label{fig:COSEBIsdata1bin_2}
 \end{figure*}

{In \fig\ref{fig:COSEBIsdata1bin_2} we replot the left hand side of \fig\ref{fig:COSEBIsdata1bin} to show the difference between the measured and the fiducial E-modes, $E_n-E_n^{\rm th}$, for a single redshift bin, while keeping the right hand side as it was since B-modes are expected to be consistent with zero. This figure has a similar format to \fig\ref{fig:COSEBIsSys}, where we showed the effect of systematics on simulations. When comparing these figures note that $E_n^{\rm th}$ in \fig\ref{fig:COSEBIsdata1bin_2} is not the input theory, but instead it is the result of a fit to the data which is inevitably affected by any systematics that may exist in the data. In contrast in \fig\ref{fig:COSEBIsSys} we know the correct values for the $E_n^{\rm fid}$, as they are measured directly from the simulations before the systematic effect is applied to them.}

